\documentclass[12pt]{elsarticle}
\usepackage{amssymb}
\usepackage{txfonts}
\usepackage{mathbbold}
\usepackage{amsfonts}
\usepackage{mathrsfs}
\usepackage{epsfig,bm,dcolumn}
\usepackage{graphicx}
\usepackage{color}
\usepackage{dcolumn}
\usepackage{overpic}

\journal{Annals of Physics (N. Y.)}
\begin{document}

\begin{frontmatter}
\title{Dynamic density and spin responses of a superfluid Fermi gas in the BCS-BEC crossover:\\
Path integral formulation and pair fluctuation theory}
\author{Lianyi He}
\address{1 Theoretical Division, Los Alamos National Laboratory, Los Alamos, NM 87545, USA\\
2 Department of Physics and Collaborative Innovation Center for Quantum Matter, Tsinghua University, Beijing 100084, China}

\date{\today}

\begin{abstract}
We present a standard field theoretical derivation of the dynamic density and spin linear response functions of a dilute superfluid Fermi gas in the BCS-BEC crossover in both three and two dimensions. The derivation of the response functions is based on the elegant functional path integral approach which allows us to calculate the density-density and spin-spin correlation functions by introducing the external sources for the density and the spin density. Since the generating functional cannot be evaluated exactly, we consider two gapless approximations which ensure a gapless collective mode (Goldstone mode) in the superfluid state: the BCS-Leggett mean-field theory and the Gaussian-pair-fluctuation (GPF) theory.  In the mean-field theory, our results of the response functions agree with the known results from the random phase approximation. We further consider the pair fluctuation effects and establish a theoretical framework for the dynamic responses within the GPF theory.  We show that the GPF response theory naturally recovers three kinds of famous  diagrammatic contributions: the Self-Energy contribution,  the Aslamazov-Lakin contribution, and the Maki-Thompson contribution. We also show that unlike the equilibrium state, in evaluating the response functions,  the linear (first-order) terms in the external sources as well as the induced order parameter perturbations should be treated carefully.  In the superfluid state, there is an additional order parameter contribution which ensures that in the static and long wavelength limit, the density response function recovers the result of the compressibility (compressibility sum rule).  We expect that the $f$-sum rule is manifested by the full number equation which includes the contribution from the Gaussian pair fluctuations. The dynamic density and spin response functions in the normal phase (above the superfluid critical temperature) are also derived within the Nozi\`{e}res- Schmitt-Rink (NSR) theory.
\end{abstract}

\end{frontmatter}

\tableofcontents

\section{Introduction}
The experimental realization of ultracold atomic Fermi gases with tunable interatomic interactions has opened a new era for the study
of some longstanding theoretical proposals in many-fermion systems.  One interesting proposal is the smooth crossover from a
Bardeen-Cooper-Schrieffer (BCS) superfluid  state with largely overlapping Cooper pairs to a Bose-Einstein condensate (BEC) of
tightly bound bosonic molecules -- a phenomenon suggested many years ago~\cite{Eagles,Leggett,NSR}. A simple but important system is
a dilute attractive Fermi gas in three dimensions (3D), where the effective range of the short-ranged interaction is much smaller
than the interparticle distance. The system can be characterized by a dimensionless gas parameter $1/(k_{\rm F}a_{\rm 3D})$, where
$a_{\rm 3D}$ is the $s$-wave scattering length of the short-ranged interaction and $k_{\rm F}$ is the Fermi momentum in the absence
of interaction. The BCS-BEC crossover occurs when the parameter $1/(k_{\rm F}a_{\rm 3D})$ is tuned from negative to positive values
~\cite{BCSBEC1,BCSBEC2,BCSBEC3,BCSBEC4,BCSBEC5}, and the BCS and BEC limits correspond to the cases 
$1/(k_{\rm F}a_{\rm 3D})\rightarrow-\infty$ and $1/(k_{\rm F}a_{\rm 3D})\rightarrow+\infty$, respectively.

The BCS-BEC crossover phenomenon in 3D dilute Fermi gases has been experimentally demonstrated by using ultracold gases of $^6$Li and
$^{40}$K atoms~\cite{BCSBECexp1,BCSBECexp2,BCSBECexp3}, where the $s$-wave scattering length and hence the gas parameter
$1/(k_{\rm F}a_{\rm 3D})$ were tuned by means of the Feshbach resonance~\cite{FR1,FR2}. The equation of state and various static and
dynamic properties of the BCS-BEC crossover have become a big challenge for quantum many-body
theory~\cite{TH01,TH02,TH03,TH04,TH05,TH06,TH07,TH08,TH09,TH10,TH11}
because the conventional perturbation theory is no longer valid. At the so-called unitary point where $a_{\rm 3D}\rightarrow\infty$, the only
length scale of the system is the inter-particle distance. Therefore, the properties of the system at the unitary point $1/(k_{\rm F}a_{\rm 3D})=0$
become universal, i.e., independent of the details of the interactions. All thermodynamic quantities, scaled by their counterparts for the
non-interacting Fermi gases, become universal constants. Determining these universal constants has been one of the most intriguing topics
in the research of the cold Fermi gases~\cite{EOSexp1,EOSexp2,EOSexp3,EOSmc1,EOSmc2,EOSmc3,EOSmc4}.  On the other hand, it was 
suggested that a 2D Fermi gas with short-ranged $s$-wave attraction can also undergo a BCS-BEC
crossover~\cite{BCSBEC2D-1,BCSBEC2D-2,BCSBEC2D-3}. Unlike 3D, a two-body bound state always exists in 2D even though the attraction is
arbitrarily weak. The BCS-BEC crossover in 2D can be realized by tuning the binding energy of the bound state. Studying the BCS-BEC crossover 
in 2D will help us understand the physics of pseudogap and Berezinskii-Kosterlitz-Thouless transitions in fermionic systems~\cite{BCSBEC2D-4}.
In recent years, quasi-2D atomic Fermi gases have been experimentally realized and studied by a number of groups~
\cite{2Dexp1,2Dexp2,2Dexp3,2Dexp4,2Dexp5,2Dexp6,2Dexp7,2Dexp8,2Dexp9,2Dexp10,2Dexp11}. 

The simplest theoretical description of the superfluid ground state of the BCS-BEC crossover is the BCS-Leggett mean-field theory~\cite{Leggett}. 
It is known that in 3D,  even the mean-field theory predicts that the system is a weakly interacting Bose condensate in the strong attraction 
limit~\cite{BCSBEC2}. The composite boson scattering length is shown to be $a_{\rm B}=2a_{\rm 3D}$~\cite{BCSBEC2}. The
inclusion of Gaussian pair fluctuations~\cite{TH03,TH04,TH05} recovers the Fermi liquid corrections in the weak attraction limit and modifies the composite boson scattering length to $a_{\rm B}\simeq0.55a_{\rm 3D}$, which is close to the exact result $a_{\rm B}\simeq0.6a_{\rm 3D}$~\cite{FB3D}.  Moreover, the equation of state (EOS) in the BCS-BEC crossover agrees excellently with the quantum Monte Carlo results and the experimental measurements if the Gaussian pair fluctuations are taken into account~\cite{TH03,TH04,TH05}. In contrast, the mean-field theory for 2D Fermi gases does not predict a weakly interacting 2D Bose condensate in the strong attraction limit~\cite{BCSBEC2D-2,BCSBEC2D-3}. The coupling constant between the composite bosons is predicted to be energy independent, which arises from the inadequacy of the Born approximation for four-body scattering in 2D. As a result, the 2D mean-field theory predicts that the pressure of a homogeneous 2D Fermi gas is equal to that of a noninteracting Fermi gas in the entire BCS-BEC crossover. However, recent experimental measurements~\cite{2Dexp2,2Dexp6} and quantum Monte Carlo simulations~\cite{QMC2D1,QMC2D2,QMC2D3,QMC2D4,QMC2D5} show that the pressure in the strong attraction limit is vanishingly small in comparison to that of a noninteracting Fermi gas, which is consistent with the picture that the system is a weakly interacting 2D Bose condensate. Recently, the inadequacy of the 2D mean-field theory has been fixed by taking into account the Gaussian pair fluctuations~\cite{GPF2D-1, GPF2D-2}. The application of the Gaussian-pair-fluctuation (GPF) theory to 2D predicts a composite boson scattering length which is very close to the exact result, and the equations of state predicted by the GPF theory agrees well with the experimental measurements and quantum Monte Carlo results~\cite{GPF2D-2}.

In addition to the equation of state and other static properties of the BCS-BEC crossover, it is also interesting to study the dynamic responses to some external perturbations. In this work, we focus on the linear responses to an inhomogeneous density perturbation or a spin density perturbation.  The responses of the system to these inhomogeneous perturbations are characterized by two dynamic response functions, the density response function $\chi_{nn}(\omega, {\bf q})$ and the spin response function  $\chi_{ss}(\omega,{\bf q})$, where $\omega$ and ${\bf q}$ are the frequency and momentum, respectively.  The spectral function of these responses, are the so-called dynamic structure factors, which are usually denoted as $S_{nn}(\omega, {\bf q})$ for the density and $S_{ss}(\omega, {\bf q})$ for the spin. The static structure factors are defined as the frequency integral of the dynamic structure factors. The dynamic structure factors for the density and spin has been experimentally determined by using Bragg spectroscopy~\cite{Bragg-1,Bragg-2,Bragg-3}. The static structure factors has been calculated by using quantum Monte Carlo simulations~\cite{SSF-QMC01,SSF-QMC02}. On the theory side, the dynamic response functions and the structure factors were calculated by using the random phase approximation (RPA)~\cite{RPA-Zwerger, RPA-Hu} and the pseudogap theory~\cite{PG-Levin}, which showed qualitative agreement with experimental measurements.  A diagrammatic approach to study the dynamic responses in the normal phase was also proposed~\cite{DT-Pieri} but so far only the static compressibility and the spin susceptibility have been reported.  In the high temperature regime, the dynamic structure factors were studied by using the virial expansion \cite{Virial01, Virial02}. On the other hand, the relation between the gauge invariance and the sum rules has been discussed~\cite{SUM-1,SUM-2,SUM-3}. The determination the spin response of strongly interacting Fermi gases is of great importance for  the study of the neutrino emissivity in neutron matter~\cite{NM-1,NM-2,NM-3}  and hence the cooling process of the compact stars.

The conventional random phase approximation, which can also be derived from a kinetic equation approach~\cite{KE}, takes into account only the contribution from the fermionic quasiparticles and the coupling to the collective modes (for the density response).  A more precise theory should consider properly the contributions from the pair fluctuations. Since the GPF theory has achieved quantitative success in describing the equation of state for both 3D and 2D systems~\cite{TH03,TH04,TH05,GPF2D-2}, we expect that its application to the dynamic responses will properly take into account the role of pair fluctuations. 
In this work, we present a standard field theoretical derivation of the dynamic response functions by using the elegant functional path integral formalism.
In the path integral formalism, the standard approach to calculate the dynamic response functions is to introduce external sources and calculate the second 
derivative of the partition function with respect to the external sources. However, since the generating functional cannot be evaluated exactly, we need to specify the approximation for the superfluid state in the absence of external sources. In this work, we consider two gapless approximations which ensures the Goldstone theorem in the superfluid state: the BCS-Leggett mean-field theory and the GPF theory. For the mean-field theory, the path integral derivation naturally recovers 
the famous RPA theory.  The response functions in the GPF theory includes not only the RPA contribution but also the contributions from the Gaussian pair fluctuations.  We show that the pair-fluctuation part naturally includes three kinds of famous contributions: the
self-energy contribution,  the Aslamazov-Lakin contribution \cite{AL}, and the Maki-Thompson contribution \cite{MT}. Unlike the equilibrium state, in evaluating the response functions,  the linear (first-order) terms in the external sources as well as the induced order parameter perturbations should be treated carefully. 
In the superfluid state, there is an order parameter contribution which ensures that in the static and long wavelength limit, the density response
function recovers the result of the compressibility (compressibility sum rule).  We expect that the $f$-sum rule is manifested by the full number equation
which includes the contribution from the pair fluctuations.

The paper is organized as follows. In Sec. \ref{s2}, we briefly review the BCS-Leggett mean-field theory of the BCS-BEC crossover, including the ground state and the collective modes, and introduce the GPF theory. In Sec. \ref{s3}, we present the general definition of the dynamic response functions in the path integral formalism. In Sec. \ref{s4}, we present the derivation of the response functions within the BCS-Leggett mean-field theory and show that it recovers the RPA. 
In Sec. \ref{s5}, we derive the response functions of within the GPF theory. Most of the results in Sec. \ref{s4} and Sec. \ref{s5} are presented for the ground state (zero temperature), however, their generalization to finite temperature is straightforward.
We also present the results of the response functions above the superfluid transition temperature in 3D systems by using the Nozi\`{e}res- Schmitt-Rink (NSR) theory in Sec. \ref{s6}. We summarize in Sec. \ref{s7}.

\section{Theory of BCS-BEC crossover: Gapless approximations}\label{s2}

\subsection{Hamiltonian and renormalization}\label{s2-1}

We consider a homogeneous spin-$1/2$ (two-component) Fermi gas with a short-ranged $s$-wave attractive interaction in the spin-singlet channel. In the dilute limit the many-body Hamiltonian of the system can be written as
\begin{eqnarray}\label{Hamiltonian}
H=\int d {\bf r}\sum_{\sigma=\uparrow,\downarrow}\psi_\sigma^{\dagger}({\bf r}) \left(-\frac{\hbar^2\nabla^2}{2m}-\mu\right) \psi_\sigma^{\phantom{\dag}}({\bf r})
-U\int d{\bf r}\ \psi^\dagger_{\uparrow}({\bf r})\psi^\dagger_{\downarrow}({\bf r})
\psi^{\phantom{\dag}}_{\downarrow}({\bf r})\psi^{\phantom{\dag}}_{\uparrow}({\bf r}),
\end{eqnarray}
where $\psi_\sigma^\dagger({\bf r})$ and $\psi_\sigma({\bf r})$ represents the creation and annihilation field operators for the two-component fermions, $m$ is the fermion mass, and $\mu$ is the chemical potential. For convenience we use the contact coupling $U$ which denotes the attractive $s$-wave interaction between unlike spins. The cost of the contact coupling is that the Lippmann-Schwinger equation for two-body scattering suffers from ultraviolet
divergence and we need to normalize the bare contact coupling $U$. The units $\hbar=k_{\rm B}=1$ will be used throughout.

With the contact interaction, the Lippmann-Schwinger equation for the two-body $T$ matrix reads
\begin{equation}
T_{\rm 2B}^{-1}(E)=-U^{-1}-{\cal B}(E),
\end{equation}
where $E=k^2/m$ is the scattering energy in the center-of-mass frame and the two-particle bubble function ${\cal B}(E)$ is given by
\begin{equation}
{\cal B}(E)=\frac{1}{V}\sum_{\bf p}\frac{1}{E+i\epsilon-2\varepsilon_{\bf p}}.
\end{equation}
Here $\epsilon=0^+$ and $\varepsilon_{\bf p}={\bf p}^2/(2m)$. We use the standard notation $\frac{1}{V}\sum_{\bf p}\equiv \int d^3{\bf p}/(2\pi)^3$ for three spatial dimensions (3D) and $\frac{1}{V}\sum_{\bf p}\equiv \int d^2{\bf p}/(2\pi)^2$ for two spatial dimensions (2D) with $V$ being the volume of the system. We see clearly that the integral over ${\bf p}$ is UV divergent. We regularize the UV divergence by introducing a hard cutoff $\Lambda$ for $|{\bf p}|$. For large $\Lambda$ we obtain
\begin{equation}
{\cal B}(E)=-\frac{m\Lambda}{2\pi^2}+\frac{m}{4\pi}\sqrt{-m(E+i\epsilon)}
\end{equation}
in 3D and
\begin{equation}
{\cal B}(E)=-\frac{m}{4\pi}\ln\frac{\Lambda^2}{m}+\frac{m}{4\pi}\ln\left(-E-i\epsilon\right)
\end{equation}
in 2D.

Next we match the scattering amplitude $f(k)=(4\pi/m) T_{\rm 2B}(E)$ to the known result. In 3D, the $s$-wave scattering amplitude is given by 
$f(k)=1/(a_{3{\rm D}}^{-1}+ik)$ where $a_{3\rm D}$ is the $s$-wave scattering length. We obtain
\begin{eqnarray}
\frac{1}{U(\Lambda)}=-\frac{m}{4\pi a_{3\rm D}}+\frac{m\Lambda}{2\pi^2}=-\frac{m}{4\pi a_{3\rm D}}+\frac{1}{V}\sum_{|{\bf p}|<\Lambda}\frac{1}{2\varepsilon_{\bf p}}.
\end{eqnarray}
In 2D, the $s$-wave scattering amplitude is given by $f(k)=1/[\ln(\varepsilon_{2\rm D}/E)+i\pi]$ \cite{BCSBEC2D-3}, where $\varepsilon_{2\rm D}$ is the binding energy of the two-body bound state which characterizes the attractive strength. We obtain
\begin{eqnarray}
\frac{1}{U(\Lambda)}=\frac{m}{4\pi}\ln\frac{\Lambda^2}{m\varepsilon_{2\rm D}}=
\frac{1}{V}\sum_{|{\bf p}|<\Lambda}\frac{1}{2\varepsilon_{\bf p}+\varepsilon_{2\rm D}}.
\end{eqnarray}
The above results should be understood in the limit $\Lambda\rightarrow\infty$. After the renormalization of the bare coupling $U$ through the physical scattering length $a_{3\rm D}$ or binding energy $\varepsilon_{2\rm D}$, the UV divergence in the many-body calculations can be eliminated and we can set $\Lambda\rightarrow\infty$ to obtain the final finite result.

\subsection{Functional path integral approach}\label{s2-2}
In the imaginary-time functional path integral formalism, the partition function of the system at finite temperature $T$ is
\begin{eqnarray}
{\cal Z} = \int [d\psi][d\bar{\psi}]\exp\left\{-{\cal S}[\psi,\bar{\psi}]\right\},
\end{eqnarray}
where the action
\begin{eqnarray}
{\cal S}[\psi,\bar{\psi}]=\int_0^\beta d\tau\int d{\bf r}
\bar{\psi}\partial_\tau \psi+\int_0^\beta d\tau H(\psi,\bar{\psi}).
\end{eqnarray}
Here $\tau$ is the imaginary time, $\beta=1/T$, and $H(\psi,\bar{\psi})$ is obtained by replacing the field operators $\psi^\dagger$ and $\psi$ with the Grassmann variables $\bar{\psi}$ and $\psi$, respectively. To decouple the interaction term we introduce the auxiliary complex pairing field $\Phi(x)$ which satisfies the equation of motion $\Phi(x) = -U\psi_\downarrow(x)\psi_\uparrow(x)$ and apply the Hubbard-Stratonovich transformation. Here and in the following $x=(\tau,{\bf r})$ and $\int dx\equiv \int_0^\beta d\tau \int d{\bf r}$. Using the Nambu-Gor'kov spinor
\begin{eqnarray}
\bar{\psi}(x)=\left(\ \bar{\psi}_\uparrow(x)\ \ \psi_\downarrow(x)\
\right),\ \ \ \ \ \
\psi(x)=\left(\begin{array}{cc} \psi_\uparrow(x)\\
\bar{\psi}_\downarrow(x)\end{array}\right),
\end{eqnarray}
we express the partition function as
\begin{eqnarray}
{\cal Z}=\int [d\psi][d\bar{\psi}][d\Phi][d\Phi^*] \exp\Big\{-{\cal S}[\psi,\bar{\psi},\Phi,\Phi^*]\Big\},
\end{eqnarray}
where the action now reads
\begin{eqnarray}
{\cal S}=\int dx\frac{|\Phi(x)|^2}{U}-\int dx\int dx^\prime\bar{\psi}(x){\bf G}^{-1}(x,x^\prime)\psi(x^\prime).
\end{eqnarray}
The inverse Nambu-Gor'kov Green's function ${\bf G}^{-1}(x,x^\prime)$ is given by
\begin{eqnarray}
{\bf G}^{-1}(x,x^\prime)=\left(\begin{array}{cc}-\partial_{\tau}+\frac{\nabla^2}{2m}+\mu &\Phi(x)\\
\Phi^*(x)& -\partial_{\tau}-\frac{\nabla^2}{2m}-\mu\end{array}\right)\delta(x-x^\prime).
\end{eqnarray}
Integrating out the fermion fields, we obtain
\begin{eqnarray}
\mathcal{Z}=\int[d\Phi][d\Phi^*] \exp \Big\{- {\cal S}_{\rm{eff}}[\Phi, \Phi^*]\Big\},
\end{eqnarray}
where the effective action reads
\begin{eqnarray}
{\cal S}_{\rm{eff}}[\Phi, \Phi^*] = \int dx \frac{|\Phi(x)|^{2}}{U} - \mbox{Trln} [{\bf G}^{-1}(x,x^\prime)].
\end{eqnarray}
Here the trace ${\rm Tr}$ is taken in the Nambu-Gor'kov space and the coordinate space.

The effective action ${\cal S}_{\rm{eff}}[\Phi, \Phi^*]$ cannot be evaluated precisely by analytical method. In this work, we consider the superfluid ground state at zero temperature ($T=0$). In the superfluid ground state, the pairing field $\Phi(x)$ acquires a static and uniform expectation value $\langle\Phi(x)\rangle=\Delta$, which serves as the order parameter of the superfluidity. Due to the U$(1)$ symmetry, we can set $\Delta$ to be real without loss of generality. Then we express the pairing field as $\Phi(x)=\Delta+\phi(x)$, where $\phi(x)$ is the fluctuation around the mean field. The effective action ${\cal S}_{\rm{eff}}[\Phi,\Phi^*]$ can be expanded in powers of the fluctuations $\phi(x)$ and $\phi^*(x)$; that is,
\begin{eqnarray}
{\cal S}_{\rm{eff}}[\Phi,\Phi^*]={\cal S}_{\rm MF}+{\cal S}_{\rm GF}[\phi,\phi^*]+\cdots,
\end{eqnarray}
where ${\cal S}_{\rm MF}\equiv{\cal S}_{\rm eff}[\Delta,\Delta]$ is the mean-field (MF) effective action and ${\cal S}_{\rm GF}[\phi,\phi^*]$ is the Gaussian fluctuation (GF) which is quadratic in $\phi$ and $\phi^*$.

\subsection{BCS-Leggett mean-field theory}\label{s2-2}
In the BCS-Leggett mean-field approximation, the contributions from the fluctuations are completely neglected and we have
\begin{eqnarray}
{\cal S}_{\rm{eff}}[\Phi,\Phi^*]\simeq{\cal S}_{\rm MF}.
\end{eqnarray}
The grand potential is given by
\begin{eqnarray}\label{MF-Omega}
\Omega_{\rm MF}=\frac{{\cal S}_{\rm MF}}{\beta V}=\frac{\Delta^2}{U}-\frac{1}{\beta V}\sum_{K}{\rm ln}{\rm det}[\beta{\cal G}^{-1}(K)]
+\frac{1}{V}\sum_{\bf k}\xi_{\bf k}
\end{eqnarray}
where the inverse Nambu-Gor'kov Green's function reads
\begin{eqnarray}
{\cal  G}^{-1}(K)=\left(\begin{array}{cc}ik_n-\xi_{\bf k} &\Delta\\
\Delta& ik_n+\xi_{\bf k}\end{array}\right).
\end{eqnarray}
Here the dispersion $\xi_{\bf k}$ is defined as $\xi_{\bf k}=\varepsilon_{\bf k}-\mu$. In this paper $K=(ik_n,{\bf k})$ denotes the energy and  momentum of fermions with $k_n=(2n+1)\pi T$ ($n$ integer) being the fermion Matsubara frequency. We use the notation $\frac{1}{\beta V}\sum_{K}=\frac{1}{\beta}\sum_n\frac{1}{V}\sum_{\bf k}$.

At $T=0$, the grand potential is explicitly given by
\begin{eqnarray}\label{MF-Omega0}
\Omega_{\rm MF}=\frac{\Delta^2}{U}+\frac{1}{V}\sum_{\bf k}\left(\xi_{\bf k}-E_{\bf k}\right),
\end{eqnarray}
where $E_{\bf k}=\sqrt{\xi_{\bf k}^2+\Delta^2}$ is the standard BCS excitation spectrum. The superfluid order parameter $\Delta$ satisfies the extreme condition $\partial\Omega_{\rm MF}/\partial\Delta=0$, which leads to the so-called gap equation
\begin{eqnarray}\label{MF-GAP}
\frac{1}{U}=\frac{1}{V}\sum_{\bf k}\frac{1}{2E_{\bf k}}.
\end{eqnarray}
Note that the UV divergence should be eliminated by using the relations (6) and (7). The total fermion density $n$ is given by $n=-\partial\Omega_{\rm MF}/\partial\mu$. We obtain the so-called number equation
\begin{eqnarray}\label{MF-NUM}
n=\frac{1}{V}\sum_{\bf k}\left(1-\frac{\xi_{\bf k}}{E_{\bf k}}\right).
\end{eqnarray}
The order parameter $\Delta$ as a functional the chemical potential $\mu$ is determined by the gap equation (\ref{MF-GAP}) and the chemical potential $\mu$ is determined by the number equation (\ref{MF-NUM}). The fermion Green's function ${\cal G}(K)$ can be expressed as
\begin{eqnarray}
{\cal G}(K)=\left(\begin{array}{cc}{\cal G}_{11}(K)&{\cal G}_{12}(K)\\
{\cal G}_{21}(K)& {\cal G}_{22}(K)\end{array}\right).
\end{eqnarray}
The elements can be evaluated as
\begin{eqnarray}
&&{\cal G}_{11}(ik_n,{\bf k})=-{\cal G}_{22}(-ik_n,{\bf k})=\frac{u_{\bf k}^2}{ik_n-E_{\bf k}}
+\frac{\upsilon_{\bf k}^2}{ik_n+E_{\bf k}},\nonumber\\
&&{\cal G}_{12}(ik_n,{\bf k})={\cal G}_{21}(-ik_n,{\bf k})=u_{\bf k}\upsilon_{\bf k}\left(\frac{1}{ik_n+E_{\bf k}}
-\frac{1}{ik_n-E_{\bf k}}\right),
\end{eqnarray}
where the BCS distribution functions are given by $u_{\bf k}^2=(1+\xi_{\bf k}/E_{\bf k})/2$ and $\upsilon_{\bf k}^2=(1-\xi_{\bf k}/E_{\bf k})/2$.

\subsection{Gaussian fluctuations: Collective modes}\label{s2-3}

The quadratic term ${\cal S}_{\rm GF}[\phi,\phi^*]$ corresponds to Gaussian fluctuations around the BCS-Leggett ground state. Physically it determines the excitation spectra of the collective modes. For convenience, we work in the momentum space by making the Fourier transformation for the quantum fluctuations
\begin{equation}
\phi(x)=\frac{1}{\sqrt{\beta V}}\sum_Q \phi(Q)e^{-iq_l\tau+i{\bf q}\cdot{\bf r}},\ \ \ \ \ \ 
\phi^*(x)=\frac{1}{\sqrt{\beta V}}\sum_Q \phi^*(-Q)e^{-iq_l\tau+i{\bf q}\cdot{\bf r}}.
\end{equation}
For convenience, we also use the decomposition $\phi(x)=\phi_1(x)+i\phi_2(x)$, where $\phi_1(x)$ and $\phi_2(x)$ are the real and imaginary parts, respectively.
We have the Fourier transformation
\begin{equation}
\phi_1(x)=\frac{1}{\sqrt{\beta V}}\sum_Q \phi_1(Q)e^{-iq_l\tau+i{\bf q}\cdot{\bf r}},\ \ \ \ \ \ 
\phi_2(x)=\frac{1}{\sqrt{\beta V}}\sum_Q \phi_2(Q)e^{-iq_l\tau+i{\bf q}\cdot{\bf r}}.
\end{equation}
The effective action can be expressed as
\begin{eqnarray}
{\cal S}_{\rm{eff}}[\Phi, \Phi^*] =\frac{1}{U}\sum_Q\phi^*(Q)\phi(Q) +\frac{\sqrt{\beta V}}{U}\sum_{Q}2\Delta\delta_{Q,0}\phi_1(Q)
-\mbox{Trln} \left[({\bf G}^{-1})_{K,K^\prime}\right].
\end{eqnarray}
Here the trace ${\rm Tr}$ is taken in the Nambu-Gor'kov space and the momentum space. To proceed the expansion in powers of the quantum fluctuations, we express the inverse Green's function ${\bf G}^{-1}$ as
\begin{equation}
({\bf G}^{-1})_{K,K^\prime}={\cal G}^{-1}(K)\delta_{K,K^\prime}-(\Sigma_\phi)_{K,K^\prime}
\end{equation}
where ${\cal G}(K)$ is the mean-field Green's function given by (23) and $\Sigma_{\phi}$ is defined as
\begin{eqnarray}
(\Sigma_\phi)_{K,K^\prime}=-\frac{1}{\sqrt{\beta V}}\left[\Gamma_+\phi(K-K^\prime)+\Gamma_-\phi^*(K^\prime-K)\right].
\end{eqnarray}
Here the matrices $\Gamma_\pm$ are defined as
\begin{eqnarray}
\Gamma_\pm=\frac{1}{2}(\sigma_1\pm i\sigma_2)
\end{eqnarray}
with $\sigma_i$ ($i=1,2,3$) being the Pauli matrices in the Nambu-Gor'kov space.

Using the derivative expansion, the linear terms in $\phi$ can be evaluated as
\begin{eqnarray}
{\cal S}_{\rm eff}^{(1)}&=&\sqrt{\beta V}\sum_Q\delta_{Q,0}\left\{\frac{2\Delta}{U}-\frac{1}{\beta V}\sum_K\left[{\cal G}_{12}(K)+{\cal G}_{21}(K)\right]\right\}\phi_1(Q)\nonumber\\
&&+\sqrt{\beta V}\sum_Q\delta_{Q,0}\left\{\frac{i}{\beta V}\sum_K\left[{\cal G}_{12}(K)-{\cal G}_{21}(K)\right]\right\}\phi_2(Q).
\end{eqnarray}
Using the mean-field Green's function ${\cal G}(K)$ and the gap equation (\ref{MF-GAP}) we can show that the linear terms vanish exactly. After some manipulations, the quadratic terms in $\phi$, corresponding to the Gaussian pair fluctuations, can be written in a compact form
\begin{eqnarray}
{\cal S}_{\rm GF}[\phi,\phi^*]=\frac{1}{2}\sum_Q\left(\begin{array}{cc}
\phi^*(Q) & \phi(-Q)\end{array}\right){\bf M}(Q)\left(\begin{array}{cc} \phi(Q)\\
\phi^*(-Q)\end{array}\right),
\end{eqnarray}
where $Q=(iq_l,{\bf q})$ with $q_l=2l\pi T$ ($l$ integer) being the boson Matsubara frequency and the inverse boson propagator ${\bf M}(Q)$
takes the form
\begin{eqnarray}
{\bf M}(Q)=\left(\begin{array}{cc}{\bf M}_{11}(Q)&{\bf M}_{12}(Q)\\
{\bf M}_{21}(Q)& {\bf M}_{22}(Q)\end{array}\right)=\left(\begin{array}{cc}{\bf M}_{-+}(Q)&{\bf M}_{--}(Q)\\
{\bf M}_{++}(Q)& {\bf M}_{+-}(Q)\end{array}\right).
\end{eqnarray}
The elements of ${\bf M}(Q)$ can be expressed in terms of the Nambu-Gor'kov Green's function ${\cal G}(K)$. We have
\begin{eqnarray}
&&{\bf M}_{11}(Q)={\bf M}_{-+}(Q)=\frac{1}{U}+\frac{1}{\beta V}\sum_K{\rm Tr}_{\rm NG}\left[{\cal G}(K)\Gamma_-{\cal G}(K+Q)\Gamma_+\right],\nonumber\\
&&{\bf M}_{22}(Q)={\bf M}_{+-}(Q)=\frac{1}{U}+\frac{1}{\beta V}\sum_K{\rm Tr}_{\rm NG}\left[{\cal G}(K)\Gamma_+{\cal G}(K+Q)\Gamma_-\right],\nonumber\\
&&{\bf M}_{12}(Q)={\bf M}_{--}(Q)=\frac{1}{\beta V}\sum_K{\rm Tr}_{\rm NG}\left[{\cal G}(K)\Gamma_-{\cal G}(K+Q)\Gamma_-\right],\nonumber\\
&&{\bf M}_{21}(Q)={\bf M}_{++}(Q)=\frac{1}{\beta V}\sum_K{\rm Tr}_{\rm NG}\left[{\cal G}(K)\Gamma_+{\cal G}(K+Q)\Gamma_+\right].
\end{eqnarray}
Here the trace ${\rm Tr}_{\rm NG}$ is taken only in the Nambu-Gor'kov space. Carrying out the trace, we obtain
\begin{eqnarray}
&&{\bf M}_{11}(iq_l,{\bf q})={\bf M}_{22}(-iq_l,{\bf q})=\frac{1}{U}+\frac{1}{\beta V}\sum_K\left[{\cal G}_{11}(K+Q){\cal G}_{22}(K)\right],\nonumber\\
&&{\bf M}_{12}(iq_l,{\bf q})={\bf M}_{21}(iq_l,{\bf q})=\frac{1}{\beta V}\sum_K\left[{\cal G}_{12}(K+Q){\cal G}_{12}(K)\right].
\end{eqnarray}
Completing the fermion Matsubara frequency sum, we obtain
\begin{eqnarray}
&&{\bf M}_{11}(iq_l,{\bf q})=\frac{1}{U}+\frac{1}{V}\sum_{\bf k}
\left(\frac{u_+^2u_-^2}{iq_l-E_+-E_-}-\frac{\upsilon_+^2\upsilon_-^2}{iq_l+E_++E_-}\right),\nonumber\\
&&{\bf M}_{12}(iq_l,{\bf q})=-\frac{1}{V}\sum_{\bf k}
\left(\frac{u_+\upsilon_+u_-\upsilon_-}{iq_l-E_+-E_-}-\frac{u_+\upsilon_+u_-\upsilon_-}{iq_l+E_++E_-}\right).
\end{eqnarray}
Here the signs $+$ and $-$ denote the momenta ${\bf k}+{\bf q}/2$ and ${\bf k}-{\bf q}/2$, respectively. We can decompose
${\bf M}_{11}(iq_l,{\bf q})$ as ${\bf M}_{11}(iq_l,{\bf q})={\bf M}_{11}^{\rm e}(iq_l,{\bf q})+{\bf M}_{11}^{\rm o}(iq_l,{\bf q})$, where ${\bf M}_{11}^{\rm e}(iq_l,{\bf q})$ 
and ${\bf M}_{11}^{\rm o}(iq_l,{\bf q})$ are even and odd functions of $iq_l$, respectively. Their explicit forms read
\begin{eqnarray}
&&{\bf M}_{11}^{\rm e}(iq_l,{\bf q})=\frac{1}{U}
+\frac{1}{4V}\sum_{\bf k}\left(1+\frac{\xi_+\xi_-}{E_+E_-}\right)
\left(\frac{1}{iq_l-E_+-E_-}-\frac{1}{iq_l+E_++E_-}\right),\nonumber\\
&&{\bf M}_{11}^{\rm o}(iq_l,{\bf q})=\frac{1}{4V}\sum_{\bf k}
\left(\frac{\xi_+}{E_+}+\frac{\xi_-}{E_-}\right)\left(\frac{1}{iq_l-E_+-E_-}+\frac{1}{iq_l+E_++E_-}\right).
\end{eqnarray}

To make the result more physical, we decompose the complex fluctuation field $\phi(x)$ into its amplitude mode $\phi_1(x)$ and
phase mode $\phi_2(x)$, $\phi(x)=\phi_1(x)+i\phi_2(x)$. Converting to the variables $\phi_1(x)$ and $\phi_2(x)$, we have
\begin{eqnarray}
&&{\cal S}_{\rm GF}[\phi_1,\phi_2]\nonumber\\
&=&\sum_Q\left(\begin{array}{cc} \phi_1(-Q)&\phi_2(-Q)\end{array}\right)\left(\begin{array}{cc} {\bf M}_{11}^{\rm e}+{\bf M}_{12}&i{\bf M}_{11}^{\rm o}\\
-i{\bf M}_{11}^{\rm o} & {\bf M}_{11}^{\rm e}-{\bf M}_{12}\end{array}\right)\left(\begin{array}{c} \phi_1(Q)\\
\phi_2(Q)\end{array}\right)\nonumber\\
&=&\frac{1}{2}\sum_Q\left(\begin{array}{cc} \phi_1(-Q)&\phi_2(-Q)\end{array}\right)\left(\begin{array}{cc}
I_{11}(Q) & q_l I_{12}(Q) \\
-q_l I_{12}(Q) & I_{22}(Q) \end{array}\right)\left(\begin{array}{c} \phi_1(Q)\\
\phi_2(Q)\end{array}\right),
\end{eqnarray}
where
\begin{eqnarray}
I_{11}(iq_l,{\bf q})&=&\frac{1}{V}\sum_{\bf k}\left[\frac{E_++E_-}{E_+E_-}\frac{E_+E_-+\xi_+\xi_--\Delta^2}{(iq_l)^2-(E_++E_-)^2}
+\frac{1}{E_{\bf k}}\right],\nonumber\\
I_{22}(iq_l,{\bf q})&=&\frac{1}{V}\sum_{\bf k}\left[\frac{E_++E_-}{E_+E_-}\frac{E_+E_-+\xi_+\xi_-+\Delta^2}{(iq_l)^2-(E_++E_-)^2}
+\frac{1}{E_{\bf k}}\right],\nonumber\\
I_{12}(iq_l,{\bf q})&=&\frac{1}{V}\sum_{\bf k}\left(\frac{\xi_+}{E_+}+\frac{\xi_-}{E_-}\right)\frac{1}{(iq_l)^2-(E_++E_-)^2}.
\end{eqnarray}
Notice that here we have used the gap equation (21) to eliminate the bare coupling $U$. To study the spectrum of the collective modes, we taking the analytical continuation to real frequency $\omega$. The dispersions $\omega({\bf q})$ of the collective modes are determined by the equation $\det{{\bf M}[\omega, {\bf q}]}=0$ for $\omega$ smaller than the two-particle continuum. We have explicitly
\begin{equation}
I_{11}(\omega,{\bf q})I_{22}(\omega,{\bf q})-\omega^2I_{12}^2(\omega,{\bf q})=0.
\end{equation}
The function $I_{22}(\omega,{\bf q})$ can be expressed as
\begin{equation}
I_{22}(\omega,{\bf q})=\frac{1}{2V}\sum_{\bf k}\frac{E_++E_-}{E_+E_-}\frac{\omega^2-({\bf k}\cdot{\bf q}/m)^2}{\omega^2-(E_++E_-)^2}.
\end{equation}
Therefore, there exist a gapless Goldstone mode associated with the superfluidity.

\subsection{Gaussian pair fluctuation (GPF) theory}
Obviously, the BCS-Leggett mean-field theory lacks the contribution from the pair fluctuations, especially the quantum fluctuations from the gapless collective mode.
There have been a number of beyond-mean-field theoretical approaches to calculate the ground-state equation of state as well as other static properties in the BCS-BEC crossover~\cite{TH01,TH02,TH03,TH04,TH05,TH06,TH07,TH08,TH09,TH10,TH11}. In accordance with the functional path integral approach, in this work we introduce the Gaussian-pair-fluctuation (GPF) theory which was first proposed by Hu, Liu, and Drummond~\cite{TH03} and was later reformulated by Diener, Sensarma, and Randeria~\cite{TH04} using the functional path integral. The equation of state predicted by the GPF theory agrees excellently with the experimental measurements and the quantum Monte Carlo calculations. Especially, in the BEC limit the GPF theory predicts a composite boson scattering length which is very close to the exact result \cite{TH03, TH04}.

In the GPF theory, the effective action is truncated at the Gaussian level so that the path integral over the fluctuations $\phi$ and $\phi^*$
can be carried out. We have
\begin{eqnarray}
{\cal S}_{\rm{eff}}[\Phi,\Phi^*]\simeq{\cal S}_{\rm MF}+{\cal S}_{\rm GF}[\phi,\phi^*].
\end{eqnarray}
After carrying out the path integral over $\phi$ and $\phi^*$, the partition function can be expressed as
\begin{eqnarray}
{\cal Z}\simeq \exp \Big[- \beta V(\Omega_{\rm MF}+\Omega_{\rm GF})\Big],
\end{eqnarray}
where $\Omega_{\rm MF}$ is the mean-field grand potential given by (\ref{MF-Omega}) and the Gaussian-fluctuation contribution $\Omega_{\rm GF}$ is formally given by
\begin{equation}\label{GPF-formal}
\Omega_{\rm GF}=\frac{1}{2\beta V}\sum_{Q}\ln\det {\bf M}(Q).
\end{equation}
Here the explicit form of ${\bf M}(Q)$ is given in Sec. \ref{s2-3}. The grand potential in the GPF approach is given by
\begin{equation}
\Omega=\Omega_{\rm MF}+\Omega_{\rm GF}.
\end{equation}

The GF grand potential (\ref{GPF-formal}) is formal because the sum over the boson Matsubara frequency is divergent. To obtain convergent and physical equation of state, we need to taken into account carefully the convergent factors \cite{TH03,TH04}. The finite expression is
\begin{eqnarray}
\Omega_{\rm GF}&=&\frac{1}{2}\frac{1}{\beta}\sum_{q_l}\frac{1}{V}\sum_{\bf q}\Bigg\{\ln\left[{\bf M}_{11}(iq_l,{\bf q})\right]e^{iq_l0^+}
+\ln\left[{\bf M}_{22}(iq_l,{\bf q})\right]e^{-iq_l0^+}\nonumber\\
&&+\ln\left[1-\frac{{\bf M}_{12}^2(iq_l,{\bf q})}{{\bf M}_{11}(iq_l,{\bf q}){\bf M}_{22}(iq_l,{\bf q})}\right]\Bigg\}.
\end{eqnarray}
The Matsubara frequency sum can be converted to a standard contour integral. We have
\begin{eqnarray}
\Omega_{\rm GF}=-\frac{1}{2}\sum_{{\bf q}}\int_{-\infty}^\infty\frac{d\omega}{\pi}\frac{1}{e^{\beta\omega}-1}\left[2\delta_{11}(\omega,{\bf q})+\delta_{\rm M}(\omega,{\bf q})\right],
\end{eqnarray}
where the phase shifts are defined as
\begin{eqnarray}
\delta_{11}(\omega,{\bf q})&=&-{\rm Im}\ln {\bf M}_{11}(\omega+i\epsilon,{\bf q}),\nonumber\\
\delta_{\rm M}(\omega,{\bf q})&=&-{\rm Im}\ln\left[1-\frac{{\bf M}_{12}^2(\omega+i\epsilon,{\bf q})}
{{\bf M}_{11}(\omega+i\epsilon,{\bf q}){\bf M}_{22}(\omega+i\epsilon,{\bf q})}\right].
\end{eqnarray}
At $T=0$, there exists a better way to evaluate $\Omega_{\rm GF}$. We define two functions ${\bf M}_{11}^{\rm C}(z,{\bf q})$ and ${\bf M}_{22}^{\rm C}(z,{\bf q})$, which are given by
\begin{eqnarray}
{\bf M}^{\rm C}_{11}(z,{\bf q})={\bf M}^{\rm C}_{22}(-z,{\bf q})
&=&\frac{1}{U}+\frac{1}{V}\sum_{\bf k}\frac{u_+^2u_-^2}{z-E_+-E_-}\nonumber\\
&=&\frac{1}{V}\sum_{\bf k}\left(\frac{u_+^2u_-^2}{z-E_+-E_-}+\frac{1}{2E_{\bf k}}\right).
\end{eqnarray}
Note that we have used the mean-field gap equation. Using the and the fact $u_{\bf k}^2<1$, we can show that ${\bf M}_{11}^{\rm C}(z,{\bf q})$ has no singularities and zeros in the left half plane (${\rm Re}z<0$). Therefore, the Matsubara sum $\sum_{q_l}\ln {\bf M}_{11}^{\rm C}(iq_l,{\bf q})$ vanishes at $T=0$ since $\ln {\bf M}_{11}^{\rm C}(iq_l,{\bf q})$ has no singularities in the left-half plane. Meanwhile, at $T=0$ we replace the Matsubara frequency sum with a continuous integral over an imaginary frequency; i.e.,
\begin{eqnarray}
T\sum_{l=-\infty}^\infty f(iq_l)\rightarrow\int_{-\infty}^\infty\frac{d\omega}{2\pi}f(i\omega).
\end{eqnarray}
The GF contribution at $T=0$ can be expressed as \cite{TH04}
\begin{eqnarray}
\Omega_{\rm GF}=\frac{1}{V}\sum_{{\bf q}}\int_0^\infty\frac{d\omega}{2\pi}\ln\left[\frac{{\bf
M}_{11}(i\omega,{\bf q}){\bf M}_{22}(i\omega,{\bf q})-{\bf M}_{12}^2(i\omega,{\bf q})}
{{\bf M}_{11}^{\rm C}(i\omega,{\bf q}){\bf M}_{22}^{\rm C}(i\omega,{\bf q})}\right].
\end{eqnarray}
Here we have used the fact that the integrand is real and even in $\omega$.

The crucial element of the GPF theory is that the relation between the order parameter $\Delta$ and the chemical potential $\mu$,
$\Delta=\Delta(\mu)$, is determined by the extreme of the mean-field grand potential $\Omega_{\rm MF}$ rather than the full grand potential
$\Omega_{\rm GPF}$. We therefore determine $\Delta(\mu)$ from the following extreme condition
\begin{equation}
\frac{\partial\Omega_{\rm MF}(\mu,\Delta)}{\partial\Delta}=0\Rightarrow \frac{1}{U}=\frac{1}{V}\sum_{\bf k}\frac{1}{2E_{\bf k}}.
\end{equation}
The use of the mean-field gap equation (\ref{MF-GAP}) ensures that the Goldstone mode is gapless, i.e., $\det {\bf M}(0,{\bf 0})=0$. Therefore, the GPF theory is 
a \emph{gapless approximation} and hence may properly take into account the contribution from the Goldstone mode fluctuation. The contribution from the
Gaussian fluctuations, $\Omega_{\rm GF}$, influences the equation of state. The chemical potential $\mu$, however, should be determined by the full
grand potential $\Omega$. The number equation is given by
\begin{equation}
n=-\frac{d\Omega(\mu)}{d\mu}=n_{\rm MF}(\mu)+n_{\rm GF}(\mu),
\end{equation}
where the mean-field contribution $n_{\rm MF}(\mu)$ reads
\begin{equation}
n_{\rm MF}(\mu)=-\frac{d \Omega_{\rm MF}}{d\mu}=\frac{1}{V}\sum_{\bf k}\left(1-\frac{\xi_{\bf k}}{E_{\bf k}}\right),
\end{equation}
and the GF contribution $n_{\rm GF}(\mu)$ is given by
\begin{equation}
n_{\rm GF}(\mu)=-\frac{d\Omega_{\rm GF}}{d\mu}.
\end{equation}
Note that the derivative of $\Omega_{\rm GF}$ with respect to $\mu$ should be evaluated as
\begin{equation}
\frac{d \Omega_{\rm GF}}{d\mu}=\frac{\partial \Omega_{\rm GF}(\mu,\Delta)}{\partial\mu}
+\frac{\partial \Omega_{\rm GF}(\mu,\Delta)}{\partial\Delta}\frac{d\Delta(\mu)}{d\mu}.
\end{equation}

\section{Dynamic density and spin responses: Definition and general formalism}\label{s3}

The dynamic density responses of the system are characterized by the Fourier transformation of the following imaginary-time-ordered density-density correlation function
\begin{equation}
\chi_{\sigma\sigma^\prime}(\tau-\tau^\prime,{\bf r}-{\bf r}^\prime)=-\langle T_\tau \hat{n}_\sigma(\tau,{\bf r})\hat{n}_{\sigma^\prime}(\tau^\prime,{\bf r}^\prime)\rangle_{\rm c},
\end{equation}
where $\sigma,\sigma^\prime=\uparrow,\downarrow$ denote the spin states and the density operators are given by
\begin{equation}
\hat{n}_\sigma(\tau,{\bf r})=\psi_\sigma^{\dagger}(\tau,{\bf r})\psi_\sigma^{\phantom{\dag}}(\tau,{\bf r}).
\end{equation}
Here $\psi_\sigma^{\phantom{\dag}}(\tau,{\bf r})$ and $\psi_\sigma^{\dagger}(\tau,{\bf r})$ are the field operators in the Heisenberg representation. The notation
$\langle \cdots\rangle_{\rm c} $ denotes the connected piece of the correlation function. In this work, we consider a spin-balanced Fermi system. Therefore, we have
\begin{eqnarray}\label{Relation-UD}
&&\chi_{\uparrow\uparrow}(\tau-\tau^\prime,{\bf r}-{\bf r}^\prime)=\chi_{\downarrow\downarrow}(\tau-\tau^\prime,{\bf r}-{\bf r}^\prime),\nonumber\\
&&\chi_{\uparrow\downarrow}(\tau-\tau^\prime,{\bf r}-{\bf r}^\prime)=\chi_{\downarrow\uparrow}(\tau-\tau^\prime,{\bf r}-{\bf r}^\prime).
\end{eqnarray}
Conventionally, we define the total density operator $\hat{n}(\tau,{\bf r})$ and the spin density operator $\hat{s}(\tau,{\bf r})$
\begin{eqnarray}
\hat{n}(\tau,{\bf r})&=&n_\uparrow(\tau,{\bf r})+n_\downarrow(\tau,{\bf r}),\nonumber\\
\hat{s}(\tau,{\bf r})&=&n_\uparrow(\tau,{\bf r})-n_\downarrow(\tau,{\bf r}).
\end{eqnarray}
Note that we consider the $z$ direction of the spin without loss of generality. Because the system is isotropic, considering other directions of the spin 
will arrive at the same result. In terms of the total density and the spin density, we can define the density response function
\begin{equation}
\chi_{nn}(\tau-\tau^\prime,{\bf r}-{\bf r}^\prime)=-\langle T_\tau \hat{n}(\tau,{\bf r})\hat{n}(\tau^\prime,{\bf r}^\prime)\rangle_{\rm c}
\end{equation}
and the spin response function
\begin{equation}
\chi_{ss}(\tau-\tau^\prime,{\bf r}-{\bf r}^\prime)=-\langle T_\tau \hat{s}(\tau,{\bf r})\hat{s}(\tau^\prime,{\bf r}^\prime)\rangle_{\rm c}.
\end{equation}
In principle, there arises two off-diagonal response functions
\begin{eqnarray}
&&\chi_{ns}(\tau-\tau^\prime,{\bf r}-{\bf r}^\prime)=-\langle T_\tau \hat{n}(\tau,{\bf r})\hat{s}(\tau^\prime,{\bf r}^\prime)\rangle_{\rm c},\nonumber\\
&&\chi_{sn}(\tau-\tau^\prime,{\bf r}-{\bf r}^\prime)=-\langle T_\tau \hat{s}(\tau,{\bf r})\hat{n}(\tau^\prime,{\bf r}^\prime)\rangle_{\rm c}.
\end{eqnarray}
Since we consider a spin-balanced system, using the relation (\ref{Relation-UD}), we have
\begin{eqnarray}
\chi_{ns}(\tau-\tau^\prime,{\bf r}-{\bf r}^\prime)=\chi_{sn}(\tau-\tau^\prime,{\bf r}-{\bf r}^\prime)=0.
\end{eqnarray}

In the functional path integral formalism, we introduce external source terms to compute the dynamic response functions. These external sources physically represents the inhomogeneous external perturbations applied to the system. We introduce two external sources $j_n(x)$ and $j_s(x)$ which are conjugate to 
the total density and the spin density respectively. Alternatively, we can also introduce $j_\uparrow(x)$ and $j_\downarrow(x)$ which are conjugate to 
the up and down spin densities respectively. In the presence of the external sources, the partition function of the system becomes
\begin{eqnarray}
{\cal Z}[J] = \int [d\psi][d\bar{\psi}]\exp\left\{-{\cal S}_J[\psi,\bar{\psi}]\right\},
\end{eqnarray}
where $J$ denotes $\{j_n,j_s\}$ or $\{j_\uparrow,j_\downarrow\}$ and the action in the presence of the external sources reads
\begin{eqnarray}
{\cal S}_J[\psi,\bar{\psi}]&=&\int_0^\beta d\tau\int d{\bf r}\bar{\psi}\partial_\tau \psi+\int_0^\beta d\tau H(\psi,\bar{\psi})+{\cal S}_{\rm source}.
\end{eqnarray}
The source term is given by
\begin{eqnarray}
{\cal S}_{\rm source}&=&\int_0^\beta d\tau\int d{\bf r}\left[j_n(\bar{\psi}_\uparrow\psi_\uparrow+\bar{\psi}_\downarrow\psi_\downarrow)
+j_s(\bar{\psi}_\uparrow\psi_\uparrow-\bar{\psi}_\downarrow\psi_\downarrow)\right],\nonumber\\
&=&\int_0^\beta d\tau\int d{\bf r}\left(j_\uparrow\bar{\psi}_\uparrow\psi_\uparrow+j_\downarrow\bar{\psi}_\downarrow\psi_\downarrow\right).
\end{eqnarray}
It is obvious that $j_\uparrow=j_n+j_s$ and $j_\downarrow=j_n-j_s$. Applying the Hubbard-Stratonovich transformation, we write the partition function as
\begin{eqnarray}
{\cal Z}[J]=\int [d\psi][d\bar{\psi}][d\Phi][d\Phi^*] \exp\Big\{-{\cal S}_J[\psi,\bar{\psi},\Phi,\Phi^*]\Big\},
\end{eqnarray}
where the action now reads
\begin{eqnarray}
{\cal S}_J=\int dx\frac{|\Phi(x)|^2}{U}-\int dx\int dx^\prime\bar{\psi}(x){\bf G}_J^{-1}(x,x^\prime)\psi(x^\prime).
\end{eqnarray}
The inverse Nambu-Gor'kov Green's function ${\bf G}_J^{-1}(x,x^\prime)$ is given by
\begin{eqnarray}
{\bf G}_J^{-1}(x,x^\prime)&=&\left(\begin{array}{cc}-\partial_{\tau}+\frac{\nabla^2}{2m}+\mu &\Phi(x)\\  \Phi^*(x)&
-\partial_{\tau}-\frac{\nabla^2}{2m}-\mu\end{array}\right)\delta(x-x^\prime)\nonumber\\
&+&\left(\begin{array}{cc}j_n(x)+j_s(x) &0\\  0&
-j_n(x)+j_s(x)\end{array}\right)\delta(x-x^\prime).
\end{eqnarray}
Integrating out the fermion fields, we obtain
\begin{eqnarray}\label{Partition-J}
{\cal Z}[J]=\int[d\Phi][d\Phi^*] \exp \Big\{- {\cal S}_{\rm{eff}}[\Phi, \Phi^*;J]\Big\},
\end{eqnarray}
where the effective action in the presence of the external sources reads
\begin{eqnarray}\label{Action-J}
{\cal S}_{\rm{eff}}[\Phi, \Phi^*;J] = \int dx \frac{|\Phi(x)|^{2}}{U} - \mbox{Trln} [{\bf G}_J^{-1}(x,x^\prime)].
\end{eqnarray}
If the partition function ${\cal Z}[J]$ can be computed exactly as a functional of the external sources, the correlation functions can be obtained.  In practice, 
we introduce the generating functional ${\cal W}[J]$, which is defined as 
\begin{eqnarray}
{\cal Z}[J] = \exp{\Big\{-{\cal W}[J]\Big\}}.
\end{eqnarray}
In the path integral formalism, the response functions are given by
\begin{eqnarray}\label{Definition01}
\chi_{\sigma\sigma^\prime}(\tau-\tau^\prime,{\bf r}-{\bf r}^\prime)=\frac{\delta^2{\cal W}[j_\uparrow,j_\downarrow]}
{\delta j_\sigma(\tau,{\bf r})\delta j_{\sigma^\prime}(\tau^\prime,{\bf r}^\prime)}\Bigg|_{j_\uparrow=j_\downarrow=0}
\end{eqnarray}
for $\sigma,\sigma^\prime=\uparrow,\downarrow$, or 
\begin{eqnarray}\label{Definition02}
\chi_{ab}(\tau-\tau^\prime,{\bf r}-{\bf r}^\prime)=\frac{\delta^2{\cal W}[j_n,j_s]}
{\delta j_a(\tau,{\bf r})\delta j_b(\tau^\prime,{\bf r}^\prime)}\Bigg|_{j_n=j_s=0}
\end{eqnarray}
for $a,b=n,s$.

In practice, we work in the momentum space by making the Fourier transformation
\begin{eqnarray}
j(x)=\frac{1}{\sqrt{\beta V}}\sum_Q j(Q)e^{-iq_l\tau+i{\bf q}\cdot{\bf r}}.
\end{eqnarray}
Here $j$ denotes $j_n,j_s$ or $j_\uparrow,j_\downarrow$. Evaluating the second order derivative in (\ref{Definition01}) and (\ref{Definition02}) is equivalent to 
expanding the generating functional ${\cal W}[J]$ up to the order $O(j^2)$. Formally, we have
\begin{eqnarray}
{\cal W}[J] = {\cal W}^{(0)}+{\cal W}^{(1)}[J]+{\cal W}^{(2)}[J]+\cdots,
\end{eqnarray}
where ${\cal W}^{(n)}$ denotes the $n$th order term in the external sources.  The zeroth order term ${\cal W}^{(0)}$ recovers the grand potential $\Omega$ in the absence of the external sources,
\begin{eqnarray}
{\cal W}^{(0)}=\beta V\Omega.
\end{eqnarray}
The first-order term ${\cal W}^{(1)}$ is related to the zero mode of the external source $j_n$,
\begin{eqnarray}
{\cal W}^{(1)}[J]= -\sqrt{\beta V} nj_n(Q=0).
\end{eqnarray}
Physically, this represents the thermodynamic relation $n=-\partial\Omega/\partial \mu$. The dynamic responses are characterized by the second-order term
${\cal W}^{(2)}$. It can be expressed as
\begin{eqnarray}
{\cal W}^{(2)}[J] =\frac{1}{2}\sum_Q\left(\begin{array}{cc} j_n(-Q)& j_s(-Q) \end{array}\right)\left(\begin{array}{cc} \chi_{nn}(Q) & \chi_{ns}(Q) \\  \chi_{sn}(Q) & \chi_{ss}(Q)\end{array}\right)\left(\begin{array}{c} j_n(Q) \\ j_s(Q) \end{array}\right),
\end{eqnarray}
or
\begin{eqnarray}
{\cal W}^{(2)}[J] =\frac{1}{2}\sum_Q\left(\begin{array}{cc} j_\uparrow(-Q)& j_\downarrow(-Q) \end{array}\right)\left(\begin{array}{cc} \chi_{\uparrow\uparrow}(Q) & \chi_{\uparrow\downarrow}(Q) \\  \chi_{\downarrow\uparrow}(Q) & \chi_{\downarrow\downarrow}(Q)\end{array}\right)
\left(\begin{array}{c} j_\uparrow(Q) \\ j_\downarrow(Q) \end{array}\right),
\end{eqnarray}
Since we consider a spin-balanced system, we have 
\begin{equation}
\chi_{ns}(Q)=\chi_{sn}(Q)=0.
\end{equation}
Using the relation between $\{j_n,j_s\}$ and $\{j_\uparrow,j_\downarrow\}$, we obtain
\begin{eqnarray}
&&\chi_{\uparrow\uparrow}(Q)=\chi_{\downarrow\downarrow}(Q)
=\frac{1}{2}\left[\chi_{nn}(Q)+\chi_{ss}(Q)\right],\nonumber\\
&&\chi_{\uparrow\downarrow}(Q)=\chi_{\downarrow\uparrow}(Q)
=\frac{1}{2}\left[\chi_{nn}(Q)-\chi_{ss}(Q)\right].
\end{eqnarray}

Finally, we define the dynamic density structure factor $S_{nn}(\omega,{\bf q})$ and the dynamic spin structure factor $S_{ss}(\omega,{\bf q})$. They are related to the density and spin response functions by the fluctuation-dissipation theorem. We have
\begin{eqnarray}
&&S_{nn}(\omega,{\bf q})=-\frac{1}{\pi}\frac{1}{1-e^{-\beta\omega}}{\rm Im}\chi_{nn}(\omega+i\epsilon,{\bf q}),\nonumber\\
&&S_{ss}(\omega,{\bf q})=-\frac{1}{\pi}\frac{1}{1-e^{-\beta\omega}}{\rm Im}\chi_{ss}(\omega+i\epsilon,{\bf q}).
\end{eqnarray}
We have similar definitions for $S_{\uparrow\uparrow}(\omega,{\bf q})=S_{\downarrow\downarrow}(\omega,{\bf q})$ and $S_{\uparrow\downarrow}(\omega,{\bf q})=S_{\downarrow\uparrow}(\omega,{\bf q})$. Here note that $Q=(iq_l,{\bf q})$ and we have made the analytical continuation from imaginary frequency $iq_l$ 
to real frequency $\omega$. The static stricture factors are defined as the frequency integral of the dynamic structure factors,
\begin{equation}
S_{nn}({\bf q})=\frac{2}{n}\int_0^\infty d\omega S_{nn}(\omega,{\bf q}),\ \ \ \ \ \ \ \ S_{ss}({\bf q})=\frac{2}{n}\int_0^\infty d\omega S_{ss}(\omega,{\bf q}).
\end{equation}

So far the above discussions are exact. However, it seems impossible to evaluate the generating functional ${\cal W}[J]$ or its second-order expansion 
${\cal W}^{(2)}[J]$ precisely. Therefore, the next task is to consider some approximations. In this work, we will consider two gapless approximations for the equilibrium state (in the absence of external sources) which have been introduced in Sec. \ref{s2}.  For a given approximation for the equilibrium state, we can construct the
corresponding response theory.

\section{BCS-Leggett response theory: Random phase approximation}\label{s4}
In this section we consider the responses of the ground state within the BCS-Leggett mean-field theory.  We will show that the BCS-Leggett response theory naturally recovers the random phase approximation (RPA).

\subsection{Responses of the BCS-Leggett ground state}
We generalize the BCS-Leggett mean-field approximation to the case with external sources. In this approximation, the partition function is given by
\begin{eqnarray}
{\cal Z}[J]\simeq \exp \Big\{- {\cal W}_{\rm MF}[J;\Delta_{\rm cl}, \Delta_{\rm cl}^*]\Big\},
\end{eqnarray}
where the mean-field generating functional ${\cal W}_{\rm MF}$ is obtained by replacing the pairing field $\Phi(x)$ with its expectation value or \emph{classical field} $\Delta_{\rm cl}(x)$, which serves as the order parameter of superfluidity. We note that the order parameter $\Delta_{\rm cl}(x)$ is no longer static and uniform in the presence of the external sources $j_n(x)$ and $j_s(x)$. It should be determined by the extreme condition
\begin{equation}\label{RPA-GAP}
\frac{\delta{\cal W}_{\rm MF}[J;\Delta_{\rm cl}, \Delta_{\rm cl}^*]}{\delta\Delta_{\rm cl}(x)}=0,\ \ \ \ \ \ 
\frac{\delta{\cal W}_{\rm MF}[J;\Delta_{\rm cl}, \Delta_{\rm cl}^*]}{\delta\Delta_{\rm cl}^*(x)}=0.
\end{equation}
This is a natural generalization of the mean-field gap equation (\ref{MF-GAP}). For infinitesimal external sources, the perturbation to the order parameter 
$\Delta_{\rm cl}(x)$ is also infinitesimal. Therefore, we write
\begin{equation}
\Delta_{\rm cl}(x)=\Delta+\eta_1(x)+i\eta_2(x),
\end{equation}
where the static and uniform part $\Delta$ is the order parameter in the absence of external sources and is determined by the mean-field gap equation 
(\ref{MF-GAP}). Then the generating functional can be expressed as
\begin{eqnarray}
{\cal W}_{\rm MF}[J;\eta_1,\eta_2] = \int dx \frac{|\Delta_{\rm cl}(x)|^{2}}{U} - \mbox{Trln}{\cal G}_J^{-1}(x,x^\prime).
\end{eqnarray}
Here we express the Nambu-Gor'kov Green's function in the presence of external sources as
\begin{equation}
{\cal G}_J^{-1}(x,x^\prime)={\cal G}^{-1}(x,x^\prime)-\Sigma_J(x,x^\prime),
\end{equation}
where the two parts are given by
\begin{eqnarray}
&&{\cal G}^{-1}(x,x^\prime)=\left(\begin{array}{cc}-\partial_{\tau}+\frac{\nabla^2}{2m}+\mu &\Delta\\  \Delta&-\partial_{\tau}-\frac{\nabla^2}{2m}-\mu\end{array}\right)\delta(x-x^\prime),\nonumber\\
&&\Sigma_J(x,x^\prime)=-\left(\begin{array}{cc}j_n(x)+j_s(x)&\eta_1(x)+i\eta_2(x)\\  \eta_1(x)-i\eta_2(x)&-j_n(x)+j_s(x)\end{array}\right)\delta(x-x^\prime).
\end{eqnarray}
For convenience, we write $\Sigma_J(x,x^\prime)$ as
\begin{eqnarray}
\Sigma_J(x,x^\prime)=-\left[\Gamma_nj_n(x)+\Gamma_sj_s(x)+\Gamma_1\eta_1(x)+\Gamma_2\eta_2(x)\right]\delta(x-x^\prime),
\end{eqnarray}
where $\Gamma_n=\sigma_3$, $\Gamma_s=1$, $\Gamma_1=\sigma_1$, and $\Gamma_2=-\sigma_2$ with $\sigma_i$ ($i=1,2,3$) being the Pauli matrices in the Nambu-Gor'kov space.

To study the linear response of the system to infinitesimal external sources, we expand the generating functional up to the quadratic order in the external sources as well as the induced perturbations $\eta_1(x)$ and $\eta_2(x)$. We shall work in the momentum space by making the Fourier transformation
\begin{eqnarray}
&&j_n(x)=\frac{1}{\sqrt{\beta V}}\sum_Q j_n(Q)e^{-iq_l\tau+i{\bf q}\cdot{\bf r}},\ \ \ \ \
j_s(x)=\frac{1}{\sqrt{\beta V}}\sum_Q j_s(Q)e^{-iq_l\tau+i{\bf q}\cdot{\bf r}},\nonumber\\
&&\eta_1(x)=\frac{1}{\sqrt{\beta V}}\sum_Q \eta_1(Q)e^{-iq_l\tau+i{\bf q}\cdot{\bf r}},\ \ \ \ \
\eta_2(x)=\frac{1}{\sqrt{\beta V}}\sum_Q \eta_2(Q)e^{-iq_l\tau+i{\bf q}\cdot{\bf r}}.
\end{eqnarray}
The expansion of the generating functional can be performed by using the momentum representation. We have
\begin{equation}
({\cal G}_J^{-1})_{K,K^\prime}={\cal G}^{-1}(K)\delta_{K,K^\prime}-(\Sigma_J)_{K,K^\prime},
\end{equation}
where
\begin{eqnarray}
(\Sigma_J)_{K,K^\prime}&=&-\frac{1}{\sqrt{\beta V}}\Big[\Gamma_nj_n(K-K^\prime)+\Gamma_sj_s(K-K^\prime)\nonumber\\
&&+\Gamma_1\eta_1(K-K^\prime)+\Gamma_2\eta_2(K-K^\prime)\Big].
\end{eqnarray}
Using the derivative expansion, we obtain the expansion of the generating functional
\begin{eqnarray}
{\cal W}_{\rm MF}[j_n,j_s;\eta_1,\eta_2] = {\cal W}_{\rm MF}^{(0)}+{\cal W}_{\rm MF}^{(1)}[j_n,j_s;\eta_1,\eta_2]+{\cal W}_{\rm MF}^{(2)}[j_n,j_s;\eta_1,\eta_2]+\cdots,
\end{eqnarray}
where ${\cal W}_{\rm MF}^{(0)}=\beta V\Omega_{\rm MF}$.

The linear term ${\cal W}_{\rm MF}^{(1)}[j_n,j_s;\eta_1,\eta_2]$ can be evaluated as
\begin{eqnarray}
{\cal W}_{\rm MF}^{(1)}&=&-\sqrt{\beta V}\sum_Q\delta_{Q,0}\left\{\frac{1}{\beta V}\sum_K\left[{\cal G}_{11}(K)e^{ik_n0^+}-{\cal G}_{22}(K)e^{-ik_n0^+}\right]\right\}
j_n(Q)\nonumber\\
&&-\sqrt{\beta V}\sum_Q\delta_{Q,0}\left\{\frac{1}{\beta V}\sum_K\left[{\cal G}_{11}(K)e^{ik_n0^+}+{\cal G}_{22}(K)e^{-ik_n0^+}\right]\right\}
j_s(Q)\nonumber\\
&&+\sqrt{\beta V}\sum_Q\delta_{Q,0}\left\{\frac{2\Delta}{U}-\frac{1}{\beta V}\sum_K\left[{\cal G}_{12}(K)+{\cal G}_{21}(K)\right]\right\}\eta_1(Q)\nonumber\\
&&+\sqrt{\beta V}\sum_Q\delta_{Q,0}\left\{\frac{i}{\beta V}\sum_K\left[{\cal G}_{12}(K)-{\cal G}_{21}(K)\right]\right\}\eta_2(Q).
\end{eqnarray}
We find that the linear contributions are only related to the zero modes of the external sources. Using the explicit form of the fermion Green's function ${\cal G}(K)$,
we obtain
\begin{eqnarray}
&&\frac{1}{\beta V}\sum_K\left[{\cal G}_{11}(K)e^{ik_n0^+}-{\cal G}_{22}(K)e^{-ik_n0^+}\right]=\sum_{\bf k}\left(1-\frac{\xi_{\bf k}}{E_{\bf k}}\right)=n,\nonumber\\
&&\frac{1}{\beta V}\sum_K\left[{\cal G}_{11}(K)e^{ik_n0^+}+{\cal G}_{22}(K)e^{-ik_n0^+}\right]=0,\nonumber\\
&&\frac{2\Delta}{U}-\frac{1}{\beta V}\sum_K\left[{\cal G}_{12}(K)+{\cal G}_{21}(K)\right]=2\Delta\left(\frac{1}{U}-\sum_{\bf k}\frac{1}{2E_{\bf k}}\right)=0,\nonumber\\
&&\frac{i}{\beta V}\sum_K\left[{\cal G}_{12}(K)-{\cal G}_{21}(K)\right]=0.
\end{eqnarray}
Therefore, the linear contributions can be simplified as
\begin{eqnarray}
\frac{{\cal W}_{\rm MF}^{(1)}}{\beta V}=-n\frac{j_n(Q=0)}{\sqrt{\beta V}}.
\end{eqnarray}
This is nothing but the thermodynamic relation $n=-\partial\Omega_{\rm MF}/\partial\mu$ in the BCS-Leggett mean-field theory. Thus the zeroth and first-order contributions just reflect the properties of the equilibrium state in the BCS-Leggett mean-field approximation. 

The dynamic density and spin responses are characterized by the quadratic term ${\cal W}_{\rm MF}^{(2)}[j_n,j_s;\eta_1,\eta_2]$. Using the derivative expansion and carrying out the trace in the momentum space, we obtain
\begin{eqnarray}
{\cal W}_{\rm MF}^{(2)}=\frac{1}{2}\sum_{K}\sum_{K^\prime}{\rm Tr}_{\rm NG}
\left[{\cal G}(K)(\Sigma_J)_{K,K^\prime}{\cal G}(K^\prime)(\Sigma_J)_{K^\prime,K}\right].
\end{eqnarray}
Defining $Q=K^\prime-K$, we have
\begin{eqnarray}
{\cal W}_{\rm MF}^{(2)}=\frac{1}{2}\sum_{a,b=n,s,1,2}\sum_{Q}\Pi_{ab}(Q)\varphi_a(-Q)\varphi_b(Q).
\end{eqnarray}
For convenience, here we have used the notation $\varphi_n(Q)=j_n(Q)$,  $\varphi_s(Q)=j_s(Q)$,  $\varphi_1(Q)=\eta_1(Q)$, and $\varphi_2(Q)=\eta_2(Q)$. 
The loop functions $\Pi_{ab}(Q)$ ($a,b=n,s,1,2$) are defined as
\begin{eqnarray}
\Pi_{ab}(Q)=\frac{1}{\beta V}\sum_{K}{\rm Tr}_{\rm NG}\left[{\cal G}(K)\Gamma_a{\cal G}(K+Q)\Gamma_b\right]+\frac{2}{U}\tilde{\delta}_{ab}.
\end{eqnarray}
Here $\tilde{\delta}_{ab}=1$ for $a=b=1$ and $a=b=2$ and $\tilde{\delta}_{ab}=0$ for all other combinations.  It is obvious that $\Pi_{ba}(Q)=\Pi_{ab}(-Q)$.
After some manipulations,  it can be written in a matrix form
\begin{eqnarray}\label{MF-Response}
{\cal W}_{\rm MF}^{(2)}=\frac{1}{2}\sum_Q\left(\begin{array}{cccc} j^{\rm T}(-Q)&\eta^{\rm T}(-Q)\end{array}\right)
\left(\begin{array}{cc}\Pi_{jj}(Q)&\Pi_{j\eta}(Q)\\
\Pi_{\eta j}(Q)&\Pi_{\eta\eta}(Q)\end{array}\right)
\left(\begin{array}{c} j(Q) \\ \eta(Q)\end{array}\right),
\end{eqnarray}
where we have defined the notations
\begin{equation}
j(Q)=\left(\begin{array}{cc}j_n(Q) \\ j_s(Q)\end{array}\right),\ \ \ \ \ \ \eta(Q)=\left(\begin{array}{cc}\eta_1(Q) \\ \eta_2(Q)\end{array}\right)
\end{equation}
The blocks $\Pi_{mn}(Q)$ ($m,n=j,\eta$) are given by
\begin{eqnarray}
\Pi_{jj}(Q)&=&\left(\begin{array}{cc}\Pi_{nn}(Q)&\Pi_{ns}(Q)\\ \Pi_{sn}(Q)&\Pi_{ss}(Q)\end{array}\right),\nonumber\\
\Pi_{j\eta}(Q)&=&\left(\begin{array}{cc}\Pi_{n1}(Q)&\Pi_{n2}(Q)\\ \Pi_{s1}(Q)&\Pi_{s2}(Q)\end{array}\right),\nonumber\\
\Pi_{\eta j}(Q)&=&\left(\begin{array}{cc}\Pi_{1n}(Q)&\Pi_{1s}(Q)\\ \Pi_{2n}(Q)&\Pi_{2s}(Q)\end{array}\right),\nonumber\\
\Pi_{\eta\eta}(Q)&=&\left(\begin{array}{cc}\Pi_{11}(Q)&\Pi_{12}(Q)\\ \Pi_{21}(Q)&\Pi_{22}(Q)\end{array}\right).
\end{eqnarray}
We have $\Pi_{j\eta}(Q)=\Pi_{\eta j}^{\rm T}(-Q)$. Completing the trace in the Nambu-Gor'kov space, we obtain
\begin{eqnarray}
\Pi_{nn}(Q)&=&\frac{1}{\beta V}\sum_{K}\Big({\cal G}_{11}^\prime{\cal G}_{11}^{\phantom{\dag}}+{\cal G}_{22}^\prime{\cal G}_{22}^{\phantom{\dag}}
-2{\cal G}_{12}^\prime{\cal G}_{12}^{\phantom{\dag}}\Big),\nonumber\\
\Pi_{ss}(Q)&=&\frac{1}{\beta V}\sum_{K}\Big({\cal G}_{11}^\prime{\cal G}_{11}^{\phantom{\dag}}+{\cal G}_{22}^\prime{\cal G}_{22}^{\phantom{\dag}}
+2{\cal G}_{12}^\prime{\cal G}_{12}^{\phantom{\dag}}\Big),\nonumber\\
\Pi_{ns}(Q)&=&\Pi_{sn}(-Q)=\frac{1}{\beta V}\sum_{K}\Big({\cal G}_{11}^\prime{\cal G}_{11}^{\phantom{\dag}}
-{\cal G}_{22}^\prime{\cal G}_{22}^{\phantom{\dag}}\Big),\nonumber\\
\Pi_{n1}(Q)&=&\Pi_{1n}(-Q)=\frac{1}{\beta V}\sum_{K}\Big[{\cal G}_{12}^\prime({\cal G}_{11}^{\phantom{\dag}}-{\cal G}_{22}^{\phantom{\dag}})
+({\cal G}_{11}^\prime-{\cal G}_{22}^\prime){\cal G}_{12}^{\phantom{\dag}}\Big],\nonumber\\
\Pi_{n2}(Q)&=&\Pi_{2n}(-Q)=-i\frac{1}{\beta V}\sum_{K}\Big[({\cal G}_{11}^\prime+{\cal G}_{22}^\prime){\cal G}_{12}^{\phantom{\dag}}
-{\cal G}_{12}^\prime({\cal G}_{11}^{\phantom{\dag}}+{\cal G}_{22}^{\phantom{\dag}})\Big],\nonumber\\
\Pi_{s1}(Q)&=&\Pi_{1s}(-Q)=\frac{1}{\beta V}\sum_{K}\Big[({\cal G}_{11}^\prime+{\cal G}_{22}^\prime){\cal G}_{12}^{\phantom{\dag}}
+{\cal G}_{12}^\prime({\cal G}_{11}^{\phantom{\dag}}+{\cal G}_{22}^{\phantom{\dag}})\Big],\nonumber\\
\Pi_{s2}(Q)&=&\Pi_{2s}(-Q)=-i\frac{1}{\beta V}\sum_{K}\Big[({\cal G}_{11}^\prime-{\cal G}_{22}^\prime]{\cal G}_{12}^{\phantom{\dag}}
-{\cal G}_{12}^\prime({\cal G}_{11}^{\phantom{\dag}}-{\cal G}_{22}^{\phantom{\dag}})\Big],\nonumber\\
\Pi_{11}(Q)&=&\frac{2}{U}+\frac{1}{\beta V}\sum_{K}\Big({\cal G}_{11}^\prime{\cal G}_{22}^{\phantom{\dag}}+{\cal G}_{22}^\prime{\cal G}_{11}^{\phantom{\dag}}
+2{\cal G}_{12}^\prime{\cal G}_{12}^{\phantom{\dag}}\Big),\nonumber\\
\Pi_{22}(Q)&=&\frac{2}{U}+\frac{1}{\beta V}\sum_{K}\Big({\cal G}_{11}^\prime{\cal G}_{22}^{\phantom{\dag}}+{\cal G}_{22}^\prime{\cal G}_{11}^{\phantom{\dag}}
-2{\cal G}_{12}^\prime{\cal G}_{12}^{\phantom{\dag}}\Big),\nonumber\\
\Pi_{12}(Q)&=&\Pi_{21}(-Q)=i\frac{1}{\beta V}\sum_{K}\Big({\cal G}_{22}^\prime{\cal G}_{11}^{\phantom{\dag}}
-{\cal G}_{11}^\prime{\cal G}_{22}^{\phantom{\dag}}\Big).
\end{eqnarray}
For convenience, here we have defined ${\cal G}_{ij}\equiv{\cal G}_{ij}(K)$ and ${\cal G}_{ij}^\prime\equiv{\cal G}_{ij}(K+Q)$.
At $T=0$, these loop functions can be evaluated as
\begin{eqnarray}
\Pi_{nn}(Q)&=&\frac{1}{V}\sum_{\bf k}\frac{E_++E_-}{E_+E_-}\frac{E_+E_--\xi_+\xi_-+\Delta^2}{(iq_l)^2-(E_++E_-)^2},\nonumber\\
\Pi_{ss}(Q)&=&\frac{1}{V}\sum_{\bf k}\frac{E_++E_-}{E_+E_-}\frac{E_+E_--\xi_+\xi_--\Delta^2}{(iq_l)^2-(E_++E_-)^2},\nonumber\\
\Pi_{ns}(Q)&=&\Pi_{sn}(-Q)=\frac{1}{V}\sum_{\bf k}\left(\frac{\xi_+}{E_+}-\frac{\xi_-}{E_-}\right)\frac{iq_l}{(iq_l)^2-(E_++E_-)^2},\nonumber\\
\Pi_{n1}(Q)&=&\Pi_{1n}(-Q)=\Delta\frac{1}{V}\sum_{\bf k}\frac{E_++E_-}{E_+E_-}\frac{\xi_++\xi_-}{(iq_l)^2-(E_++E_-)^2},\nonumber\\
\Pi_{n2}(Q)&=&\Pi_{2n}(-Q)=-i\Delta\frac{1}{V}\sum_{\bf k}\frac{E_++E_-}{E_+E_-}\frac{iq_l}{(iq_l)^2-(E_++E_-)^2},\nonumber\\
\Pi_{s1}(Q)&=&\Pi_{1s}(-Q)=-\Delta\frac{1}{V}\sum_{\bf k}\left(\frac{1}{E_+}-\frac{1}{E_-}\right)\frac{iq_l}{(iq_l)^2-(E_++E_-)^2},\nonumber\\
\Pi_{s2}(Q)&=&\Pi_{2s}(-Q)=i\Delta\frac{1}{V}\sum_{\bf k}\frac{E_++E_-}{E_+E_-}\frac{\xi_+-\xi_-}{(iq_l)^2-(E_++E_-)^2},\nonumber\\
\Pi_{11}(Q)&=&\frac{1}{V}\sum_{\bf k}\left[\frac{E_++E_-}{E_+E_-}\frac{E_+E_-+\xi_+\xi_--\Delta^2}{(iq_l)^2-(E_++E_-)^2}
+\frac{1}{E_{\bf k}}\right],\nonumber\\
\Pi_{22}(Q)&=&\frac{1}{V}\sum_{\bf k}\left[\frac{E_++E_-}{E_+E_-}\frac{E_+E_-+\xi_+\xi_-+\Delta^2}{(iq_l)^2-(E_++E_-)^2}
+\frac{1}{E_{\bf k}}\right],\nonumber\\
\Pi_{12}(Q)&=&\Pi_{21}(-Q)=-i\frac{1}{V}\sum_{\bf k}\left(\frac{\xi_+}{E_+}+\frac{\xi_-}{E_-}\right)\frac{iq_l}{(iq_l)^2-(E_++E_-)^2}.
\end{eqnarray}
Completing the integral over the angle between ${\bf k}$ and ${\bf q}$, we can show that
\begin{eqnarray}
\Pi_{ns}(Q)=\Pi_{sn}(Q)=0,\ \ \ \ \Pi_{s1}(Q)=\Pi_{1s}(Q)=\Pi_{s2}(Q)=\Pi_{2s}(Q)=0.
\end{eqnarray}
Note that these results can also be proven by using the fact ${\cal G}_{22}(K)=-{\cal G}_{11}(-K)$ and hence hold for arbitrary temperature $T$.

Note that the perturbations $\eta_1$ and $\eta_2$ are induced by the external sources. They should be determined by the extreme condition (\ref{RPA-GAP}).  
We have
\begin{equation}
\frac{\delta{\cal W}_{\rm MF}^{(2)}[j;\eta]}{\delta \eta(Q)}=0.
\end{equation}
Using the expression ({\ref{MF-Response}}), to the lowest order in $j_n$ and $j_s$ we obtain 
\begin{eqnarray}\label{RPA-GAP2}
\eta(Q)&=&-\Pi_{\eta\eta}^{-1}(Q)\Pi_{\eta j}(Q)j(Q)+O(j^2),\nonumber\\
\eta^{\rm T}(-Q)&=&-j^{\rm T}(-Q)\Pi_{j\eta}(Q)\Pi_{\eta\eta}^{-1}(Q)+O(j^2).
\end{eqnarray}
Substituting these results into the expression (\ref{MF-Response}) and eliminating the induced perturbations $\eta_1$ and $\eta_2$, we obtain
\begin{eqnarray}
{\cal W}_{\rm MF}^{(2)}[j_n,j_s] =\frac{1}{2}\sum_Q\left(\begin{array}{cc} j_n(-Q)& j_s(-Q) \end{array}\right)\left(\begin{array}{cc} \chi_{nn}(Q) & \chi_{ns}(Q) \\  \chi_{sn}(Q) & \chi_{ss}(Q)\end{array}\right)\left(\begin{array}{c} j_n(Q) \\ j_s(Q) \end{array}\right),
\end{eqnarray}
where the dynamic susceptibility matrix can be expressed as
\begin{eqnarray}\label{MF-RPA}
\left(\begin{array}{cc} \chi_{nn} & \chi_{ns} \\  \chi_{sn} & \chi_{ss}\end{array}\right)
=\left(\begin{array}{cc} \Pi_{nn} & \Pi_{ns} \\  \Pi_{sn} & \Pi_{ss}\end{array}\right)-\left(\begin{array}{cc} \Pi_{n1} & \Pi_{n 2} \\  \Pi_{s1} & \Pi_{s2}\end{array}\right)\left(\begin{array}{cc} \Pi_{11} & \Pi_{12} \\  \Pi_{21} & \Pi_{22}\end{array}\right)^{-1}\left(\begin{array}{cc} \Pi_{1n} & \Pi_{1 s} \\  \Pi_{2n} & \Pi_{2s}\end{array}\right)
\end{eqnarray}
or in short
\begin{equation}
\chi(Q)=\Pi_{jj}(Q)-\Pi_{j\eta}(Q)\Pi_{\eta\eta}^{-1}(Q)\Pi_{\eta j}(Q).
\end{equation}
Note that $\Pi_{\eta\eta}(Q)$ is precisely the inverse propagator of the collective modes; i.e.,
\begin{eqnarray}
\Pi_{\eta\eta}(Q)=\left(\begin{array}{cc} \Pi_{11}(Q) & \Pi_{12}(Q) \\  \Pi_{21}(Q) & \Pi_{22}(Q)\end{array}\right)
=\left(\begin{array}{cc} I_{11}(Q) & q_l I_{12}(Q) \\ -q_l I_{12}(Q) & I_{22}(Q)\end{array}\right).
\end{eqnarray}

For convenience we make the analytical continuation to real frequency $\omega$ and define the following functions
\begin{eqnarray}
I_{nn}(\omega,q)&=&\frac{1}{V}\sum_{\bf k}\frac{E_++E_-}{E_+E_-}
\frac{E_+E_--\xi_+\xi_-+\Delta^2}{\omega^2-(E_++E_-)^2}=\Pi_{nn}(\omega,q),\nonumber\\
I_{ss}(\omega,q)&=&\frac{1}{V}\sum_{\bf k}\frac{E_++E_-}{E_+E_-}
\frac{E_+E_--\xi_+\xi_--\Delta^2}{\omega^2-(E_++E_-)^2}=\Pi_{ss}(\omega,q),\nonumber\\
A(\omega,q)&=&\frac{1}{V}\sum_{\bf k}\frac{E_++E_-}{E_+E_-}\frac{\xi_++\xi_-}{\omega^2-(E_++E_-)^2}
=\frac{\Pi_{n1}(\omega,q)}{\Delta},\nonumber\\
B(\omega,q)&=&\frac{1}{V}\sum_{\bf k}\frac{E_++E_-}{E_+E_-}\frac{1}{\omega^2-(E_++E_-)^2}
=\frac{\Pi_{n2}(\omega,q)}{-i\omega\Delta}.
\end{eqnarray}
Here and in the following $q\equiv|{\bf q}|$. Using the result (\ref{MF-RPA}), we obtain
\begin{eqnarray}
\left(\begin{array}{cc} \chi_{nn} & \chi_{ns} \\  \chi_{sn} & \chi_{ss}\end{array}\right)
&=&\left(\begin{array}{cc} I_{nn} & 0 \\  0 & I_{ss}\end{array}\right)-\frac{\Delta^2}{I_{11}I_{22}-\omega^2I_{12}^2}\left(\begin{array}{cc}  A & -i\omega B \\  0 & 0\end{array}\right)\nonumber\\
&&\times\left(\begin{array}{cc}I_{22} & i\omega I_{12} \\
-i\omega I_{12} & I_{11}\end{array}\right)\left(\begin{array}{cc} A & 0 \\  i\omega B & 0\end{array}\right),
\end{eqnarray}
Working out the product of matrices, we find that the off-diagonal components of the susceptibility matrix $\chi$ vanish. We have
\begin{equation}
\chi_{ns}(\omega,q)=\chi_{sn}(\omega,q)=0.
\end{equation}
The diagonal components $\chi_{nn}(\omega,q)$ and  $\chi_{ss}(\omega,q)$ correspond to the dynamic density and spin response functions, respectively. The density response function can be expressed as
\begin{eqnarray}\label{RPA-D}
\chi_{nn}(\omega,q)=I_{nn}(\omega,q)-\frac{\Delta^2C(\omega,q)}{I_{11}(\omega,q)I_{22}(\omega,q)-\omega^2I_{12}^2(\omega,q)},
\end{eqnarray}
where
\begin{equation}
C(\omega,q)=A^2(\omega,q)I_{22}(\omega,q)+\omega^2B^2(\omega,q)I_{11}(\omega,q)
-2\omega^2A(\omega,q)B(\omega,q)I_{12}(\omega,q).
\end{equation}
It is clear that the density response couples to the collective modes through the nonzero couplings $A(\omega,q)$ and $B(\omega,q)$.
The spin response function is simply given by
\begin{eqnarray}\label{RPA-S}
\chi_{ss}(\omega,q)=I_{ss}(\omega,q).
\end{eqnarray}
Therefore, the spin response does not couple to the collective modes. Using the above expressions for $\chi_{nn}(\omega,q)$ and $\chi_{ss}(\omega,q)$, we can calculate the dynamical structure factors $S_{nn}(\omega,q)$ and $S_{ss}(\omega,q)$. 

We note that the above results (\ref{RPA-D}) and (\ref{RPA-S}), is nothing but the dynamic density and spin response functions from the random phase approximation \cite{RPA-Zwerger, RPA-Hu, KE}. Therefore, we have shown that the generalized mean-field approximation in the presence of external sources, Eq. (\ref{RPA-GAP}), leads to the famous RPA theory for dynamic density and spin responses.

In Fig. 1 we show some numerical results for the density and spin structure factors of a 3D Fermi gas in the BCS-BEC crossover at large momentum 
$q=4.5k_{\rm F}$. The density structure factor is a clear signature of the BCS-BEC crossover.

\begin{figure*}[!htb]
\begin{center}
\includegraphics[width=8cm]{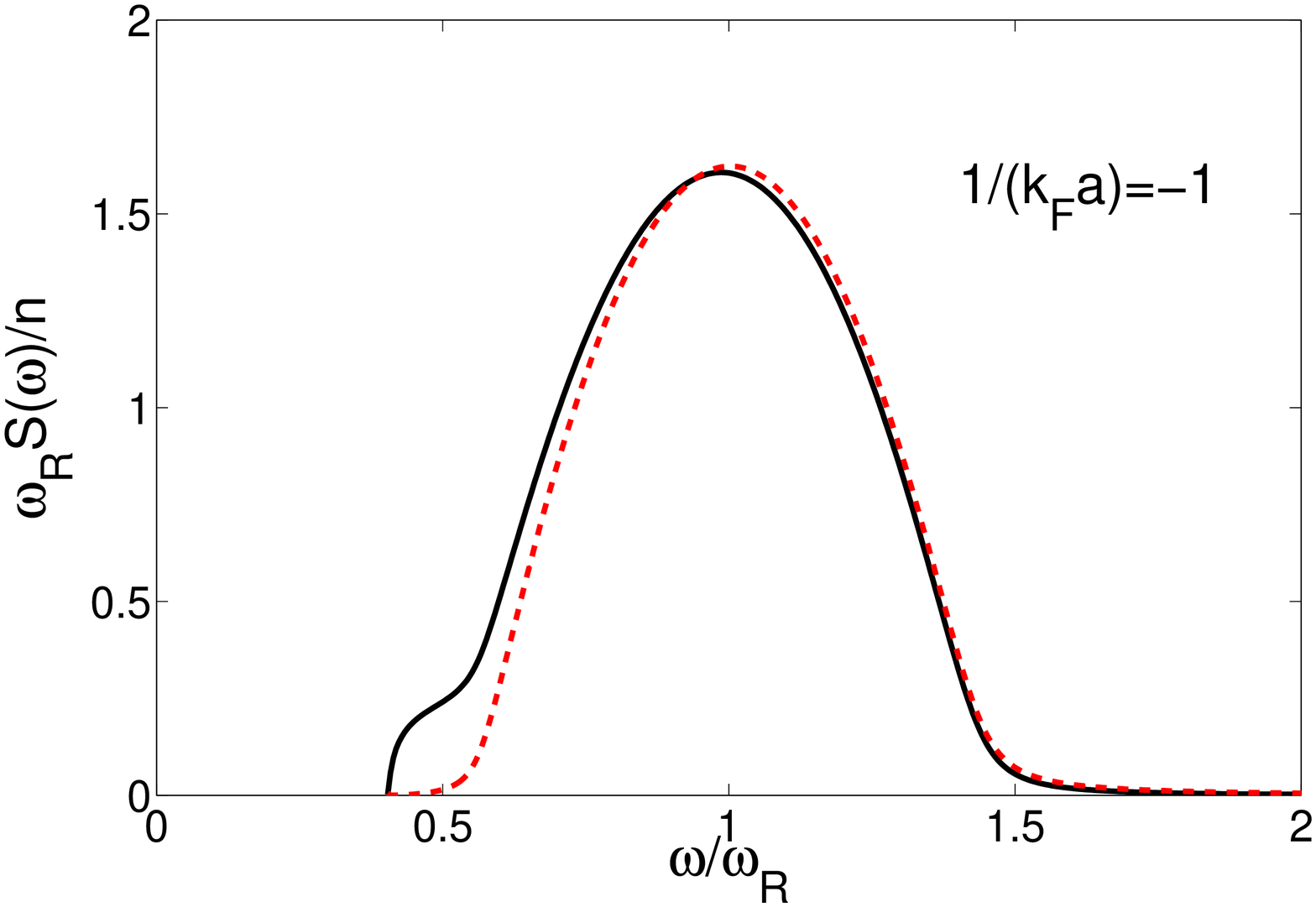}
\includegraphics[width=8cm]{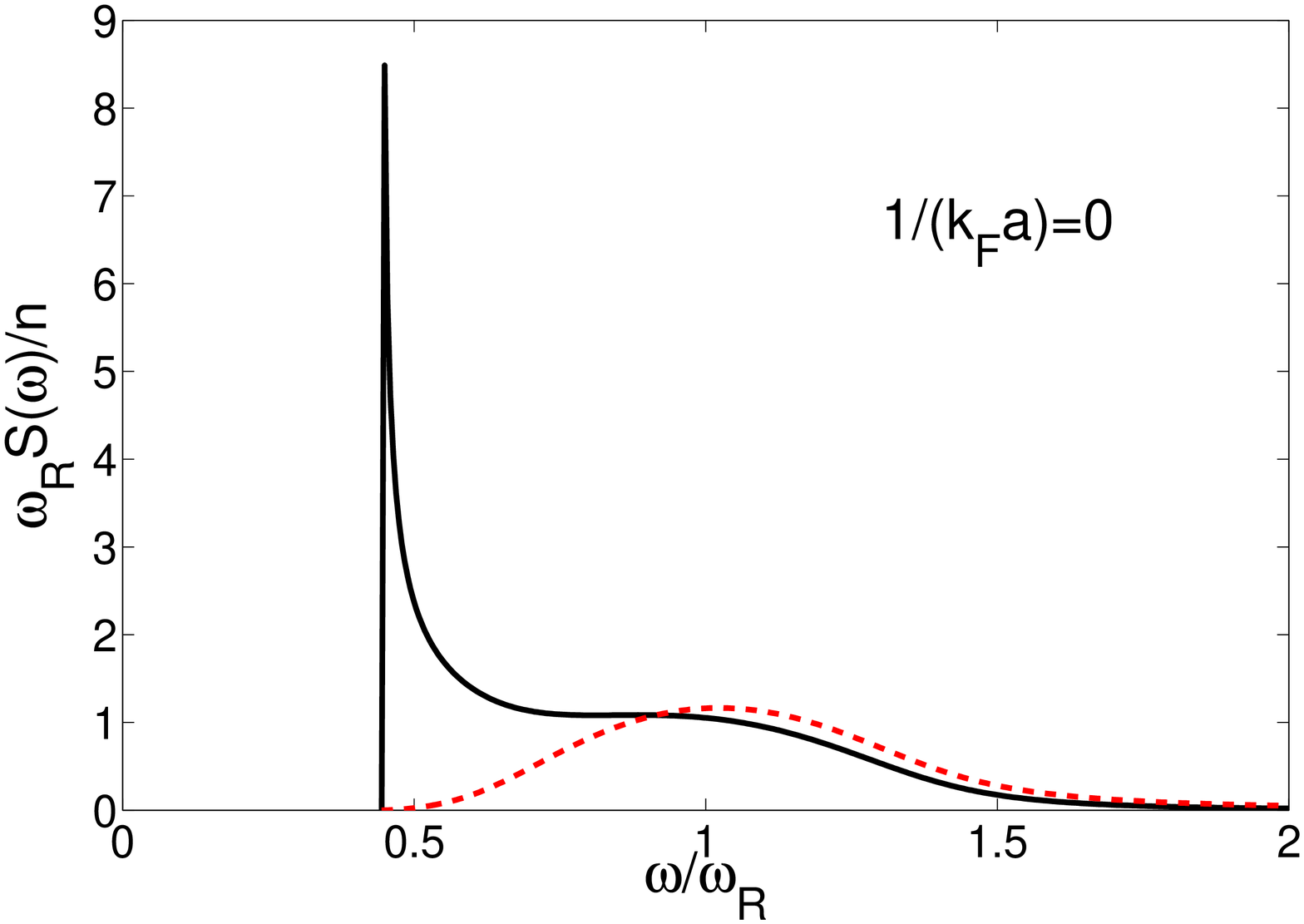}
\includegraphics[width=8cm]{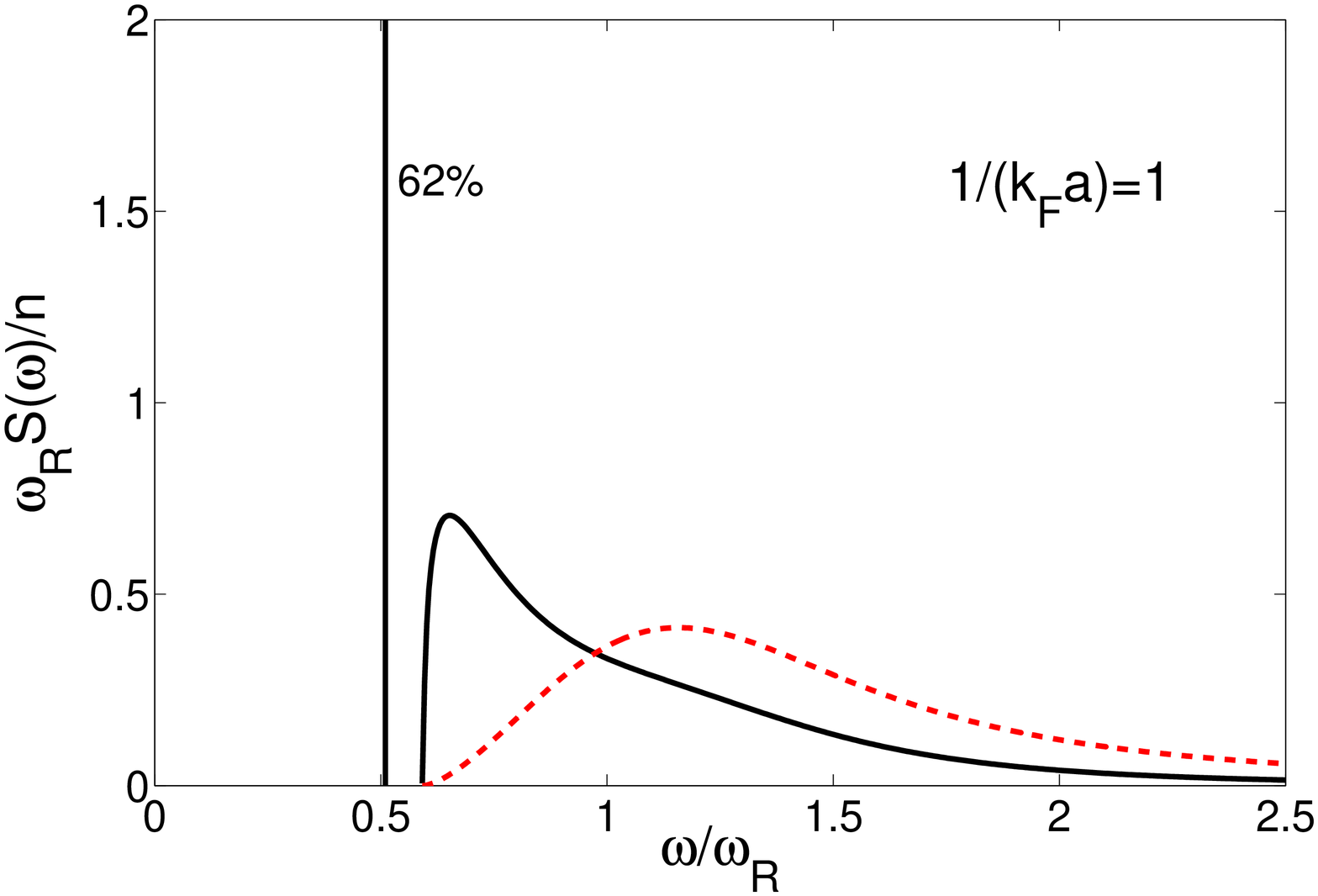}
\caption{(Color-online) Density and spin structure factors of a 3D Fermi gas in the BCS-BEC crossover at zero temperature. The solid and the dashed lines 
denote the density and spin structure factors, respectively. The momentum transfer is fixed at $q=4.5k_{\rm F}$ and $\omega_{\rm R}=q^2/(2m)$. \label{fig1}}
\end{center}
\end{figure*}

\subsection{Static and long wavelength limit}
In the static and long wavelength limit, $\omega=q=0$, the density and spin response functions are related to the compressibility and the spin susceptibility, respectively. For the density response function, we can show that
\begin{equation}\label{STATIC-D}
\chi_{nn}(0,0)=\lim_{q\rightarrow0}\chi_{nn}(\omega=0,q)=\frac{d^2\Omega_{\rm MF}}{d\mu^2},
\end{equation}
where $\Omega_{\rm MF}$ is the grand potential in the BCS-Leggett mean-field approximation. To prove this relation, we write $\Omega_{\rm MF}=\Omega_{\rm MF}(\mu,\Delta)$
with $\Delta=\Delta(\mu)$ being an implicit function determined by the gap equation (21). We have
\begin{equation}
\frac{d^2\Omega_{\rm MF}}{d\mu^2}=\frac{\partial^2\Omega_{\rm MF}(\mu,\Delta)}{\partial\mu^2}
+\frac{\partial^2\Omega_{\rm MF}(\mu,\Delta)}{\partial\mu\partial\Delta}\frac{d\Delta(\mu)}{d\mu}.
\end{equation}
Using the gap equation for $\Delta$, we obtain
\begin{equation}
\frac{d\Delta(\mu)}{d\mu}=-\frac{\partial^2\Omega_{\rm MF}(\mu,\Delta)}{\partial\mu\partial\Delta}
\left[\frac{\partial^2\Omega_{\rm MF}(\mu,\Delta)}{\partial\Delta^2}\right]^{-1}.
\end{equation}
Therefore, we have
\begin{equation}
\frac{d^2\Omega_{\rm MF}}{d\mu^2}=\frac{\partial^2\Omega_{\rm MF}(\mu,\Delta)}{\partial\mu^2}
-\left[\frac{\partial^2\Omega_{\rm MF}(\mu,\Delta)}{\partial\mu\partial\Delta}\right]^2
\left[\frac{\partial^2\Omega_{\rm MF}(\mu,\Delta)}{\partial\Delta^2}\right]^{-1}.
\end{equation}
On the other hand, using the expression (\ref{RPA-D}), we obtain
\begin{equation}
\chi_{nn}(0,0)=I_{nn}(0,0)-\frac{\Delta^2A^2(0,0)}{I_{11}(0,0)}.
\end{equation}
It is straightforward to show that
\begin{eqnarray}
&&I_{nn}(0,0)=-\frac{1}{V}\sum_{\bf k}\frac{\Delta^2}{E_{\bf k}^3}=\frac{\partial^2\Omega_{\rm MF}(\mu,\Delta)}{\partial\mu^2},\nonumber\\
&&I_{11}(0,0)=\frac{1}{V}\sum_{\bf k}\frac{\Delta^2}{E_{\bf k}^3}=\frac{\partial^2\Omega_{\rm MF}(\mu,\Delta)}{\partial\Delta^2},\nonumber\\
&&\Delta A(0,0)=-\Delta\frac{1}{V}\sum_{\bf k}\frac{\xi_{\bf k}}{E_{\bf k}^3}=\frac{\partial^2\Omega_{\rm MF}(\mu,\Delta)}{\partial\mu\partial\Delta}.
\end{eqnarray}
Therefore, we have proven the relation (\ref{STATIC-D}).

For the spin response function, we find that $\chi_{ss}(0,0)=0$, which is consistent with the fact that the spin susceptibility vanishes for
an $s$-wave Fermi superfluid at $T=0$. In fact, if we introduce a Zeeman field $h$ and add a term $h(\bar{\psi}_\uparrow\psi_\uparrow-\bar{\psi}_\downarrow\psi_\downarrow)$ to the Hamiltonian, we can show that
\begin{equation}
\chi_{ss}(0,0)=\frac{\partial^2\Omega_{\rm MF}(\mu,h)}{\partial h^2}\bigg|_{h=0}=0.
\end{equation}

\subsection{$f$-sum rule}\label{fsum}
It is well known that the dynamic structure factors should satisfy precisely the so-called $f$-sum rules \cite{SSF-QMC01},
\begin{eqnarray}
\int_0^\infty d\omega \omega S_{nn}(\omega,q)=\frac{nq^2}{2m},\ \ \ \ \ \ \ \ \int_0^\infty d\omega \omega S_{ss}(\omega,q)=\frac{nq^2}{2m},
\end{eqnarray}
or equivalently,
\begin{eqnarray}
\int_{-\infty}^\infty d\omega \omega S_{nn}(\omega,q)=\frac{nq^2}{m},\ \ \ \ \ \ \ \
\int_{-\infty}^\infty d\omega \omega S_{ss}(\omega,q)=\frac{nq^2}{m}.
\end{eqnarray}
Here we check these sum rules and find the key facts that guarantee them.

We first check the $f$-sum rule for the spin structure factor. We have explicitly
\begin{equation}
S_{ss}(\omega,q)=\frac{1}{V}\sum_{\bf k}\frac{E_+E_--\xi_+\xi_--\Delta^2}{2E_+E_-}\left[\delta(\omega-E_+-E_-)-\delta(\omega+E_++E_-)\right].
\end{equation}
Applying the $f$-sum, we obtain
\begin{eqnarray}
\int_0^\infty d\omega \omega S_{ss}(\omega,q)=\frac{1}{V}\sum_{\bf k}\frac{E_++E_-}{2E_+E_-}\left(E_+E_--\xi_+\xi_--\Delta^2\right).
\end{eqnarray}
Using the following identities
\begin{eqnarray}
&&\xi_+\xi_-+\Delta^2=\frac{E_+^2+E_-^2}{2}-\frac{1}{2}\left(\frac{{\bf k}\cdot{\bf q}}{m}\right)^2,
\ \ \ \ \ \ \xi_+-\xi_-=\frac{{\bf k}\cdot{\bf q}}{m},\nonumber\\
&&E_+^2-E_-^2=\xi_+^2-\xi_-^2=\left(\xi_++\xi_-\right)\frac{{\bf k}\cdot{\bf q}}{m},
\end{eqnarray}
we obtain
\begin{eqnarray}
&&\frac{1}{V}\sum_{\bf k}\frac{E_++E_-}{2E_+E_-}\left(E_+E_--\xi_+\xi_--\Delta^2\right)\nonumber\\
&=&\frac{1}{V}\sum_{\bf k}\left[\frac{{\bf k}\cdot{\bf q}}{2m}\left(1-\frac{\xi_-}{E_-}\right)
-\frac{{\bf k}\cdot{\bf q}}{2m}\left(1-\frac{\xi_+}{E_+}\right)\right]\nonumber\\
&=&\frac{q^2}{2m}\frac{1}{V}\sum_{\bf k}\left(1-\frac{\xi_{\bf k}}{E_{\bf k}}\right).
\end{eqnarray}
Here we make the shifts ${\bf k}\rightarrow{\bf k}+{\bf q}/2$ and ${\bf k}\rightarrow{\bf k}-{\bf q}/2$ for the first and second terms in the second line to obtain the expression in the last line. Therefore, we have
\begin{eqnarray}
\int_0^\infty d\omega \omega S_{ss}(\omega,q)=\frac{q^2}{2m}\frac{1}{V}\sum_{\bf k}\left(1-\frac{\xi_{\bf k}}{E_{\bf k}}\right).
\end{eqnarray}
Now it becomes clear that the spin $f$-sum rule is guaranteed by the number equation (\ref{MF-NUM}) in the BCS-Leggett mean-field theory.

To prove the $f$-sum rule for the density structure factor, we note that $\chi_{nn}(\omega,q)=\chi_{ss}(\omega,q)+2\chi_{\uparrow\downarrow}(\omega,q)$. Therefore, we only need to show that
\begin{eqnarray}\label{FSUM-UD}
\int_0^\infty d\omega \omega S_{\uparrow\downarrow}(\omega,q)=0,
\end{eqnarray}
where $S_{\uparrow\downarrow}(\omega,q)=-(1/\pi){\rm Im}\chi_{\uparrow\downarrow}(\omega+i\epsilon,q)$. The explicit form of $\chi_{\uparrow\downarrow}(\omega,q)$ reads
\begin{eqnarray}
\chi_{\uparrow\downarrow}(\omega,q)=\frac{\Delta^2}{2}\left[2B(\omega,q)
-\frac{C(\omega,q)}{I_{11}(\omega,q)I_{22}(\omega,q)-\omega^2I_{12}^2(\omega,q)}\right].
\end{eqnarray}
For the proof of (\ref{FSUM-UD}), we refer to Ref. \cite{SUM-1}.  It is shown in \cite{SUM-1} that the $f$-sum rule of the density structure factor is related to the Ward identity and hence gauge invariance of the response theory. We note that in deriving the expressions of $I_{11}$, $I_{22}$, and $I_{12}$, the mean-field
gap equation is used. Therefore, the density $f$-sum rule is ensured by both the gap and number equations in the mean-field theory.

\section{GPF response theory: beyond the RPA}\label{s5}

In the RPA theory, the response functions and the structure factors include only the contribution from the fermionic quasiparticles, except for the fact that the density
response couples to the collective modes. At low temperature, it is known that the quantum fluctuations are rather important \cite{TH02,TH03,TH04} in the BCS-BEC crossover. Therefore, it is interesting to develop a response theory beyond the RPA to include properly the contribution from the quantum fluctuations. In this part, we 
consider another gapless approximation for the BCS-BEC crossover, the GPF theory introduced in Sec. \ref{s2}. We construct the response theory corresponding to the GPF theory in equilibrium. We expect that the GPF response theory quantitatively improves the RPA theory.  It is intuitive to take a look at the static and long wavelength limit of the density response function $\chi_{nn}(\omega,q)$, which is related to the compressibility of the equilibrium state. At unitarity in 3D, the equation of state at $T=0$ is universally given by
\begin{equation}
\Omega(\mu)=\xi^{-3/2}\Omega_0(\mu),
\end{equation}
where $\Omega_0(\mu)=-2(2m)^{3/2}\mu^{5/2}/(15\pi^2)$ is the equation of state of a noninteracting Fermi gas. Therefore, we have
\begin{equation}
\frac{\chi_{nn}(0,0)}{\chi_{nn}^0(0,0)}=\frac{1}{\xi},
\end{equation}
where $\chi^0_{nn}(\omega,q)$ is the density response function of a noninteracting Fermi gas. In the BCS-Leggett mean-field theory, we have $\xi=0.5906$, while the GPF theory gives $\xi=0.40(1)$ \cite{TH03,TH04}. On the other hand, recent quantum Monte Carlo simulations have determined that $\xi\simeq0.37-0.38$ \cite{EOSmc3,EOSmc4}. It is obvious that the GPF theory predicts better result than the BCS-Leggett mean-field theory.

\subsection{Dynamic responses of GPF ground state: Idea and sketch }\label{s5-1}

The naive thought is that we still employ the formalism of the response functions $\chi_{nn}^{\rm MF}(\omega,q)$ and $\chi_{ss}^{\rm MF}(\omega,q)$ derived in Sec. \ref{s4} but with the order parameter $\Delta$ and the chemical potential $\mu$ replaced by the values obtained in the GPF theory. Here and in the following
we use MF to denote the mean-field results for the response functions obtained in Sec. \ref{s4}. According to the discussions in Sec. \ref{fsum}, it is easy to show that the mean-field results of the structure factors do not satisfy the $f$-sum rules in the GPF theory. We have
\begin{eqnarray}
\int_0^\infty d\omega \omega S_{nn}^{\rm MF}(\omega,q)=\int_0^\infty d\omega \omega S_{ss}^{\rm MF}(\omega,q)=\frac{n_{\rm MF}q^2}{2m}<\frac{nq^2}{2m}.
\end{eqnarray}
Here the total particle density $n=n_{\rm MF}+n_{\rm GF}$, where $n_{\rm MF}$ is the mean-field contribution to the particle density and $n_{\rm GF}$ is the particle  density coming from the Gaussian pair fluctuations.. Therefore, to restore the $f$-sum rules, we should include the contributions $\chi_{nn}^{\rm GF}(\omega,q)$ and $\chi_{ss}^{\rm GF}(\omega,q)$ coming from the Gaussian fluctuations. 
Here we use GF to denote the contributions from the Gaussian fluctuations. The total response functions in the GPF response theory can be expressed as
\begin{eqnarray}
\chi_{nn}(\omega,q)&=&\chi_{nn}^{\rm MF}(\omega,q)+\chi_{nn}^{\rm GF}(\omega,q),\nonumber\\
\chi_{ss}(\omega,q)&=&\chi_{ss}^{\rm MF}(\omega,q)+\chi_{ss}^{\rm GF}(\omega,q).
\end{eqnarray}
We expect that the $f$-sum rules are restored by the contributions from the Gaussian pair fluctuations; that is,
\begin{eqnarray}
\int_0^\infty d\omega \omega S_{nn}^{\rm GF}(\omega,q)=\int_0^\infty d\omega \omega S_{ss}^{\rm GF}(\omega,q)=\frac{n_{\rm GF}q^2}{2m}.
\end{eqnarray}

In the presence of the external sources, the partition function in the GPF theory is given by
\begin{eqnarray}
{\cal Z}[J]\simeq \exp \Big\{-{\cal W}_{\rm MF}[j_n,j_s;\eta_1,\eta_2]-{\cal W}_{\rm GF}[j_n,j_s;\eta_1,\eta_2]\Big\},
\end{eqnarray}
where the BCS-Leggett mean-field contribution ${\cal W}_{\rm MF}[j_n,j_s;\eta_1,\eta_2]$ has been evaluated in Sec. \ref{s4} and
${\cal W}_{\rm GF}[j_n,j_s;\eta_1,\eta_2]$ is the contribution from the Gaussian pair fluctuations. For each part of the generating functional, we
expand it up to the quadratic order in the external sources $j_n$ and $j_s$ as well as the induced perturbations $\eta_1$ and $\eta_2$.
We have
\begin{eqnarray}
&&{\cal W}_{\rm MF}[j_n,j_s,\eta_1,\eta_2]={\cal W}_{\rm MF}^{(0)}+{\cal W}_{\rm MF}^{(1)}[j_n,j_s;\eta_1,\eta_2]
+{\cal W}_{\rm MF}^{(2)}[j_n,j_s;\eta_1,\eta_2]+\cdots,\nonumber\\
&&{\cal W}_{\rm GF}[j_n,j_s,\eta_1,\eta_2]={\cal W}_{\rm GF}^{(0)}+{\cal W}_{\rm GF}^{(1)}[j_n,j_s;\eta_1,\eta_2]
+{\cal W}_{\rm GF}^{(2)}[j_n,j_s;\eta_1,\eta_2]+\cdots,\nonumber\\
\end{eqnarray}
where ${\cal W}_{\rm MF}^{(0)}=\beta V\Omega_{\rm MF}$ and ${\cal W}_{\rm GF}^{(0)}=\beta V\Omega_{\rm GF}$ are the actions in the absence of the 
external sources. 

The linear terms are related only to the zero modes of the perturbations. For the mean-field part, we have
\begin{eqnarray}
{\cal W}_{\rm MF}^{(1)}=-\sqrt{\beta V}n_{\rm MF}j_n(0).
\end{eqnarray}
For the Gaussian-fluctuation part, we will show that
\begin{eqnarray}\label{linear-term}
{\cal W}_{\rm GF}^{(1)}=\sqrt{\beta V}\left[{\cal C}_nj_n(0)+{\cal C}_sj_s(0)+{\cal C}_1\eta_1(0)+{\cal C}_2\eta_2(0)\right],
\end{eqnarray}
where the coefficients ${\cal C}_a$ ($a=n,s,1,2$) will be determined later. We note that ${\cal C}_1$ does not vanish because the
order parameter $\Delta(\mu)$ is determined by the mean-field gap equation. Later we will show that ${\cal C}_s=0$ because the system is spin-balanced 
and ${\cal C}_2=0$ because we have chosen $\Delta(\mu)$ to be real. Therefore,  to eliminate the induced perturbation $\eta_1$, we should expand it up to the second order in $j_n$ and $j_s$. One key problem we encounter here is how to eliminate the induced perturbations $\eta_1$ and $\eta_2$ when going beyond mean field. Keep in mind that in the GPF theory, the order parameter is determined only by minimizing the mean-field effective action ${\cal S}_{\rm MF}$. Therefore, we determine the induced perturbations $\eta_1$ and $\eta_2$ through by using the extreme condition (\ref{RPA-GAP}). We have
\begin{eqnarray}
\eta_1(0)&=&{\cal R}_nj_n(0)+{\cal R}_sj_s(0)\nonumber\\
&+&\frac{1}{2\sqrt{\beta V}}\sum_Q\left(\begin{array}{cccc} j^{\rm T}(-Q)&\eta^{\rm T}(-Q)\end{array}\right)
\left(\begin{array}{cc}{\cal U}_{jj}(Q)&{\cal U}_{j\eta}(Q)\\
{\cal U}_{\eta j}(Q)&{\cal U}_{\eta\eta}(Q)\end{array}\right)
\left(\begin{array}{c} j(Q) \\ \eta(Q)\end{array}\right)+\cdots.
\end{eqnarray}
Later we will show that ${\cal R}_s=0$.  Note that $\eta(Q)$ and $\eta(-Q)$ should be eliminated by using the extreme condition (\ref{RPA-GAP}) or explicitly Eq. (\ref{RPA-GAP2}). However, it is convenient to keep this form to combine other contributions. Therefore, we have
\begin{eqnarray}
{\cal W}_{\rm GF}^{(1)}&=&\sqrt{\beta V}\left({\cal C}_n+{\cal C}_1{\cal R}_n\right)j_n(0)\nonumber\\
&+&\frac{1}{2}\sum_Q\left(\begin{array}{cccc} j^{\rm T}(-Q)&\eta^{\rm T}(-Q)\end{array}\right)
\left(\begin{array}{cc}\Xi_{jj}^{(1)}(Q)&\Xi_{j\eta}^{(1)}(Q)\\
\Xi_{\eta j}^{(1)}(Q)&\Xi_{\eta\eta}^{(1)}(Q)\end{array}\right)
\left(\begin{array}{c} j(Q) \\ \eta(Q)\end{array}\right)+\cdots.
\end{eqnarray}
The coefficient ${\cal C}_n+{\cal C}_1{\cal R}_n$ is related to nothing but the Gaussian-fluctuation contribution to the particle density,
\begin{eqnarray}
{\cal C}_n+{\cal C}_1{\cal R}_n=-n_{\rm GF}.
\end{eqnarray}
The elements $\Xi^{(1)}_{mn}$ ($m,n=j,\eta$) is given by $\Xi^{(1)}_{mn}={\cal C}_1{\cal U}_{mn}$. 

Because of the translational invariance, the quadratic terms ${\cal W}_{\rm MF}^{(2)}[j,\eta]$ and ${\cal W}_{\rm GF}^{(2)}[j,\eta]$ can be expressed as
\begin{eqnarray}
{\cal W}_{\rm MF}^{(2)}&=&\frac{1}{2}\sum_Q\left(\begin{array}{cccc} j^{\rm T}(-Q)&\eta^{\rm T}(-Q)\end{array}\right)
\left(\begin{array}{cc}\Pi_{jj}(Q)&\Pi_{j\eta}(Q)\\
\Pi_{\eta j}(Q)&\Pi_{\eta\eta}(Q)\end{array}\right)
\left(\begin{array}{c} j(Q) \\ \eta(Q)\end{array}\right),\nonumber\\
{\cal W}_{\rm GF}^{(2)}&=&\frac{1}{2}\sum_Q\left(\begin{array}{cccc} j^{\rm T}(-Q)&\eta^{\rm T}(-Q)\end{array}\right)
\left(\begin{array}{cc}\Xi_{jj}^{(2)}(Q)&\Xi_{j\eta}^{(2)}(Q)\\
\Xi_{\eta j}^{(2)}(Q)&\Xi_{\eta\eta}^{(2)}(Q)\end{array}\right)
\left(\begin{array}{c} j(Q) \\ \eta(Q)\end{array}\right).
\end{eqnarray}
Here the mean-field susceptibility functions $\Pi_{mn}(Q)$ ($m,n=j,\eta$) have been evaluated in Sec. \ref{s4}.
We will evaluate the fluctuation contributions $\Xi_{mn}^{(1)}(Q)$ and $\Xi_{mn}^{(2)}(Q)$ in the following subsection.
In summary, the expansion of the generating functional takes the following form
\begin{eqnarray}
{\cal W}_{\rm MF}[J;\eta]&=&\beta V\Omega_{\rm MF}-\sqrt{\beta V}n_{\rm MF}j_n(0)\nonumber\\
&+&\frac{1}{2}\sum_Q\left(\begin{array}{cccc} j^{\rm T}(-Q)&\eta^{\rm T}(-Q)\end{array}\right)
\left(\begin{array}{cc}\Pi_{jj}(Q)&\Pi_{j\eta}(Q)\\
\Pi_{\eta j}(Q)&\Pi_{\eta\eta}(Q)\end{array}\right)
\left(\begin{array}{c} j(Q) \\ \eta(Q)\end{array}\right)+\cdots,\nonumber\\
{\cal W}_{\rm GF}[J;\eta]&=&\beta V\Omega_{\rm GF}-\sqrt{\beta V}n_{\rm GF}j_n(0)\nonumber\\
&+&\frac{1}{2}\sum_Q\left(\begin{array}{cccc} j^{\rm T}(-Q)&\eta^{\rm T}(-Q)\end{array}\right)
\left(\begin{array}{cc}\Xi_{jj}(Q)&\Xi_{j\eta}(Q)\\
\Xi_{\eta j}(Q)&\Xi_{\eta\eta}(Q)\end{array}\right)
\left(\begin{array}{c} j(Q) \\ \eta(Q)\end{array}\right)+\cdots.
\end{eqnarray}
Here $\Xi_{mn}(Q)=\Xi_{mn}^{(1)}(Q)+\Xi_{mn}^{(2)}(Q)$.

Using the extreme condition (\ref{RPA-GAP}) or explicitly Eq. (\ref{RPA-GAP2}) to eliminate the induced perturbations $\eta_1$ and $\eta_2$, we obtain
\begin{eqnarray}
&&{\cal W}_{\rm MF}[J]=\beta V\Omega_{\rm MF}-\sqrt{\beta V}n_{\rm MF}j_n(0)+\frac{1}{2}\sum_Qj^{\rm T}(-Q)\chi_{\rm MF}(Q)j(Q)+O(j^3),\nonumber\\
&&{\cal W}_{\rm GF}[J]=\beta V\Omega_{\rm GF}-\sqrt{\beta V}n_{\rm GF}j_n(0)+\frac{1}{2}\sum_Qj^{\rm T}(-Q)\chi_{\rm GF}(Q)j(Q)+O(j^3),
\end{eqnarray}
where the mean-field contribution $\chi_{\rm MF}(Q)$ is given by
\begin{equation}
\chi_{\rm MF}(Q)=\Pi_{jj}(Q)-\Pi_{j\eta}(Q)\Pi_{\eta\eta}^{-1}(Q)\Pi_{\eta j}(Q)
\end{equation}
and the pair-fluctuation contribution $\chi_{\rm GF}(Q)$ reads
\begin{eqnarray}
\chi_{\rm GF}(Q)&=&\Xi_{jj}(Q)-\Pi_{j\eta}(Q)\Pi_{\eta\eta}^{-1}(Q)\Xi_{\eta j}(Q)-\Xi_{j\eta}(Q)\Pi_{\eta\eta}^{-1}(Q)\Pi_{\eta j}(Q)\nonumber\\
&&+\ \Pi_{j\eta}(Q)\Pi_{\eta\eta}^{-1}(Q)\Xi_{\eta\eta}(Q)\Pi_{\eta\eta}^{-1}(Q)\Pi_{\eta j}(Q).
\end{eqnarray}
Note that the mean-field contribution $\chi_{\rm MF}(Q)$ takes the same form as that derived in Sec. \ref{s4}. Finally, the total generating functional reads 
\begin{eqnarray}
{\cal W}[J]=\beta V\Omega-\sqrt{\beta V}nj_n(0)+\frac{1}{2}\sum_Qj^{\rm T}(-Q)\chi(Q)j(Q)+O(j^3),
\end{eqnarray}
where the full dynamic response function $\chi(Q)$ is 
\begin{equation}
\chi(Q)=\chi_{\rm MF}(Q)+\chi_{\rm GF}(Q).
\end{equation}

The above prescription for eliminating the induced perturbations $\eta_1$ and $\eta_2$ is a natural generalization of the gapless approximation in the GPF theory for equilibrium state. Since the mean-field gap equation (\ref{MF-GAP}) guarantees that $\Pi_{\eta\eta}(0,{\bf 0})=0$, the use of the extreme condition 
(\ref{RPA-GAP}) or explicitly Eq. (\ref{RPA-GAP2}) ensures that the low-energy collective mode which couples to the density response is gapless in the static and long wavelength limit. Moreover, the use of the extreme condition (\ref{RPA-GAP}) or explicitly Eq. (\ref{RPA-GAP2}) also ensures that we recover the correct limit in the static and long wavelength limit, $\omega\rightarrow0$ and $q\rightarrow0$.  Eq. (\ref{RPA-GAP2}) can be explicitly expressed as
\begin{eqnarray}
\eta_1(Q)&=&-\Delta\frac{A(\omega,q)I_{22}(\omega,q)-\omega^2B(\omega,q)I_{12}(\omega,q)}
{I_{11}(\omega,q)I_{22}(\omega,q)-\omega^2I_{12}^2(\omega,q)}j_n(Q)+O(j^2),\nonumber\\
\eta_2(Q)&=&i\omega\Delta\frac{A(\omega,q)I_{12}(\omega,q)-B(\omega,q)I_{11}(\omega,q)}
{I_{11}(\omega,q)I_{22}(\omega,q)-\omega^2I_{12}^2(\omega,q)}j_n(Q)+O(j^2).
\end{eqnarray}
In the static and long wavelength limit, we find
\begin{eqnarray}
\lim_{Q\rightarrow0}\frac{\delta\eta_1(Q)}{\delta j_n(Q)}\bigg|_{j\rightarrow0}&=&-\frac{\Delta A(0,0)}{I_{11}(0,0)}
=\frac{d\Delta(\mu)}{d\mu},\nonumber\\
\lim_{Q\rightarrow0}\frac{\delta\eta_2(Q)}{\delta j_n(Q)}\bigg|_{j\rightarrow0}&=&0.
\end{eqnarray}
On the other hand, we expect that in the static and long wavelength limit the density response function $\chi_{nn}(\omega,q)$ satisfies the relation
\begin{equation}
\chi_{nn}(0,0)=\frac{d^2\Omega(\mu)}{d\mu^2}=\frac{d^2\Omega_{\rm MF}}{d\mu^2}
+\frac{d^2\Omega_{\rm GF}}{d\mu^2},
\end{equation}
where the mean-field contribution is simply given by
\begin{equation}
\frac{d^2\Omega_{\rm MF}}{d\mu^2}=\frac{\partial^2\Omega_{\rm MF}(\mu,\Delta)}{\partial\mu^2}
+\frac{\partial^2\Omega_{\rm MF}(\mu,\Delta)}{\partial\mu\partial\Delta}\frac{d\Delta(\mu)}{d\mu}.
\end{equation}
The pair-fluctuation contribution includes a number of terms. We have
\begin{eqnarray}
\frac{d^2\Omega_{\rm GF}}{d\mu^2}&=&\frac{\partial^2\Omega_{\rm GF}(\mu,\Delta)}{\partial\mu^2}
+2\frac{\partial^2\Omega_{\rm GF}(\mu,\Delta)}{\partial\mu\partial\Delta}\frac{d\Delta(\mu)}{d\mu}
+\frac{\partial^2\Omega_{\rm GF}(\mu,\Delta)}{\partial\Delta^2}\left[\frac{d\Delta(\mu)}{d\mu}\right]^2\nonumber\\
&&+\frac{\partial\Omega_{\rm GF}(\mu,\Delta)}{\partial\Delta}\frac{d^2\Delta(\mu)}{d\mu^2}
\end{eqnarray}
Note that the last term corresponds to the contribution $\Xi^{(1)}(Q)$ in the limit $Q\rightarrow0$. Therefore, the careful treatment of the linear term (\ref{linear-term}) ensures the correct static and long wavelength limit. Similarly, by introducing a Zeeman field $h$, we have
\begin{equation}
\chi_{ss}(0,0)=\frac{\partial^2\Omega(\mu,h)}{\partial h^2}\bigg|_{h=0}.
\end{equation}
We expect that $\chi_{ss}(0,0)=0$ holds in the GPF theory.

\subsection{Deriving the GPF generating functional}

Now we derive the Gaussian-fluctuation contribution to the generating functional, ${\cal W}_{\rm GF}[J]$. We start from the partition function (\ref{Partition-J}) with the effective action (\ref{Action-J}). We write
\begin{equation}
\Phi(x)=\Delta_{\rm cl}(x)+\phi(x),
\end{equation}
where $\Delta_{\rm cl}(x)=\Delta+\eta_1(x)+i\eta_2(x)$ is the classic field in the presence of the external sources and $\phi(x)=\phi_1(x)+i\phi_2(x)$ is the quantum fluctuation around the classical field. The effective action is given by
\begin{eqnarray}
{\cal S}_{\rm{eff}}[\Phi, \Phi^*;J] = \int dx \frac{|\Phi(x)|^{2}}{U} - \mbox{Trln} [{\bf G}_J^{-1}(x,x^\prime)].
\end{eqnarray}
According to the GPF theory, we expand the above effective action to the quadratic terms in the quantum fluctuations $\phi$ and $\phi^*$. We have
\begin{eqnarray}
{\cal S}_{\rm{eff}}[\Phi, \Phi^*;J] = {\cal S}_{\rm MF}[\Delta_{\rm cl},\Delta_{\rm cl}^*;J]+{\cal S}_{\rm GF}[\phi,\phi^*;J]+\cdots.
\end{eqnarray}
Note that the classical field $\Delta_{\rm cl}(x)$ is determined by the mean-field part,
\begin{equation}
\frac{\delta{\cal S}_{\rm MF}[\Delta_{\rm cl}, \Delta_{\rm cl}^*;J]}{\delta\Delta_{\rm cl}(x)}=0,\ \ \ \ \ \ 
\frac{\delta{\cal S}_{\rm MF}[\Delta_{\rm cl}, \Delta_{\rm cl}^*;J]}{\delta\Delta_{\rm cl}^*(x)}=0.
\end{equation}
Hence we can show that the linear terms in $\phi$ and $\phi^*$ vanish exactly. The Gaussian fluctuation contribution ${\cal S}_{\rm GF}$ can be evaluated by using the derivative expansion. It is convenient to work in the momentum space. For the Nambu-Gor'kov Green's function ${\bf G}_J$, it is a matrix in the momentum space as well as in the Nambu-Gor'kov space. We write
\begin{equation}
({\bf G}_J^{-1})_{K,K^\prime}=({\cal G}_J^{-1})_{K,K^\prime}-(\Sigma_\phi)_{K,K^\prime}
\end{equation}
where ${\cal G}_J$ is the mean-field Green's function in the presence of the external sources and $\Sigma_{\phi}$ is given by
\begin{eqnarray}
(\Sigma_\phi)_{K,K^\prime}=\frac{-1}{\sqrt{\beta V}}\Big[\Gamma_+\phi(K-K^\prime)+\Gamma_-\phi^*(K^\prime-K)\Big].
\end{eqnarray}
After some manipulations, the quadratic terms in $\phi$ and $\phi^*$, corresponding to the Gaussian pair fluctuations, can be written in a compact form
\begin{eqnarray}
{\cal S}_{\rm GF}[\phi,\phi^*;J]=\frac{1}{2}\sum_{Q,Q^\prime}\left(\begin{array}{cc}
\phi^*(Q) & \phi(-Q)\end{array}\right)({\bf M}_J)_{Q,Q^\prime}\left(\begin{array}{cc} \phi(Q^\prime)\\
\phi^*(-Q^\prime)\end{array}\right),
\end{eqnarray}
where the inverse boson propagator ${\bf M}_J$ in the presence of the external sources takes the form
\begin{eqnarray}\label{Boson-J}
({\bf M}_J)_{Q,Q^\prime}=\left(\begin{array}{cc}({\bf M}_J^{11})_{Q,Q^\prime}&({\bf M}_J^{12})_{Q,Q^\prime}\\
({\bf M}_J^{21})_{Q,Q^\prime}& ({\bf M}_J^{22})_{Q,Q^\prime}\end{array}\right)
=\left(\begin{array}{cc}({\bf M}_J^{-+})_{Q,Q^\prime}&({\bf M}_J^{--})_{Q,Q^\prime}\\
({\bf M}_J^{++})_{Q,Q^\prime}& ({\bf M}_J^{+-})_{Q,Q^\prime}\end{array}\right).
\end{eqnarray}
The elements of ${\bf M}_J$ can be expressed in terms of the mean-field Green's function ${\cal G}_J$. We have
\begin{eqnarray}
&&({\bf M}_J^{11})_{Q,Q^\prime}=({\bf M}_J^{-+})_{Q,Q^\prime}=\frac{\delta_{Q,Q^\prime}}{U}+\frac{1}{\beta V}\sum_{K,K^\prime}{\rm Tr}_{\rm NG}
\left[({\cal G}_J)_{K,K^\prime-Q}\Gamma_-({\cal G}_J)_{K^\prime,K+Q^\prime}\Gamma_+\right],\nonumber\\
&&({\bf M}_J^{22})_{Q,Q^\prime}=({\bf M}_J^{+-})_{Q,Q^\prime}=\frac{\delta_{Q,Q^\prime}}{U}+\frac{1}{\beta V}\sum_{K,K^\prime}{\rm Tr}_{\rm NG}
\left[({\cal G}_J)_{K,K^\prime-Q}\Gamma_+({\cal G}_J)_{K^\prime,K+Q^\prime}\Gamma_-\right],\nonumber\\
&&({\bf M}_J^{12})_{Q,Q^\prime}=({\bf M}_J^{--})_{Q,Q^\prime}=\frac{1}{\beta V}\sum_{K,K^\prime}{\rm Tr}_{\rm NG}
\left[({\cal G}_J)_{K,K^\prime-Q}\Gamma_-({\cal G}_J)_{K^\prime,K+Q^\prime}\Gamma_-\right],\nonumber\\
&&({\bf M}_J^{21})_{Q,Q^\prime}=({\bf M}_J^{++})_{Q,Q^\prime}=\frac{1}{\beta V}\sum_{K,K^\prime}{\rm Tr}_{\rm NG}
\left[({\cal G}_J)_{K,K^\prime-Q}\Gamma_+({\cal G}_J)_{K^\prime,K+Q^\prime}\Gamma_+\right].
\end{eqnarray}
It is easy to show that in the absence of the external sources (homogeneous case),  we recover the results obtained in Sec. \ref{s2}. Carrying out the trace in 
the Nambu-Gor'kov space, we obtain
\begin{eqnarray}
&&({\bf M}_J^{11})_{Q,Q^\prime}=\frac{\delta_{Q,Q^\prime}}{U}+\frac{1}{\beta V}\sum_{K,K^\prime}
\left[({\cal G}_J^{22})_{K,K^\prime-Q}({\cal G}_J^{11})_{K^\prime,K+Q^\prime}\right],\nonumber\\
&&({\bf M}_J^{22})_{Q,Q^\prime}=\frac{\delta_{Q,Q^\prime}}{U}+\frac{1}{\beta V}\sum_{K,K^\prime}
\left[({\cal G}_J^{11})_{K,K^\prime-Q}({\cal G}_J^{22})_{K^\prime,K+Q^\prime}\right],\nonumber\\
&&({\bf M}_J^{12})_{Q,Q^\prime}=\frac{1}{\beta V}\sum_{K,K^\prime}
\left[({\cal G}_J^{12})_{K,K^\prime-Q}({\cal G}_J^{12})_{K^\prime,K+Q^\prime}\right],\nonumber\\
&&({\bf M}_J^{21})_{Q,Q^\prime}=\frac{1}{\beta V}\sum_{K,K^\prime}
\left[({\cal G}_J^{21})_{K,K^\prime-Q}({\cal G}_J^{21})_{K^\prime,K+Q^\prime}\right].
\end{eqnarray}
It is easy to show that
\begin{eqnarray}
({\bf M}_J^{11})_{Q,Q^\prime}=({\bf M}_J^{22})_{-Q^\prime,-Q}.
\end{eqnarray}
The path integral over the quantum fluctuations $\phi$ and $\phi^*$ is Gaussian and can be carried out. The partition function can be expressed as
\begin{eqnarray}
{\cal Z}[J]\simeq \exp \Big\{- {\cal W}_{\rm MF}[j_n,j_s;\eta_1,\eta_2]-{\cal W}_{\rm GF}[j_n,j_s;\eta_1,\eta_2]\Big\},
\end{eqnarray}
where the mean-field contribution ${\cal W}_{\rm MF}$ is given in Sec. \ref{s4} and the contribution from the Gaussian fluctuations can be expressed as
\begin{eqnarray}
{\cal W}_{\rm GF}[j_n,j_s;\eta_1,\eta_2]=\frac{1}{2}{\rm Tr}\ln\left[({\bf M}_J)_{Q,Q^\prime}\right].
\end{eqnarray}
Here the ${\rm Tr}\ln$ acts not only on the two-dimensional space shown in (\ref{Boson-J}) but also on the momentum space indexed by $Q$ and $Q^\prime$.

\subsection{Expansion of the GPF generating functional}

The next step is to expand the generating functional order by order in the external sources. The expansion of the mean-field has been completed in Sec. \ref{s4}. To expand the Gaussian-fluctuation part,  we first need to expand ${\bf M}_J$ in powers of the external sources $j_n$ and $j_s$ and the induced perturbations $\eta_1$ and $\eta_2$. In general, the expansion can be expressed as
\begin{equation}
({\bf M}_J)_{Q,Q^\prime}={\bf M}(Q)\delta_{Q,Q^\prime}+\Sigma^{(1)}_{Q,Q^\prime}+\Sigma^{(2)}_{Q,Q^\prime}+\cdots.
\end{equation}
Here ${\bf M}(Q)$ is the collective mode propagator given in Sec. \ref{s2}, and $\Sigma^{(1)}$ and $\Sigma^{(2)}$ are the first and second order expansions in the external sources $j_n$ and $j_s$ and the induced perturbations $\eta_1$ and $\eta_2$. The higher order expansions are irrelevant to the study of the density and spin linear responses.  Like ${\bf M}_J$ and ${\bf M}$, $\Sigma^{(1)}$ and $\Sigma^{(2)}$ are
$2\times2$ matrices. They can be expressed as
\begin{eqnarray}\label{Sigma-Matrix}
\Sigma^{(1)}_{Q,Q^\prime}=\left(\begin{array}{cc}(\Sigma^{(1)}_{11})_{Q,Q^\prime}&(\Sigma^{(1)}_{12})_{Q,Q^\prime}\\
(\Sigma^{(1)}_{21})_{Q,Q^\prime}& (\Sigma^{(1)}_{22})_{Q,Q^\prime}\end{array}\right)
=\left(\begin{array}{cc}(\Sigma^{(1)}_{-+})_{Q,Q^\prime}&(\Sigma^{(1)}_{--})_{Q,Q^\prime}\\
(\Sigma^{(1)}_{++})_{Q,Q^\prime}& (\Sigma^{(1)}_{+-})_{Q,Q^\prime}\end{array}\right),\nonumber\\
\Sigma^{(2)}_{Q,Q^\prime}=\left(\begin{array}{cc}(\Sigma^{(2)}_{11})_{Q,Q^\prime}&(\Sigma^{(2)}_{12})_{Q,Q^\prime}\\
(\Sigma^{(2)}_{21})_{Q,Q^\prime}& (\Sigma^{(2)}_{22})_{Q,Q^\prime}\end{array}\right)
=\left(\begin{array}{cc}(\Sigma^{(1)}_{-+})_{Q,Q^\prime}&(\Sigma^{(1)}_{--})_{Q,Q^\prime}\\
(\Sigma^{(2)}_{++})_{Q,Q^\prime}& (\Sigma^{(2)}_{+-})_{Q,Q^\prime}\end{array}\right).
\end{eqnarray}

To obtain $\Sigma^{(1)}$ and $\Sigma^{(2)}$, we express the inverse of the mean-field Green's function ${\cal G}_J^{-1}$ in the presence of the external sources as
\begin{equation}
({\cal G}_J^{-1})_{K,K^\prime}={\cal G}^{-1}(K)\delta_{K,K^\prime}-(\Sigma_J)_{K,K^\prime}
\end{equation}
where
\begin{eqnarray}
(\Sigma_J)_{K,K^\prime}&=&-\frac{1}{\sqrt{\beta V}}\Big[\Gamma_nj_n(K-K^\prime)+\Gamma_sj_s(K-K^\prime)\nonumber\\
&&+\Gamma_1\eta_1(K-K^\prime)+\Gamma_2\eta_2(K-K^\prime)\Big].
\end{eqnarray}
Using the Taylor expansion of matrix functions, we expand the Green's function ${\cal G}_J$ to the second order in $\Sigma_J$,
\begin{equation}
{\cal G}_J={\cal G}+{\cal G}\Sigma_J{\cal G}+{\cal G}\Sigma_J{\cal G}\Sigma_J{\cal G}+\cdots.
\end{equation}
This expansion should me understood simultaneously in the momentum space and the Nambu-Gor'kov space. In the momentum space, we write
\begin{eqnarray}
({\cal G}_J)_{K,K^\prime}&=&{\cal G}_{K,K^\prime}+\sum_{K_1,K_2}{\cal G}_{K,K_1}(\Sigma_J)_{K_1,K_2}{\cal G}_{K_2,K^\prime}\nonumber\\
&+&\sum_{K_1,K_2,K_3,K_4}{\cal G}_{K,K_1}(\Sigma_J)_{K_1,K_2}{\cal G}_{K_2,K_3}(\Sigma_J)_{K_3,K_4}{\cal G}_{K_4,K^\prime}+\cdots.
\end{eqnarray}
Using the fact that ${\cal G}_{K,K^\prime}={\cal G}(K)\delta_{K,K^\prime}$ and $(\Sigma_J)_{K,K^\prime}=\Sigma_J(K-K^\prime)$, we obtain
\begin{eqnarray}
({\cal G}_J)_{K,K^\prime}&=&{\cal G}(K)\delta_{K,K^\prime}+{\cal G}(K)\Sigma_J(K-K^\prime){\cal G}(K^\prime)\nonumber\\
&+&\sum_{K^{\prime\prime}}{\cal G}(K)\Sigma_J(K-K^{\prime\prime}){\cal G}(K^{\prime\prime})\Sigma_J(K^{\prime\prime}-K^\prime){\cal G}(K^{\prime})+\cdots.
\end{eqnarray}

Using the above expansion for ${\cal G}_J$, we can derive the explicit form of $\Sigma^{(1)}$ and $\Sigma^{(2)}$. $\Sigma^{(1)}$ is composed of one 
leading-order expansion of ${\cal G}_J$ and one next-to-leading-order expansion of ${\cal G}_J$. We obtain (${\rm s,t}=+,-$)
\begin{eqnarray}
(\Sigma^{(1)}_{\rm st})_{Q,Q^\prime}&=&-\frac{1}{\sqrt{\beta V}}\Big[X^n_{\rm st}(Q,Q^\prime)j_n(Q-Q^\prime)+X^s_{\rm st}(Q,Q^\prime)j_s(Q-Q^\prime)\nonumber\\
&&+X^1_{\rm st}(Q,Q^\prime)\eta_1(Q-Q^\prime)+X^2_{\rm st}(Q,Q^\prime)\eta_2(Q-Q^\prime)\Big],
\end{eqnarray}
where ($a=n,s,1,2$)
\begin{eqnarray}
X^a_{\rm st}(Q,Q^\prime)&=&\frac{1}{\beta V}\sum_K{\rm Tr}_{\rm NG}\left[{\cal G}(K)\Gamma_{\rm s}{\cal G}(K+Q)\Gamma_a
{\cal G}(K+Q^\prime)\Gamma_{\rm t}\right]\nonumber\\
&+&\frac{1}{\beta V}\sum_K{\rm Tr}_{\rm NG}\left[{\cal G}(K)\Gamma_a{\cal G}(K+Q^\prime-Q)\Gamma_{\rm s}{\cal G}(K+Q^\prime)\Gamma_{\rm t}\right].
\end{eqnarray}
The second-order term $\Sigma^{(2)}$ is composed of two types of contributions. We write
\begin{eqnarray}
\Sigma^{(2)}=\Sigma^{(2A)}+\Sigma^{(2B)}.
\end{eqnarray}
$\Sigma^{(2A)}$ is composed of one leading-order expansion of ${\cal G}_J$ and one next-to-next-to-leading order expansion of ${\cal G}_J$. We have
\begin{eqnarray}
(\Sigma^{(2A)}_{\rm st})_{Q,Q^\prime}&=&\frac{1}{(\beta V)^2}\sum_{K,K^\prime}\left(\begin{array}{cccc} j^{\rm T}(Q_1)&\eta^{\rm T}(Q_1)\end{array}\right)
Y_{\rm st}(Q,Q^\prime;K,K^\prime)\left(\begin{array}{c} j(Q_2) \\ \eta(Q_2)\end{array}\right)\nonumber\\
&+&\frac{1}{(\beta V)^2}\sum_{K,K^\prime}\left(\begin{array}{cccc} j^{\rm T}(Q_3)&\eta^{\rm T}(Q_3)\end{array}\right)Z_{\rm st}(Q,Q^\prime;K,K^\prime)
\left(\begin{array}{c} j(Q_4) \\ \eta(Q_4)\end{array}\right).
\end{eqnarray}
Here the definitions of $j(Q)$ and $\eta(Q)$ is given in (104) and the momenta $Q_1,Q_2,Q_3$ and $Q_4$ are defined as
\begin{eqnarray}
&&Q_1=K-K^\prime+Q,\ \ \ \ \ \ Q_2=K^\prime-K-Q^\prime,\nonumber\\
&&Q_3=K-K^\prime,\ \ \ \ \ \ Q_4=K^\prime-K+Q-Q^\prime.
\end{eqnarray}
The matrix $Y_{\rm st}$ is defined as 
\begin{eqnarray}
Y_{\rm st}=\left(\begin{array}{cc}Y_{\rm st}^{jj}&Y_{\rm st}^{j\eta}\\ Y_{\rm st}^{\eta j}&Y_{\rm st}^{\eta\eta}\end{array}\right)
\end{eqnarray}
with the four blocks given by
\begin{eqnarray}
Y_{\rm st}^{jj}=\left(\begin{array}{cc}Y_{\rm st}^{nn}&Y_{\rm st}^{ns}\\ Y_{\rm st}^{sn}&Y_{\rm st}^{ss}\end{array}\right),\ \ \ \ \ \
Y_{\rm st}^{j\eta}=\left(\begin{array}{cc}Y_{\rm st}^{n1}&Y_{\rm st}^{n2}\\ Y_{\rm st}^{s1}&Y_{\rm st}^{s2}\end{array}\right),\nonumber\\
Y_{\rm st}^{\eta j}=\left(\begin{array}{cc}Y_{\rm st}^{1n}&Y_{\rm st}^{1s}\\ Y_{\rm st}^{2n}&Y_{\rm st}^{2s}\end{array}\right),\ \ \ \ \ \
Y_{\rm st}^{\eta\eta}=\left(\begin{array}{cc}Y_{\rm st}^{11}&Y_{\rm st}^{12}\\ Y_{\rm st}^{21}&Y_{\rm st}^{22}\end{array}\right).
\end{eqnarray}
The matrix elements $Y_{\rm st}^{ab}$ ($a,b=n,s,1,2$) are given by
\begin{eqnarray}
Y_{\rm st}^{ab}(Q,Q^\prime;K,K^\prime)={\rm Tr}_{\rm NG}\left[{\cal G}(K)\Gamma_{\rm s}{\cal G}(K+Q)\Gamma_a{\cal G}(K^\prime)\Gamma_b{\cal G}(K+Q^\prime)\Gamma_{\rm t}\right]
\end{eqnarray}
The matrix $Z_{\rm st}$ is defined as 
\begin{eqnarray}
Z_{\rm st}=\left(\begin{array}{cc}Z_{\rm st}^{jj}&Z_{\rm st}^{j\eta}\\ Z_{\rm st}^{\eta j}&Z_{\rm st}^{\eta\eta}\end{array}\right)
\end{eqnarray}
with the four blocks given by
\begin{eqnarray}
Z_{\rm st}^{jj}=\left(\begin{array}{cc}Z_{\rm st}^{nn}&Z_{\rm st}^{ns}\\ Z_{\rm st}^{sn}&Z_{\rm st}^{ss}\end{array}\right),\ \ \ \ \ \
Z_{\rm st}^{j\eta}=\left(\begin{array}{cc}Z_{\rm st}^{n1}&Z_{\rm st}^{n2}\\ Z_{\rm st}^{s1}&Z_{\rm st}^{s2}\end{array}\right),\nonumber\\
Z_{\rm st}^{\eta j}=\left(\begin{array}{cc}Z_{\rm st}^{1n}&Z_{\rm st}^{1s}\\ Z_{\rm st}^{2n}&Z_{\rm st}^{2s}\end{array}\right),\ \ \ \ \ \
Z_{\rm st}^{\eta\eta}=\left(\begin{array}{cc}Z_{\rm st}^{11}&Z_{\rm st}^{12}\\ Z_{\rm st}^{21}&Z_{\rm st}^{22}\end{array}\right).
\end{eqnarray}
The matrix elements $Z_{\rm st}^{ab}$ ($a,b=n,s,1,2$) are given by
\begin{eqnarray}
Z_{\rm st}^{ab}(Q,Q^\prime;K,K^\prime)={\rm Tr}_{\rm NG}\left[{\cal G}(K)\Gamma_a{\cal G}(K^\prime)\Gamma_b{\cal G}(K+Q^\prime-Q)
\Gamma_{\rm s}{\cal G}(K+Q^\prime)\Gamma_{\rm t}\right]
\end{eqnarray}
$\Sigma^{(2B)}$ is composed of two next-to-leading-order expansions of ${\cal G}_J$. We have
\begin{eqnarray}
(\Sigma^{(2B)}_{\rm st})_{Q,Q^\prime}=\frac{1}{(\beta V)^2}\sum_{K,K^\prime}\left(\begin{array}{cccc} j^{\rm T}(Q_1)&\eta^{\rm T}(Q_1)\end{array}\right)
W_{\rm st}(Q,Q^\prime;K,K^\prime)\left(\begin{array}{c} j(Q_2) \\ \eta(Q_2)\end{array}\right).
\end{eqnarray}
The matrix $W_{\rm st}$ is defined as 
\begin{eqnarray}
W_{\rm st}=\left(\begin{array}{cc}W_{\rm st}^{jj}&W_{\rm st}^{j\eta}\\ W_{\rm st}^{\eta j}&W_{\rm st}^{\eta\eta}\end{array}\right)
\end{eqnarray}
with the four blocks given by
\begin{eqnarray}
W_{\rm st}^{jj}=\left(\begin{array}{cc}W_{\rm st}^{nn}&W_{\rm st}^{ns}\\ W_{\rm st}^{sn}&W_{\rm st}^{ss}\end{array}\right),\ \ \ \ \ \
W_{\rm st}^{j\eta}=\left(\begin{array}{cc}W_{\rm st}^{n1}&W_{\rm st}^{n2}\\ W_{\rm st}^{s1}&W_{\rm st}^{s2}\end{array}\right),\nonumber\\
W_{\rm st}^{\eta j}=\left(\begin{array}{cc}W_{\rm st}^{1n}&W_{\rm st}^{1s}\\ W_{\rm st}^{2n}&W_{\rm st}^{2s}\end{array}\right),\ \ \ \ \ \
W_{\rm st}^{\eta\eta}=\left(\begin{array}{cc}W_{\rm st}^{11}&W_{\rm st}^{12}\\ W_{\rm st}^{21}&W_{\rm st}^{22}\end{array}\right).
\end{eqnarray}
The matrix elements $W_{\rm st}^{ab}$ ($a,b=n,s,1,2$) are given by
\begin{eqnarray}
W_{\rm st}^{ab}(Q,Q^\prime;K,K^\prime)={\rm Tr}_{\rm NG}\left[{\cal G}(K)\Gamma_a{\cal G}(K^\prime-Q)\Gamma_{\rm s}{\cal G}(K^\prime)\Gamma_b{\cal G}(K+Q^\prime)\Gamma_{\rm t}\right]
\end{eqnarray}

Finally, the Gaussian-fluctuation contribution to the generating functional can be expressed as
\begin{eqnarray}
{\cal W}_{\rm GF}[j_n,j_s;\eta_1,\eta_2]=\frac{1}{2}{\rm Tr}\ln\left[{\bf M}(Q)\delta_{Q,Q^\prime}+\Sigma^{(1)}_{Q,Q^\prime}+\Sigma^{(2A)}_{Q,Q^\prime}
+\Sigma^{(2B)}_{Q,Q^\prime}+\cdots\right].
\end{eqnarray}
Using the derivative expansion, we can expand ${\cal W}_{\rm GF}$ to the second order in the external sources $j_n$ and $j_s$ as well as the induced perturbations $\eta_1$ and $\eta_2$. We have
\begin{eqnarray}
{\cal W}_{\rm GF}[j_n,j_s;\eta_1,\eta_2]={\cal W}_{\rm GF}^{(0)}+{\cal W}_{\rm GF}^{(1)}[j_n,j_s;\eta_1,\eta_2]+{\cal W}_{\rm GF}^{(2)}[j_n,j_s;\eta_1,\eta_2]+\cdots,
\end{eqnarray}
where ${\cal W}_{\rm GF}^{(0)}=\beta V\Omega_{\rm GF}$. The first-order expansion is given by
\begin{eqnarray}\label{W-linear}
{\cal W}_{\rm GF}^{(1)}[j_n,j_s;\eta_1,\eta_2]=\frac{1}{2}\sum_{Q}{\rm Tr}_{2\rm D}\left[{\bf D}(Q)\Sigma^{(1)}_{Q,Q}\right].
\end{eqnarray}
Here the trace ${\rm Tr}_{2\rm D}$ is now taken only in the two-dimensional space defined in (\ref{Sigma-Matrix}) and ${\bf D}(Q)={\bf M}^{-1}(Q)$ is explicitly 
given by
\begin{eqnarray}
{\bf D}(Q)=\frac{1}{{\bf M}_{11}(Q){\bf M}_{22}(Q)-{\bf M}_{12}(Q){\bf M}_{21}(Q)}\left(\begin{array}{cc}{\bf M}_{22}(Q)&-{\bf M}_{12}(Q)\\ -{\bf M}_{21}(Q)&{\bf M}_{11}(Q)\end{array}\right)
\end{eqnarray}
The second-order expansion reads
\begin{eqnarray}
{\cal W}_{\rm GF}^{(2)}[j_n,j_s;\eta_1,\eta_2]={\cal W}_{\rm GF}^{({\rm AL})}+{\cal W}_{\rm GF}^{({\rm SE})}+{\cal W}_{\rm GF}^{({\rm MT})},
\end{eqnarray}
where the three contributions are given by
\begin{eqnarray}
&&{\cal W}_{\rm GF}^{({\rm AL})}[j_n,j_s;\eta_1,\eta_2]=-\frac{1}{4}\sum_{Q,Q^\prime}
{\rm Tr}_{2\rm D}\left[{\bf D}(Q)\Sigma^{(1)}_{Q,Q^\prime}{\bf D}(Q^\prime)\Sigma^{(1)}_{Q^\prime,Q}\right],\nonumber\\
&&{\cal W}_{\rm GF}^{({\rm SE})}[j_n,j_s;\eta_1,\eta_2]=\frac{1}{2}\sum_{Q}{\rm Tr}_{2\rm D}\left[{\bf D}(Q)\Sigma^{(2A)}_{Q,Q}\right],\nonumber\\
&&{\cal W}_{\rm GF}^{({\rm MT})}[j_n,j_s;\eta_1,\eta_2]=\frac{1}{2}\sum_{Q}{\rm Tr}_{2\rm D}\left[{\bf D}(Q)\Sigma^{(2B)}_{Q,Q}\right].
\end{eqnarray}
As we will show below, the above terms with the notations AL, SE, and MT diagrammatically correspond to the Aslamazov-Lakin, Self-Energy, and Maki-Thompson contributions in the broken-symmetry (superfluid) phase.

\subsection{The GPF response functions}

Now we discuss the expansions ${\cal W}_{\rm GF}^{(1)}$ and ${\cal W}_{\rm GF}^{(2)}$.  As mentioned in Sec. \ref{s5-1}, the linear term ${\cal W}_{\rm GF}^{(1)}$ has a nontrivial contribution to the response function. Therefore, the GPF response functions includes four kinds of contributions from the Gaussian pair fluctuations.

\subsubsection{Superfluid order parameter induced contribution}
This contribution comes from the linear term, Eq. (\ref{W-linear}), which can be expressed as
\begin{eqnarray}\label{W-linear2}
{\cal W}_{\rm GF}^{(1)}=\sqrt{\beta V}\left[{\cal C}_nj_n(0)+{\cal C}_sj_s(0)+{\cal C}_1\eta_1(0)+{\cal C}_2\eta_2(0)\right],
\end{eqnarray}
where the coefficients read
\begin{eqnarray}
{\cal C}_a=-\frac{1}{2\beta V}\sum_{Q}{\rm Tr}_{2\rm D}\left[{\bf D}(Q)X^a(Q,Q)\right],\ \ \ \ \ a=n,s,1,2.
\end{eqnarray}
Using the explicit form of $X_{\rm st}^a(Q,Q)$, we can show that ${\cal C}_s=0$ and ${\cal C}_2=0$.  It is obvious to identify 
\begin{equation}
{\cal C}_n=\frac{\partial \Omega_{\rm GF}(\mu,\Delta)}{\partial\mu},\ \ \ \ \ \ \ \ \ \ \ \ 
{\cal C}_1=\frac{\partial \Omega_{\rm GF}(\mu,\Delta)}{\partial\Delta}.
\end{equation}
The fact that ${\cal C}_1\neq0$ means that we cannot simply neglect the linear term ${\cal W}_{\rm GF}^{(1)}$ when evaluating the response functions.
To eliminate the induced perturbation $\eta_1(0)$, we need to expand $\eta_1(0)$ up to the second order in $j$. To this end, we should expand 
the mean-field  action  ${\cal W}_{\rm MF}[j;\eta]$ up to the third order in $j$ and $\eta$. We have
\begin{eqnarray}
{\cal W}_{\rm MF}^{(3)}=\frac{1}{3}\sum_{K}\sum_{K^\prime}\sum_{K^{\prime\prime}}{\rm Tr}_{\rm NG}\left[{\cal G}(K)(\Sigma_J)_{K,K^\prime}{\cal G}(K^\prime)
(\Sigma_J)_{K^\prime,K^{\prime\prime}}{\cal G}(K^{\prime\prime})(\Sigma_J)_{K^{\prime\prime},K}\right].
\end{eqnarray}
For convenience, we define $K^\prime=K+Q$ and $K^{\prime\prime}=K+Q^\prime$. We obtain
\begin{eqnarray}
{\cal W}_{\rm MF}^{(3)}=-\frac{1}{3\sqrt{\beta V}}\sum_{a,b,c=n,s,1,2}\sum_{Q}\sum_{Q^\prime}F_{abc}(Q,Q^\prime)
\varphi_a(-Q)\varphi_b(Q-Q^\prime)\varphi_c(Q^\prime),
\end{eqnarray}
where the function $F_{abc}(Q,Q^\prime)$ is defined as
\begin{eqnarray}
F_{abc}(Q,Q^\prime)=\frac{1}{\beta V}\sum_{K}{\rm Tr}_{\rm NG}\left[{\cal G}(K)\Gamma_a{\cal G}(K+Q)\Gamma_b{\cal G}(K+Q^\prime)\Gamma_c\right],
\end{eqnarray}
Using the saddle point equation
\begin{eqnarray}
\frac{\delta{\cal W}_{\rm MF}[j;\eta]}{\delta\eta(Q)}=0
\end{eqnarray}
with ${\cal W}_{\rm MF}={\cal W}_{\rm MF}^{(0)}+{\cal W}_{\rm MF}^{(1)}+{\cal W}_{\rm MF}^{(2)}+{\cal W}_{\rm MF}^{(3)}+\cdots$, we obtain
\begin{eqnarray}\label{Exp-eta}
\eta_1(0)={\cal R}_nj_n(0)+\frac{1}{2\sqrt{\beta V}}\sum_{a,b=n,s,1,2}\sum_Q{\cal U}_{ab}(Q)\varphi_a(-Q)\varphi_b(Q)+\cdots,
\end{eqnarray}
where ${\cal R}_n=d\Delta(\mu)/d\mu$ and the function ${\cal U}_{ab}(Q)$ is given by
\begin{eqnarray}
{\cal U}_{ab}(Q)=\frac{4}{3}\frac{F_{1ab}(0,Q)}{I_{11}(0,0)}.
\end{eqnarray}

Substituting the expansion (\ref{Exp-eta}) into (\ref{W-linear2}), we obtain
\begin{eqnarray}
{\cal W}_{\rm GF}^{(1)}=-\sqrt{\beta V}n_{\rm GF}j_n(0)+\frac{1}{2}\sum_{a,b=n,s,1,2}\sum_Q\Xi^{\rm OP}_{ab}(Q)\varphi_a(-Q)\varphi_b(Q)+\cdots,
\end{eqnarray}
where the function $\Xi^{\rm OP}_{ab}(Q)$ is given by
\begin{eqnarray}
\Xi_{ab}^{\rm OP}(Q)={\cal C}_1{\cal U}_{ab}(Q)=\frac{4{\cal C}_1}{3I_{11}(0,0)}F_{1ab}(0,Q).
\end{eqnarray}
Therefore, the linear term has a nontrivial contribution to the response function.  We note that it is because of the nonvanishing order parameter in the superfluid phase. Hence it can be called the order parameter induced contribution to the response, which is given by
\begin{eqnarray}
{\cal W}_{\rm GF}^{({\rm OP})}=\frac{1}{2}\sum_Q\left(\begin{array}{cccc} j^{\rm T}(-Q)&\eta^{\rm T}(-Q)\end{array}\right)
\left(\begin{array}{cc}\Xi_{jj}^{\rm OP}(Q)&\Xi_{j\eta}^{\rm OP}(Q)\\
\Xi_{\eta j}^{\rm OP}(Q)&\Xi_{\eta\eta}^{\rm OP}(Q)\end{array}\right)
\left(\begin{array}{c} j(Q) \\ \eta(Q)\end{array}\right)
\end{eqnarray}

\subsubsection{Superfluid Aslamazov-Lakin contribution}
The Aslamazov-Lakin contribution in the superfluid phase is given by
\begin{eqnarray}
{\cal W}_{\rm GF}^{({\rm AL})}=-\frac{1}{4}\sum_{Q,Q^\prime}
{\rm Tr}_{\rm 2D}\left[{\bf D}(Q)\Sigma^{(1)}_{Q,Q^\prime}{\bf D}(Q^\prime)\Sigma^{(1)}_{Q^\prime,Q}\right].
\end{eqnarray}
After some manipulations, it can be expressed as
\begin{eqnarray}
{\cal W}_{\rm GF}^{({\rm AL})}=\frac{1}{2}\sum_{a,b=n,s,1,2}\sum_{Q}\Xi_{ab}^{\rm AL}(Q)\varphi_a(-Q)\varphi_b(Q),
\end{eqnarray}
where the function $\Xi_{ab}^{\rm AL}(Q)$ is given by
\begin{eqnarray}
\Xi_{ab}^{\rm AL}(Q)=-\frac{1}{2}\frac{1}{\beta V}\sum_{P}{\rm Tr}_{\rm 2D}\left[{\bf D}(P)X^a(P,P+Q){\bf D}(P+Q)X^b(P+Q,P)\right].
\end{eqnarray}
The matrices $X^a(P,P+Q)$ and $X^b(P+Q,P)$ are explicitly given by
\begin{eqnarray}
X^a_{\rm st}(P,P+Q)&=&\frac{1}{\beta V}\sum_K{\rm Tr}_{\rm NG}\left[{\cal G}(K)\Gamma_{\rm s}{\cal G}(K+P)\Gamma_a
{\cal G}(K+P+Q)\Gamma_{\rm t}\right]\nonumber\\
&+&\frac{1}{\beta V}\sum_K{\rm Tr}_{\rm NG}\left[{\cal G}(K)\Gamma_a{\cal G}(K+Q)\Gamma_{\rm s}{\cal G}(K+P+Q)\Gamma_{\rm t}\right],\nonumber\\
X^b_{\rm st}(P+Q,P)&=&\frac{1}{\beta V}\sum_K{\rm Tr}_{\rm NG}\left[{\cal G}(K)\Gamma_{\rm s}{\cal G}(K+P+Q)\Gamma_b
{\cal G}(K+P)\Gamma_{\rm t}\right]\nonumber\\
&+&\frac{1}{\beta V}\sum_K{\rm Tr}_{\rm NG}\left[{\cal G}(K)\Gamma_b{\cal G}(K-Q)\Gamma_{\rm s}{\cal G}(K+P)\Gamma_{\rm t}\right].
\end{eqnarray}
Therefore, in the superfluid phase, there is an Aslamazov-Lakin contribution to the response, which is given by
\begin{eqnarray}
{\cal W}_{\rm GF}^{({\rm AL})}=\frac{1}{2}\sum_Q\left(\begin{array}{cccc} j^{\rm T}(-Q)&\eta^{\rm T}(-Q)\end{array}\right)
\left(\begin{array}{cc}\Xi_{jj}^{\rm AL}(Q)&\Xi_{j\eta}^{\rm AL}(Q)\\
\Xi_{\eta j}^{\rm AL}(Q)&\Xi_{\eta\eta}^{\rm AL}(Q)\end{array}\right)
\left(\begin{array}{c} j(Q) \\ \eta(Q)\end{array}\right)
\end{eqnarray}

\subsubsection{Superfluid Self-Energy contribution}
The Self-Energy contribution in the superfluid phase is given by
\begin{eqnarray}
{\cal W}_{\rm GF}^{({\rm SE})}=\frac{1}{2}\sum_{Q}{\rm Tr}_{\rm 2D}\left[{\bf D}(Q)\Sigma^{(2A)}_{Q,Q}\right],
\end{eqnarray}
After some manipulations, it can be expressed as
\begin{eqnarray}
{\cal W}_{\rm GF}^{({\rm SE})}=\frac{1}{2}\sum_{a,b=n,s,1,2}\sum_{Q}\Xi_{ab}^{\rm SE}(Q)\varphi_a(-Q)\varphi_b(Q),
\end{eqnarray}
where the function $\Xi_{ab}^{\rm SE}(Q)$ is given by
\begin{eqnarray}
\Xi_{ab}^{\rm SE}(Q)&=&\frac{1}{\beta V}\sum_{P}{\rm Tr}_{2\rm D}\left[{\bf D}(P){\bf Y}^{ab}(P,Q)\right]\nonumber\\
&+&\frac{1}{\beta V}\sum_{P}{\rm Tr}_{\rm 2D}\left[{\bf D}(P){\bf Z}^{ab}(P,Q)\right].
\end{eqnarray}
Here the matrices ${\bf Y}^{ab}(P,Q)$ and ${\bf Z}^{ab}(P,Q)$ are given by
\begin{eqnarray}
{\bf Y}_{\rm st}^{ab}(P,Q)&=&\frac{1}{\beta V}\sum_{K}{\rm Tr}_{\rm NG}\left[{\cal G}(K-P)\Gamma_{\rm s}{\cal G}(K)\Gamma_a{\cal G}(K+Q)\Gamma_b{\cal G}(K)\Gamma_{\rm t}\right],\nonumber\\
{\bf Z}_{\rm st}^{ab}(P,Q)&=&\frac{1}{\beta V}\sum_{K}{\rm Tr}_{\rm NG}\left[{\cal G}(K)\Gamma_a{\cal G}(K+Q)\Gamma_b{\cal G}(K)
\Gamma_{\rm s}{\cal G}(K+P)\Gamma_{\rm t}\right].
\end{eqnarray}
Therefore, in the superfluid phase, there is a Self-Energy contribution to the response, which is given by
\begin{eqnarray}
{\cal W}_{\rm GF}^{({\rm SE})}=\frac{1}{2}\sum_Q\left(\begin{array}{cccc} j^{\rm T}(-Q)&\eta^{\rm T}(-Q)\end{array}\right)
\left(\begin{array}{cc}\Xi_{jj}^{\rm SE}(Q)&\Xi_{j\eta}^{\rm SE}(Q)\\
\Xi_{\eta j}^{\rm SE}(Q)&\Xi_{\eta\eta}^{\rm SE}(Q)\end{array}\right)
\left(\begin{array}{c} j(Q) \\ \eta(Q)\end{array}\right)
\end{eqnarray}

\subsubsection{Superfluid Maki-Thompson contribution}
The Maki-Thompson contribution in the superfluid phase is given by
\begin{eqnarray}
{\cal W}_{\rm GF}^{({\rm MT})}=\frac{1}{2}\sum_{Q}{\rm Tr}_{2\rm D}\left[{\bf D}(Q)\Sigma^{(2B)}_{Q,Q}\right].
\end{eqnarray}
After some manipulations, it can be expressed as
\begin{eqnarray}
{\cal W}_{\rm GF}^{({\rm MT})}=\frac{1}{2}\sum_{a,b=n,s,1,2}\sum_{Q}\Xi_{ab}^{\rm MT}(Q)\varphi_a(-Q)\varphi_b(Q),
\end{eqnarray}
where the function $\Xi_{ab}^{\rm MT}(Q)$ is given by
\begin{eqnarray}
\Xi_{ab}^{\rm MT}(Q)=\frac{1}{\beta V}\sum_{P}{\rm Tr}_{2\rm D}\left[{\bf D}(P){\bf W}^{ab}(P,Q)\right].
\end{eqnarray}
Here the matrix ${\bf W}^{ab}(P,Q)$ is given by
\begin{eqnarray}
{\bf W}_{\rm st}^{ab}(P,Q)=\frac{1}{\beta V}\sum_{K}{\rm Tr}_{\rm NG}\left[{\cal G}(K)\Gamma_a{\cal G}(K+Q)\Gamma_{\rm s}{\cal G}(K+P+Q)\Gamma_b{\cal G}(K+P)\Gamma_{\rm t}\right].
\end{eqnarray}
Therefore, in the superfluid phase, there is a Maki-Thompson contribution to the response, which is given by
\begin{eqnarray}
{\cal W}_{\rm GF}^{({\rm MT})}=\frac{1}{2}\sum_Q\left(\begin{array}{cccc} j^{\rm T}(-Q)&\eta^{\rm T}(-Q)\end{array}\right)
\left(\begin{array}{cc}\Xi_{jj}^{\rm MT}(Q)&\Xi_{j\eta}^{\rm MT}(Q)\\
\Xi_{\eta j}^{\rm MT}(Q)&\Xi_{\eta\eta}^{\rm MT}(Q)\end{array}\right)
\left(\begin{array}{c} j(Q) \\ \eta(Q)\end{array}\right)
\end{eqnarray}

In summary, we have shown that the GPF generating functional can be expanded as
\begin{eqnarray}
{\cal W}_{\rm GF}[J;\eta]&=&\beta V\Omega_{\rm GF}-\sqrt{\beta V}n_{\rm GF}j_n(0)\nonumber\\
&+&\frac{1}{2}\sum_Q\left(\begin{array}{cccc} j^{\rm T}(-Q)&\eta^{\rm T}(-Q)\end{array}\right)
\left(\begin{array}{cc}\Xi_{jj}(Q)&\Xi_{j\eta}(Q)\\
\Xi_{\eta j}(Q)&\Xi_{\eta\eta}(Q)\end{array}\right)
\left(\begin{array}{c} j(Q) \\ \eta(Q)\end{array}\right)+\cdots,
\end{eqnarray}
where 
\begin{eqnarray}
\Xi_{ab}(Q)=\Xi_{ab}^{\rm OP}(Q)+\Xi_{ab}^{\rm AL}(Q)+\Xi_{ab}^{\rm SE}(Q)+\Xi_{ab}^{\rm MT}(Q).
\end{eqnarray}
The final task is to eliminate the induced perturbations by using Eq. (\ref{RPA-GAP2}). We obtain
\begin{eqnarray}
{\cal W}_{\rm GF}[J]=\beta V\Omega_{\rm GF}-\sqrt{\beta V}n_{\rm GF}j_n(0)+\frac{1}{2}\sum_Qj^{\rm T}(-Q)\chi_{\rm GF}(Q)j(Q)+\cdots
\end{eqnarray}
where
\begin{eqnarray}
\chi_{\rm GF}(Q)&=&\Xi_{jj}(Q)-\Pi_{j\eta}(Q)\Pi_{\eta\eta}^{-1}(Q)\Xi_{\eta j}(Q)-\Xi_{j\eta}(Q)\Pi_{\eta\eta}^{-1}(Q)\Pi_{\eta j}(Q)\nonumber\\
&&+\ \Pi_{j\eta}(Q)\Pi_{\eta\eta}^{-1}(Q)\Xi_{\eta\eta}(Q)\Pi_{\eta\eta}^{-1}(Q)\Pi_{\eta j}(Q).
\end{eqnarray}
The pair-fluctuation contribution to the dynamic response function, $\chi_{\rm GF}(Q)$, can be expressed as
\begin{eqnarray}
\chi_{\rm GF}(Q)=\left(\begin{array}{cc}\chi^{\rm GF}_{nn}(Q)&\chi^{\rm GF}_{ns}(Q)\\ \chi^{\rm GF}_{sn}(Q)&\chi^{\rm GF}_{ss}(Q)\end{array}\right).
\end{eqnarray}
While it is rather tedious, using the facts ${\cal G}_{22}(K)=-{\cal G}_{11}(-K)$ and ${\cal G}_{21}(K)={\cal G}_{12}(K)$ which indicate the spin balance, we can show that
\begin{eqnarray}
\Xi_{ns}(Q)=\Xi_{sn}(Q)=\Xi_{1s}(Q)=\Xi_{2s}(Q)=\Xi_{s1}(Q)=\Xi_{s2}(Q)=0.
\end{eqnarray}
Therefore, the off-diagonal components of the dynamic response function $\chi_{\rm GF}(Q)$ vanish; i.e.,
\begin{eqnarray}
\chi^{\rm GF}_{ns}(Q)=\chi^{\rm GF}_{sn}(Q)=0.
\end{eqnarray}
The diagonal components,  $\chi^{\rm GF}_{nn}(Q)$ and $\chi^{\rm GF}_{ss}(Q)$, corresponds to the GPF contributions to the density response and spin response, respectively.

\section{Response functions above $T_c$: NSR response theory}\label{s6}
In this section, we present the results of the dynamic response functions in the normal state for the 3D system, namely above the superfluid transition 
temperature $T_c$. In this case, the derivation of the response functions becomes much simpler.  The fermion Green's function becomes diagonal and we do not
need to introduce the induced perturbations $\eta_1$ and $\eta_2$.  

\subsection{NSR theory in equilibrium}

Let us first review the NSR theory \cite{NSR,BCSBEC1} without the external sources. In the NSR theory, the effective action is given by 
\begin{eqnarray}
{\cal S}_{\rm{eff}}[\Phi,\Phi^*]={\cal S}_{\rm MF}+{\cal S}_{\rm GF}[\phi,\phi^*],
\end{eqnarray}
where the mean-field effective action or grand potential reads
\begin{eqnarray}
\Omega_{\rm MF}=\frac{{\cal S}_{\rm MF}}{\beta V}=-\frac{2}{\beta}\sum_{\bf k}\ln\left(1+e^{-\beta\xi_{\bf k}}\right).
\end{eqnarray}
The mean-field contribution to the number density reads
\begin{eqnarray}
n_{\rm MF}=2\sum_{\bf k}f(\xi_{\bf k}),
\end{eqnarray}
where $f(x)=1/(1+e^{\beta x})$ is the Fermi-Dirac distribution function. In the normal phase, the bare fermion Green's function is given by
\begin{eqnarray}
{\cal G}(K)=\left(\begin{array}{cc}{\cal G}_\uparrow(K)&0\\
0& {\cal G}_\downarrow(K)\end{array}\right).
\end{eqnarray}
The elements reads
\begin{eqnarray}
{\cal G}_\uparrow(K)={\cal G}_0(K)=\frac{1}{ik_n-\xi_{\bf k}},\ \ \ \ \ \ \ {\cal G}_\downarrow(K)=-{\cal G}_0(-K)=\frac{1}{ik_n+\xi_{\bf k}}.
\end{eqnarray}
The Gaussian part is given by
\begin{eqnarray}
{\cal S}_{\rm GF}[\phi,\phi^*]=\sum_{Q}{\bf M}(Q)\phi^*(Q)\phi(Q).
\end{eqnarray}
Here ${\bf M}(Q)$ is no longer a matrix and is given by
\begin{eqnarray}
{\bf M}(Q)=\frac{1}{U}+\frac{1}{\beta V}\sum_{K}\left[{\cal G}_\uparrow(K+Q){\cal G}_\downarrow(K)\right].
\end{eqnarray}
It explicit form in 3D can be evaluated as
\begin{eqnarray}\label{MQ3D}
{\bf M}(iq_l,{\bf q})=-\frac{m}{4\pi a_{3\rm D}}+\frac{1}{V}\sum_{\bf k}\left[\frac{1-f(\xi_+)-f(\xi_-)}{iq_l-\xi_+-\xi_-}+\frac{1}{2\varepsilon_{\bf k}}\right].
\end{eqnarray}
Carrying out the path integral over $\phi^*$ and $\phi$, we obtain the Gaussian contribution to the grand potential,
\begin{eqnarray}
\Omega_{\rm GF}=\frac{1}{\beta}\sum_{q_l}\frac{1}{V}\sum_{\bf q}\ln\left[{\bf M}(iq_l,{\bf q})\right]e^{iq_l0^+}
\end{eqnarray}
The Matsubara frequency sum can be converted to a standard contour integral.  We have
\begin{eqnarray}
\Omega_{\rm GF}=\sum_{{\bf q}}\int_{-\infty}^\infty \frac{d\omega}{\pi}b(\omega)\delta(\omega,{\bf q}),
\end{eqnarray}
where $b(\omega)=1/(e^{\beta\omega}-1)$ is the Bose-Einstein distribution function and the phase shifts is defined as
\begin{eqnarray}
\delta(\omega,{\bf q})=-{\rm Im}\ln {\bf M}(\omega+i\epsilon,{\bf q}).
\end{eqnarray}
The Gaussian fluctuation contribution to the number density is given by
\begin{eqnarray}
n_{\rm GF}=\sum_{{\bf q}}\int_{-\infty}^\infty \frac{d\omega}{\pi}b(\omega)\frac{\partial\delta(\omega,{\bf q})}{\partial\mu}.
\end{eqnarray}

In the NSR theory, the chemical potential $\mu$ is determined by the full number equation
\begin{eqnarray}
n=n_{\rm MF}(\mu)+n_{\rm GF}(\mu).
\end{eqnarray}
The superfluid transition temperature is determined by the above number equation together with the so-called Thouless criterion
${\bf M}(0,{\bf 0})=0$, i.e., 
\begin{eqnarray}
-\frac{m}{4\pi a_{3\rm D}}+\frac{1}{V}\sum_{\bf k}\left[\frac{1}{2\varepsilon_{\bf k}}-\frac{1-2f(\xi_{\bf k})}{2\xi_{\bf k}}\right]=0,
\end{eqnarray}
which is actually the BCS gap equation at the superfluid transition temperature.

\subsection{NSR response theory}
The dynamic response functions in the NSR theory can be obtained by taking vanishing order parameter $\Delta$ in the GPF response theory.  In the normal phase,
the derivation of the response functions becomes much simpler because we do not need to introduce the induced perturbations $\eta_1$ and $\eta_2$.  In the presence of external sources $j_n$ and $j_s$, the generating in the NSR theory is given by
\begin{eqnarray}
{\cal W}_{\rm NSR}[j_n,j_s]={\cal W}_{\rm MF}[j_n,j_s]+{\cal W}_{\rm GF}[j_n,j_s],
\end{eqnarray}
where the mean-field and Gaussian-fluctuation parts can be expanded as
\begin{eqnarray}
{\cal W}_{\rm MF}[j_n,j_s]&=&{\cal W}_{\rm MF}^{(0)}+{\cal W}_{\rm MF}^{(1)}[j_n,j_s]
+{\cal W}_{\rm MF}^{(2)}[j_n,j_s]+\cdots,\nonumber\\
{\cal W}_{\rm GF}[j_n,j_s]&=&{\cal W}_{\rm GF}^{(0)}+{\cal W}_{\rm GF}^{(1)}[j_n,j_s]
+{\cal W}_{\rm GF}^{(2)}[j_n,j_s]+\cdots,
\end{eqnarray}

The mean-field part is quite simple. We have
\begin{eqnarray}
{\cal W}_{\rm MF}^{(1)}[j_n,j_s]=-\sqrt{\beta V}n_{\rm MF}j_n(0)
\end{eqnarray}
and
\begin{eqnarray}
{\cal W}_{\rm MF}^{(2)}[j_n,j_s]=\frac{1}{2}\sum_Q\left(\begin{array}{cccc} j_n(-Q)&j_s(-Q)\end{array}\right)
\left(\begin{array}{cc}\Pi_{nn}(Q)&\Pi_{ns}(Q)\\ \Pi_{sn}(Q)&\Pi_{ss}(Q)\end{array}\right)
\left(\begin{array}{c} j_n(Q) \\ j_s(Q)\end{array}\right).
\end{eqnarray}
Here the loop functions can be evaluated as
\begin{eqnarray}
\Pi_{nn}(Q)=\Pi_{ss}(Q)=\frac{1}{\beta V}\sum_{K}\Big[{\cal G}_\uparrow(K+Q){\cal G}_\uparrow(K)
+{\cal G}_\downarrow(K+Q){\cal G}_\downarrow(K)\Big],\nonumber\\
\Pi_{ns}(Q)=\Pi_{sn}(-Q)=\frac{1}{\beta V}\sum_{K}\Big[{\cal G}_\uparrow(K+Q){\cal G}_\uparrow(K)
-{\cal G}_\downarrow(K+Q){\cal G}_\downarrow(K)\Big].
\end{eqnarray}
It is easy to show that $\Pi_{sn}(Q)=\Pi_{ns}(Q)=0$. The diagonal components can be evaluated as
\begin{eqnarray}
\Pi_{nn}(Q)=\Pi_{ss}(Q)=\frac{2}{V}\sum_{\bf k}\frac{f(\xi_+)-f(\xi_-)}{iq_l+\xi_+-\xi_-}
\end{eqnarray}
Therefore, in the mean-field approximation, the density and spin response functions becomes degenerate in the normal phase. We have
\begin{eqnarray}
\chi_{nn}^{\rm MF}(\omega,{\bf q})=\chi_{ss}^{\rm MF}(\omega,{\bf q})=\frac{2}{V}\sum_{\bf k}\frac{f(\xi_+)-f(\xi_-)}{\omega+\xi_+-\xi_-}
\end{eqnarray}
It is therefore quite necessary to include the beyond-mean-field contributions.

The Gaussian-fluctuation part in the normal phase is given by
\begin{eqnarray}
{\cal W}_{\rm GF}[j_n,j_s]=\frac{1}{2}{\rm Tr}\ln \left[{\bf M}^J_{Q,Q^\prime}\right]+\frac{1}{2}{\rm Tr}\ln \left[\tilde{\bf M}^J_{Q,Q^\prime}\right],
\end{eqnarray}
where ${\bf M}^J$ and $\tilde{\bf M}^J$ are now only matrices in the momentum space,
\begin{eqnarray}
{\bf M}^J_{Q,Q^\prime}&=&\frac{\delta_{Q,Q^\prime}}{U}+\frac{1}{\beta V}\sum_{K,K^\prime}{\rm Tr}_{\rm NG}
\left[({\cal G}_J)_{K,K^\prime-Q}\Gamma_-({\cal G}_J)_{K^\prime,K+Q^\prime}\Gamma_+\right],\nonumber\\
\tilde{\bf M}^J_{Q,Q^\prime}&=&\frac{\delta_{Q,Q^\prime}}{U}+\frac{1}{\beta V}\sum_{K,K^\prime}{\rm Tr}_{\rm NG}
\left[({\cal G}_J)_{K,K^\prime-Q}\Gamma_+({\cal G}_J)_{K^\prime,K+Q^\prime}\Gamma_-\right].
\end{eqnarray}
We can therefore expand ${\bf M}^J$ and  $\tilde{\bf M}^J$ to the second order in the external sources and obtain
\begin{eqnarray}
{\bf M}^J_{Q,Q^\prime}&=&{\bf M}(Q)\delta_{Q,Q^\prime}+\Sigma^{(1)}_{Q,Q^\prime}+\Sigma^{(2)}_{Q,Q^\prime}+\cdots,\nonumber\\
\tilde{\bf M}^J_{Q,Q^\prime}&=&\tilde{\bf M}(Q)\delta_{Q,Q^\prime}+\tilde{\Sigma}^{(1)}_{Q,Q^\prime}+\tilde{\Sigma}^{(2)}_{Q,Q^\prime}+\cdots,
\end{eqnarray}
Here $\tilde{\bf M}(Q)={\bf M}(-Q)$ with ${\bf M}(Q)$ explicitly given by (\ref{MQ3D}). Note that here $\Sigma^{(1)}$, $\Sigma^{(2)}$,  $\tilde{\Sigma}^{(1)}$, and 
$\tilde{\Sigma}^{(2)}$ are no longer $2\times 2$ matrices. $\Sigma^{(1)}_{Q,Q^\prime}$ and  $\tilde{\Sigma}^{(1)}_{Q,Q^\prime}$  can be expressed as
\begin{eqnarray}
\Sigma^{(1)}_{Q,Q^\prime}&=&-\frac{1}{\sqrt{\beta V}}\Big[X_n(Q,Q^\prime)j_n(Q-Q^\prime)+X_s(Q,Q^\prime)j_s(Q-Q^\prime)\Big],\nonumber\\
\tilde{\Sigma}^{(1)}_{Q,Q^\prime}&=&-\frac{1}{\sqrt{\beta V}}\Big[\tilde{X}_n(Q,Q^\prime)j_n(Q-Q^\prime)+\tilde{X}_s(Q,Q^\prime)j_s(Q-Q^\prime)\Big],
\end{eqnarray}
where 
\begin{eqnarray}
X_n(Q,Q^\prime)&=&\frac{1}{\beta V}\sum_K\left[{\cal G}_\downarrow(K){\cal G}_\uparrow(K+Q){\cal G}_\uparrow(K+Q^\prime)\right]\nonumber\\
&-&\frac{1}{\beta V}\sum_K\left[{\cal G}_\downarrow(K){\cal G}_\downarrow(K+Q^\prime-Q){\cal G}_\uparrow(K+Q^\prime)\right],\nonumber\\
X_s(Q,Q^\prime)&=&\frac{1}{\beta V}\sum_K\left[{\cal G}_\downarrow(K){\cal G}_\uparrow(K+Q){\cal G}_\uparrow(K+Q^\prime)\right]\nonumber\\
&+&\frac{1}{\beta V}\sum_K\left[{\cal G}_\downarrow(K){\cal G}_\downarrow(K+Q^\prime-Q){\cal G}_\uparrow(K+Q^\prime)\right],\nonumber\\
\tilde{X}_n(Q,Q^\prime)&=&X_n(-Q^\prime,-Q),\nonumber\\
\tilde{X}_s(Q,Q^\prime)&=&X_s(-Q^\prime,-Q).
\end{eqnarray}
Using the fact ${\cal G}_\downarrow(K)=-{\cal G}_\uparrow(-K)$, we can show that
\begin{eqnarray}
X_s(Q,Q^\prime)=0,\ \ \ \ \ \ \ \ \tilde{X}_s(Q,Q^\prime)=0.
\end{eqnarray}
The second-order terms $\Sigma^{(2)}_{Q,Q^\prime}$ and $\tilde{\Sigma}^{(2)}_{Q,Q^\prime}$ can be decomposed as
\begin{eqnarray}
\Sigma^{(2)}=\Sigma^{(2A)}+\Sigma^{(2B)},\ \ \ \ \ \ \ \ \tilde{\Sigma}^{(2)}=\tilde{\Sigma}^{(2A)}+\tilde{\Sigma}^{(2B)}.
\end{eqnarray}
$\Sigma^{(2A)}$ and $\tilde{\Sigma}^{(2A)}$ are composed of one leading-order expansion of ${\cal G}_J$ and one next-to-next-to-leading order expansion of 
${\cal G}_J$. We have
\begin{eqnarray}
\Sigma^{(2A)}_{Q,Q^\prime}&=&\frac{1}{(\beta V)^2}\sum_{K,K^\prime} j^{\rm T}(Q_1)Y(Q,Q^\prime;K,K^\prime) j(Q_2) \nonumber\\
&+&\frac{1}{(\beta V)^2}\sum_{K,K^\prime} j^{\rm T}(Q_3)Z(Q,Q^\prime;K,K^\prime)j(Q_4),\nonumber\\
\tilde{\Sigma}^{(2A)}_{Q,Q^\prime}&=&\frac{1}{(\beta V)^2}\sum_{K,K^\prime} j^{\rm T}(Q_1)\tilde{Y}(Q,Q^\prime;K,K^\prime) j(Q_2) \nonumber\\
&+&\frac{1}{(\beta V)^2}\sum_{K,K^\prime} j^{\rm T}(Q_3)\tilde{Z}(Q,Q^\prime;K,K^\prime)j(Q_4) .
\end{eqnarray}
The matrices $Y$, $Z$, $\tilde{Y}$, and $\tilde{Z}$ are defined as 
\begin{eqnarray}
Y=\left(\begin{array}{cc}Y_{nn}&Y_{ns}\\ Y_{sn}&Y_{ss}\end{array}\right),\ \  Z=\left(\begin{array}{cc}Z_{nn}&Z_{ns}\\ Z_{sn}&Z_{ss}\end{array}\right),\ \ 
\tilde{Y}=\left(\begin{array}{cc}\tilde{Y}_{nn}&\tilde{Y}_{ns}\\ \tilde{Y}_{sn}&\tilde{Y}_{ss}\end{array}\right),\ \ 
\tilde{Z}=\left(\begin{array}{cc}\tilde{Z}_{nn}&\tilde{Z}_{ns}\\ \tilde{Z}_{sn}&\tilde{Z}_{ss}\end{array}\right).
\end{eqnarray}
The elements are given by ($a,b=n,s$)
\begin{eqnarray}
Y_{ab}(Q,Q^\prime;K,K^\prime)&=&{\rm Tr}_{\rm NG}\left[{\cal G}(K)\Gamma_-{\cal G}(K+Q)\Gamma_a{\cal G}(K^\prime)\Gamma_b
{\cal G}(K+Q^\prime)\Gamma_+\right],\nonumber\\
Z_{ab}(Q,Q^\prime;K,K^\prime)&=&{\rm Tr}_{\rm NG}\left[{\cal G}(K)\Gamma_a{\cal G}(K^\prime)\Gamma_b{\cal G}(K+Q^\prime-Q)
\Gamma_-{\cal G}(K+Q^\prime)\Gamma_+\right],\nonumber\\
\tilde{Y}_{ab}(Q,Q^\prime;K,K^\prime)&=&{\rm Tr}_{\rm NG}\left[{\cal G}(K)\Gamma_+{\cal G}(K+Q)\Gamma_a{\cal G}(K^\prime)\Gamma_b
{\cal G}(K+Q^\prime)\Gamma_-\right],\nonumber\\
\tilde{Z}_{ab}(Q,Q^\prime;K,K^\prime)&=&{\rm Tr}_{\rm NG}\left[{\cal G}(K)\Gamma_a{\cal G}(K^\prime)\Gamma_b{\cal G}(K+Q^\prime-Q)
\Gamma_+{\cal G}(K+Q^\prime)\Gamma_-\right].\nonumber\\
\end{eqnarray}
$\Sigma^{(2B)}$ and $\tilde{\Sigma}^{(2B)}$ are composed of two next-to-leading-order expansions of ${\cal G}_J$. We have
\begin{eqnarray}
\Sigma^{(2B)}_{Q,Q^\prime}&=&\frac{1}{(\beta V)^2}\sum_{K,K^\prime} j^{\rm T}(Q_1)W(Q,Q^\prime;K,K^\prime) j(Q_2),\nonumber\\
\tilde{\Sigma}^{(2B)}_{Q,Q^\prime}&=&\frac{1}{(\beta V)^2}\sum_{K,K^\prime} j^{\rm T}(Q_1)\tilde{W}(Q,Q^\prime;K,K^\prime) j(Q_2).
\end{eqnarray}
The matrices $W$ and $\tilde{W}$ are defined as 
\begin{eqnarray}
W=\left(\begin{array}{cc}W_{nn}&W_{ns}\\ W_{sn}&W_{ss}\end{array}\right),\ \ \ \  
\tilde{W}=\left(\begin{array}{cc}\tilde{W}_{nn}&\tilde{W}_{ns}\\ \tilde{W}_{sn}&\tilde{W}_{ss}\end{array}\right)
\end{eqnarray}
where the elements are given by
\begin{eqnarray}
W_{ab}(Q,Q^\prime;K,K^\prime)&=&{\rm Tr}_{\rm NG}\left[{\cal G}(K)\Gamma_a{\cal G}(K^\prime-Q)\Gamma_-{\cal G}(K^\prime)\Gamma_b{\cal G}(K+Q^\prime)\Gamma_+\right],\nonumber\\
\tilde{W}_{ab}(Q,Q^\prime;K,K^\prime)&=&{\rm Tr}_{\rm NG}\left[{\cal G}(K)\Gamma_a{\cal G}(K^\prime-Q)\Gamma_+{\cal G}(K^\prime)\Gamma_b{\cal G}(K+Q^\prime)\Gamma_-\right].
\end{eqnarray}

Using the expressions of $\Sigma^{(1)}$ and $\Sigma^{(2)}$, the generating functional ${\cal W}_{\rm GF}[j_n,j_s]$ can be expanded as
\begin{eqnarray}
{\cal W}_{\rm GF}[j_n,j_s]={\cal W}_{\rm GF}^{(0)}+{\cal W}_{\rm GF}^{(1)}[j_n,j_s]+{\cal W}_{\rm GF}^{(2)}[j_n,j_s]+\cdots,
\end{eqnarray}
where ${\cal W}_{\rm GF}^{(0)}=\beta V\Omega_{\rm GF}$. The first-order expansion is given by
\begin{eqnarray}\label{W-linear}
{\cal W}_{\rm GF}^{(1)}[j_n,j_s]=\frac{1}{2}\sum_{Q}\left[{\bf D}(Q)\Sigma^{(1)}_{Q,Q}+{\bf D}(-Q)\tilde{\Sigma}^{(1)}_{Q,Q}\right],
\end{eqnarray}
where ${\bf D}(Q)={\bf M}^{-1}(Q)$. It can be explicitly expressed as
\begin{eqnarray}\label{W-linear2}
{\cal W}_{\rm GF}^{(1)}=\sqrt{\beta V}\left[{\cal C}_nj_n(0)+{\cal C}_sj_s(0)\right],
\end{eqnarray}
where the coefficients read
\begin{eqnarray}
{\cal C}_a=-\frac{1}{\beta V}\sum_{Q}{\bf D}(Q)X_a(Q,Q),\ \ \ \ \ a=n,s.
\end{eqnarray}
Since $X_s(Q,Q)=0$, we have  ${\cal C}_s=0$.  It is obvious to identify 
\begin{equation}
{\cal C}_n=\frac{\partial \Omega_{\rm GF}(\mu)}{\partial\mu}=-n_{\rm GF}.
\end{equation}
Therefore, in the normal phase, the linear term does not contribute to the dynamic response. The second-order expansion reads
\begin{eqnarray}
{\cal W}_{\rm GF}^{(2)}[j_n,j_s]={\cal W}_{\rm GF}^{({\rm AL})}+{\cal W}_{\rm GF}^{({\rm SE})}+{\cal W}_{\rm GF}^{({\rm MT})},
\end{eqnarray}
where the three contributions are given by
\begin{eqnarray}
&&{\cal W}_{\rm GF}^{({\rm AL})}[j_n,j_s]=-\frac{1}{4}\sum_{Q,Q^\prime}
\left[{\bf D}(Q)\Sigma^{(1)}_{Q,Q^\prime}{\bf D}(Q^\prime)\Sigma^{(1)}_{Q^\prime,Q}+
{\bf D}(-Q)\tilde{\Sigma}^{(1)}_{Q,Q^\prime}{\bf D}(-Q^\prime)\tilde{\Sigma}^{(1)}_{Q^\prime,Q}\right],\nonumber\\
&&{\cal W}_{\rm GF}^{({\rm SE})}[j_n,j_s]=\frac{1}{2}\sum_{Q}\left[{\bf D}(Q)\Sigma^{(2A)}_{Q,Q}+{\bf D}(-Q)\tilde{\Sigma}^{(2A)}_{Q,Q}\right],\nonumber\\
&&{\cal W}_{\rm GF}^{({\rm MT})}[j_n,j_s]=\frac{1}{2}\sum_{Q}\left[{\bf D}(Q)\Sigma^{(2B)}_{Q,Q}+{\bf D}(-Q)\tilde{\Sigma}^{(2B)}_{Q,Q}\right].
\end{eqnarray}

\subsubsection{Aslamazov-Lakin contribution}

\begin{figure*}[!htb]
\begin{center}
\includegraphics[width=8cm]{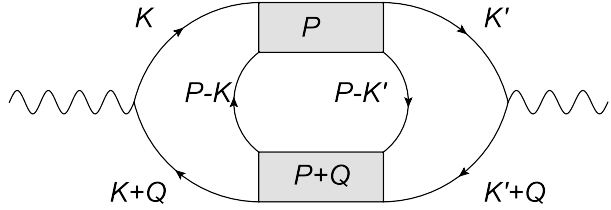}
\caption{Diagrammatic representation of the Aslamazov-Lakin contribution in Eq. (\ref{NSR-AL}).  The solid lines with arrows denote the fermion propagators,
the gray boxes denote the collective mode propagators, and the wave lines represent the external sources. \label{fig2}}
\end{center}
\end{figure*}

The Aslamazov-Lakin contribution in the normal phase is given by
\begin{eqnarray}
{\cal W}_{\rm GF}^{({\rm AL})}=\frac{1}{2}\sum_Q\left(\begin{array}{cccc} j_n(-Q)&j_s(-Q)\end{array}\right)
\left(\begin{array}{cc}\Xi_{nn}^{\rm AL}(Q)&\Xi_{ns}^{\rm AL}(Q)\\
\Xi_{sn}^{\rm AL}(Q)&\Xi_{ss}^{\rm AL}(Q)\end{array}\right)
\left(\begin{array}{c} j_n(Q) \\ j_s(Q)\end{array}\right).
\end{eqnarray}
Since $X_s(Q,Q^\prime)=\tilde{X}_s(Q,Q^\prime)=0$, it is easy to show that
\begin{eqnarray}
\Xi_{ns}^{\rm AL}(Q)=\Xi_{sn}^{\rm AL}(Q)=\Xi_{ss}^{\rm AL}(Q)=0.
\end{eqnarray}
Therefore, the spin response has no Aslamazov-Lakin contribution. The Aslamazov-Lakin  contribution to the density response is given by
\begin{eqnarray}\label{NSR-AL}
\Xi_{nn}^{\rm AL}(Q)=-\frac{4}{\beta V}\sum_P{\bf D}(P){\bf D}(P+Q)\left[\frac{1}{\beta V}\sum_{K}{\cal G}_0(K){\cal G}_0(K+Q){\cal G}_0(P-K)\right]^2
\end{eqnarray}
The Aslamazov-Lakin contribution can be diagrammatically represented in Fig. \ref{fig2}.

\subsubsection{Self-Energy contribution}
\begin{figure*}[!htb]
\begin{center}
\includegraphics[width=8cm]{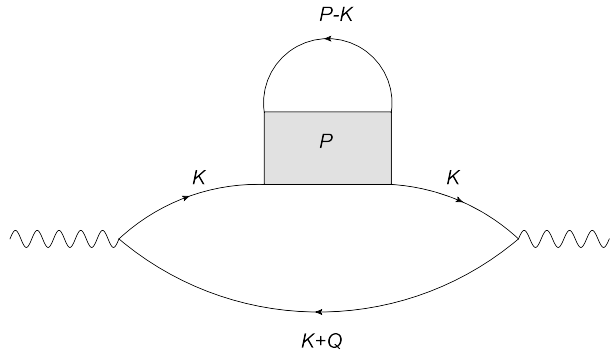}
\caption{Diagrammatic representation of the Self-Energy contribution in Eq. (\ref{NSR-SE}). \label{fig3}}
\end{center}
\end{figure*}

The Self-Energy contribution in the normal phase is given by
\begin{eqnarray}
{\cal W}_{\rm GF}^{({\rm SE})}=\frac{1}{2}\sum_Q\left(\begin{array}{cccc} j_n(-Q)&j_s(-Q)\end{array}\right)
\left(\begin{array}{cc}\Xi_{nn}^{\rm SE}(Q)&\Xi_{ns}^{\rm SE}(Q)\\
\Xi_{sn}^{\rm SE}(Q)&\Xi_{ss}^{\rm SE}(Q)\end{array}\right)
\left(\begin{array}{c} j_n(Q) \\ j_s(Q)\end{array}\right),
\end{eqnarray}
where the function $\Xi_{ab}^{\rm SE}(Q)$ ($a,b=n,s$) is given by
\begin{eqnarray}
\Xi_{ab}^{\rm SE}(Q)&=&\frac{1}{\beta V}\sum_{P}{\bf D}(P)\left[{\bf Y}_{ab}(P,Q)+{\bf Z}_{ab}(P,Q)\right]\nonumber\\
&+&\frac{1}{\beta V}\sum_{P}{\bf D}(-P)\left[\tilde{\bf Y}_{ab}(P,Q)+\tilde{\bf Z}_{ab}(P,Q)\right].
\end{eqnarray}
Here the matrices ${\bf Y}_{ab}$, ${\bf Z}_{ab}$,  $\tilde{\bf Y}_{ab}$, and $\tilde{\bf Z}_{ab}$ are given by
\begin{eqnarray}
{\bf Y}_{ab}(P,Q)&=&\frac{1}{\beta V}\sum_{K}{\rm Tr}_{\rm NG}\left[{\cal G}(K-P)\Gamma_-{\cal G}(K)\Gamma_a{\cal G}(K+Q)\Gamma_b{\cal G}(K)\Gamma_+\right],\nonumber\\
{\bf Z}_{ab}(P,Q)&=&\frac{1}{\beta V}\sum_{K}{\rm Tr}_{\rm NG}\left[{\cal G}(K)\Gamma_a{\cal G}(K+Q)\Gamma_b{\cal G}(K)
\Gamma_-{\cal G}(K+P)\Gamma_+\right],\nonumber\\
\tilde{\bf Y}_{ab}(P,Q)&=&\frac{1}{\beta V}\sum_{K}{\rm Tr}_{\rm NG}\left[{\cal G}(K-P)\Gamma_+{\cal G}(K)\Gamma_a{\cal G}(K+Q)\Gamma_b{\cal G}(K)\Gamma_-\right],\nonumber\\
\tilde{\bf Z}_{ab}(P,Q)&=&\frac{1}{\beta V}\sum_{K}{\rm Tr}_{\rm NG}\left[{\cal G}(K)\Gamma_a{\cal G}(K+Q)\Gamma_b{\cal G}(K)
\Gamma_+{\cal G}(K+P)\Gamma_-\right].
\end{eqnarray}
Completing the trace in the Nambu-Gor'kov space and using the fact ${\cal G}_\downarrow(K)=-{\cal G}_\uparrow(-K)$, we can show that
\begin{eqnarray}
\Xi_{ns}^{\rm SE}(Q)=\Xi_{sn}^{\rm SE}(Q)=0.
\end{eqnarray}
The density and spin responses have equal Self-Energy contributions, which are explicitly given by
\begin{eqnarray}\label{NSR-SE}
&&\Xi_{nn}^{\rm SE}(Q)=\Xi_{ss}^{\rm SE}(Q)\nonumber\\
&=&-\frac{4}{\beta V}\sum_P{\bf D}(P)\left[\frac{1}{\beta V}\sum_{K}{\cal G}_0(K+Q){\cal G}_0(K){\cal G}_0(P-K){\cal G}_0(K)\right].
\end{eqnarray}
The Self-Energy contributions can be diagrammatically represented in Fig. \ref{fig3}.

\subsubsection{Maki-Thompson contribution}
\begin{figure*}[!htb]
\begin{center}
\includegraphics[width=8cm]{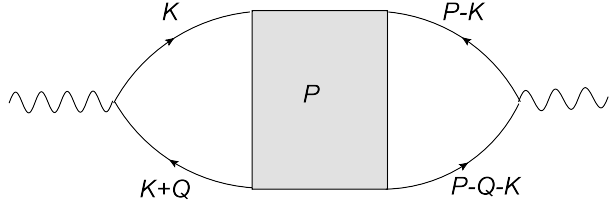}
\caption{Diagrammatic representation of the Maki-Thompson contribution in Eq. (\ref{NSR-MT}). \label{fig4}}
\end{center}
\end{figure*}

The Maki-Thompson contribution in the normal phase is given by
\begin{eqnarray}
{\cal W}_{\rm GF}^{({\rm MT})}=\frac{1}{2}\sum_Q\left(\begin{array}{cccc} j_n(-Q)&j_s(-Q)\end{array}\right)
\left(\begin{array}{cc}\Xi_{nn}^{\rm MT}(Q)&\Xi_{ns}^{\rm MT}(Q)\\
\Xi_{sn}^{\rm MT}(Q)&\Xi_{ss}^{\rm MT}(Q)\end{array}\right)
\left(\begin{array}{c} j_n(Q) \\ j_s(Q)\end{array}\right),
\end{eqnarray}
where the function $\Xi_{ab}^{\rm MT}(Q)$ ($a,b=n,s$) is given by
\begin{eqnarray}
\Xi_{ab}^{\rm MT}(Q)=\frac{1}{\beta V}\sum_{P}\left[{\bf D}(P){\bf W}_{ab}(P,Q)+{\bf D}(-P)\tilde{\bf W}_{ab}(P,Q)\right].
\end{eqnarray}
Here the matrices ${\bf W}_{ab}$ and $\tilde{\bf W}_{ab}$ are given by
\begin{eqnarray}
{\bf W}_{ab}(P,Q)&=&\frac{1}{\beta V}\sum_{K}{\rm Tr}_{\rm NG}\left[{\cal G}(K)\Gamma_a{\cal G}(K+Q)\Gamma_-
{\cal G}(K+P+Q)\Gamma_b{\cal G}(K+P)\Gamma_+\right],\nonumber\\
\tilde{\bf W}_{ab}(P,Q)&=&\frac{1}{\beta V}\sum_{K}{\rm Tr}_{\rm NG}\left[{\cal G}(K)\Gamma_a{\cal G}(K+Q)\Gamma_+
{\cal G}(K+P+Q)\Gamma_b{\cal G}(K+P)\Gamma_-\right].\nonumber\\
\end{eqnarray}
Completing the trace in the Nambu-Gor'kov space and using the fact ${\cal G}_\downarrow(K)=-{\cal G}_\uparrow(-K)$, we can show that
\begin{eqnarray}
\Xi_{ns}^{\rm MT}(Q)=\Xi_{sn}^{\rm MT}(Q)=0.
\end{eqnarray}
The density and spin responses have unequal Maki-Thompson contributions, which are explicitly given by
\begin{eqnarray}\label{NSR-MT}
\Xi_{nn}^{\rm MT}(Q)&=&-\frac{2}{\beta V}\sum_P{\bf D}(P)\left[\frac{1}{\beta V}\sum_{K}{\cal G}_0(K+Q){\cal G}_0(K){\cal G}_0(P-K){\cal G}_0(P-Q-K)\right],\nonumber\\
\Xi_{ss}^{\rm MT}(Q)&=&\frac{2}{\beta V}\sum_P{\bf D}(P)\left[\frac{1}{\beta V}\sum_{K}{\cal G}_0(K+Q){\cal G}_0(K){\cal G}_0(P-K){\cal G}_0(P-Q-K)\right].\nonumber\\
\end{eqnarray}
The Maki-Thompson contributions can be diagrammatically represented in Fig. \ref{fig4}.

In summary,  within the NSR theory, the dynamic density and spin response functions are given by
\begin{eqnarray}
\chi_{nn}(Q)&=&\Pi_{nn}(Q)+\Xi_{nn}^{\rm AL}(Q)+\Xi_{nn}^{\rm SE}(Q)+\Xi_{nn}^{\rm MT}(Q),\nonumber\\
\chi_{ss}(Q)&=&\Pi_{ss}(Q)+\Xi_{ss}^{\rm SE}(Q)+\Xi_{ss}^{\rm MT}(Q).
\end{eqnarray}
It is obvious that the dynamic density and spin response functions becomes different when we include the Gaussian pair fluctuations. 

\section{Summary and outlook}\label{s7}
In summary, we have presented a standard field theoretical derivation of the dynamic density and spin response functions of a dilute superfluid Fermi gas in the BCS-BEC crossover. The functional path integral approach provides an elegant way to study the dynamic responses in both the BCS-Leggett mean-field theory
and the GPF theory.  In the mean-field theory, our results of the response functions agree with the known results from the random phase approximation.  We have established a theoretical framework for the dynamic responses in the GPF theory.  We show that the GPF response theory naturally recover three kinds of famous  diagrammatic contributions: the Self-Energy contribution,  the Aslamazov-Lakin contribution, and the Maki-Thompson contribution. In the superfluid state, there is an additional order parameter induced contribution which ensures that in the static and long wavelength limit, the density response function recovers the result of the static compressibility (the compressibility sum rule as pointed out in \cite{SUM-2,SUM-3}).  

An important issue which has not been solved in this work is the $f$-sum rule. It is interesting to verify in the future that the $f$-sum rule is manifested by the full number equation which includes the contribution from the Gaussian pair fluctuations; i.e.,
\begin{eqnarray}
\int_0^\infty d\omega \omega S_{nn}^{\rm GF}(\omega,q)=\int_0^\infty d\omega \omega S_{ss}^{\rm GF}(\omega,q)=\frac{n_{\rm GF}q^2}{2m}.
\end{eqnarray}
It has been shown that the $f$-sum rule is manifested by the gauge invariance in the BCS-Leggett mean-field theory \cite{SUM-1}.  Recently it was also shown that the gauge invariance is generally satisfied in the functional path integral approach, including the GPF response theory established in this work. Therefore, we expect that the $f$-sum rule is precisely satisfied in the GPF response theory. The explicit proof will be published elsewhere.

The dynamic structure factors for the density and the spin for a resonantly interacting Fermi gas has been experimentally measured  by using Bragg spectroscopy~\cite{Bragg-1,Bragg-2,Bragg-3}. The static structure factors has been calculated by using quantum Monte Carlo simulations~\cite{SSF-QMC01,SSF-QMC02}. The dynamic structure factors may also be calculated by using the quantum Monte Carlo simulations in the future. Therefore, it is interesting to perform numerical calculations of the dynamic and static structure factors and compare our theoretical results with experimental measurements and quantum Monte Carlo results. Our theory could also be applied to other strongly interacting systems, such as dense QCD matter \cite{BCSBEC-NM01,BCSBEC-NM02,BCSBEC-QM01,BCSBEC-QM02,BCSBEC-QM03,BCSBEC-QM04,BCSBEC-QM05,BCSBEC-QM06,BCSBEC-QM07,BCSBEC-QM08,BCSBEC-QM09} and spin-orbit coupled atomic Fermi gases \cite{SOC-01,SOC-02,SOC-03,SOC-04,SOC-05,SOC-06,SOC-07,SOC-08,SOC-09,SOC-10,SOC-11,SOC-12,SOC-13,SOC-14,SOC-15}.

\section*{Acknowledgments} The author thanks Joseph Carlson and Stefano Gandolfi for guiding him to the topic of the dynamic density and spin density responses.  He also thanks Hui Hu for useful discussions. The Feynman diagrams were plotted by Yin Jiang. The work is supported by the US Department of Energy Nuclear Physics Office, Los Alamos National Laboratory, Tsinghua University, and Thousand Young Talent Program in China.

\bibliographystyle{model1-num-names}

\end{document}